\documentclass[twocolumn]{aastex}
\usepackage{comment}
\usepackage{graphicx,floatrow,amsmath,gensymb,rotating}
\usepackage{subfigure}
\usepackage{natbib}
\usepackage{xcolor}
\usepackage{threeparttable,booktabs}
\usepackage{etoolbox}

\begin{document}

\title{The Active Galactic Nuclei in the Hobby-Eberly Telescope Dark Energy Experiment Survey (HETDEX) I. Sample Selection}

\author[0000-0001-5561-2010]{Chenxu Liu}
\affiliation{Department of Astronomy, The University of Texas at Austin, 2515 Speedway Boulevard, Austin, TX 78712, USA}
\correspondingauthor{Chenxu Liu}
\email{lorenaustc@gmail.com}

\author[0000-0002-8433-8185]{Karl Gebhardt}
\affiliation{Department of Astronomy, The University of Texas at Austin, 2515 Speedway Boulevard, Austin, TX 78712, USA}

\author[0000-0002-2307-0146]{Erin Mentuch Cooper}
\affiliation{Department of Astronomy, The University of Texas at Austin, 2515 Speedway Boulevard, Austin, TX 78712, USA}
\affiliation{McDonald Observatory, The University of Texas at Austin, 2515 Speedway Boulevard, Austin, TX 78712, USA}

\author[0000-0002-8925-9769]{Dustin Davis}
\affiliation{Department of Astronomy, The University of Texas at Austin, 2515 Speedway Boulevard, Austin, TX 78712, USA}

\author{Donald P. Schneider}
\affiliation{Department of Astronomy \& Astrophysics, The Pennsylvania State University, University Park, PA 16802, USA}
\affiliation{Institute for Gravitation and the Cosmos, The Pennsylvania State University, University Park, PA 16802, USA}

\author[0000-0002-1328-0211]{Robin Ciardullo}
\affiliation{Department of Astronomy \& Astrophysics, The Pennsylvania State University, University Park, PA 16802, USA}
\affiliation{Institute for Gravitation and the Cosmos, The Pennsylvania State University, University Park, PA 16802, USA}

\author[0000-0003-2575-0652]{Daniel J. Farrow}  
\affiliation{Max-Planck Institut f\"ur extraterrestrische Physik, Giessenbachstrasse 1, 85748 Garching, Germany}
\affiliation{University Observatory, Fakult\"at f\"ur Physik, Ludwig-Maximilians University Munich, Scheiner Strasse 1, 81679 Munich, Germany}

\author[0000-0001-8519-1130]{Steven L. Finkelstein}
\affiliation{Department of Astronomy, The University of Texas at Austin, 2515 Speedway Boulevard, Austin, TX 78712, USA}

\author{Caryl Gronwall}
\affiliation{Department of Astronomy \& Astrophysics, The Pennsylvania State University, University Park, PA 16802, USA}
\affiliation{Institute for Gravitation and the Cosmos, The Pennsylvania State University, University Park, PA 16802, USA}

\author{Yuchen Guo}
\affiliation{Department of Astronomy, The University of Texas at Austin, 2515 Speedway Boulevard, Austin, TX 78712, USA}

\author[0000-0001-6717-7685]{Gary J. Hill}
\affiliation{McDonald Observatory, The University of Texas at Austin, 2515 Speedway Boulevard, Austin, TX 78712, USA}
\affiliation{Department of Astronomy, The University of Texas at Austin, 2515 Speedway Boulevard, Austin, TX 78712, USA}

\author{Lindsay House}
\affiliation{Department of Astronomy, The University of Texas at Austin, 2515 Speedway Boulevard, Austin, TX 78712, USA}

\author[0000-0002-8434-979X]{Donghui Jeong}
\affiliation{Department of Astronomy \& Astrophysics, The Pennsylvania State University, University Park, PA 16802, USA}
\affiliation{Institute for Gravitation and the Cosmos, The Pennsylvania State University, University Park, PA 16802, USA}

\author{Shardha Jogee}
\affiliation{Department of Astronomy, The University of Texas at Austin, 2515 Speedway Boulevard, Austin, TX 78712, USA}

\author{Wolfram Kollatschny}
\affiliation{Institut f\"ur Astrophysik, Universit\"at G\"ottingen, Friedrich-Hund Platz 1, 37077 G\"ottingen, Germany}

\author{Mirko Krumpe}
\affiliation{Leibniz-Institut für Astrophysik Potsdam (AIP) An der Sternwarte 16, 14482 Potsdam, Germany}

\author[0000-0003-1838-8528]{Martin Landriau}
\affiliation{Lawrence Berkeley National Laboratory, 1 Cyclotron Road, Berkeley, CA 94720, USA}

\author{Oscar A Chavez Ortiz}
\affiliation{Department of Astronomy, The University of Texas at Austin, 2515 Speedway Boulevard, Austin, TX 78712, USA}

\author[0000-0003-3817-8739]{Yechi Zhang}
\affiliation{Institute for Cosmic Ray Research, The University of Tokyo, 5-1-5 Kashiwanoha, Kashiwa, Chiba 277-8582, Japan}
\affiliation{Department of Astronomy, Graduate School of Science, the University of Tokyo, 7-3-1 Hongo, Bunkyo, Tokyo 113-0033, Japan}

\author{(The HETDEX Collaboration)}

\begin{abstract}

We present the first Active Galactic Nuclei (AGN) catalog of the Hobby-Eberly Telescope Dark Energy Experiment Survey (HETDEX) observed between January 2017 and June 2020. HETDEX is an ongoing spectroscopic survey (3500\,\AA\ - 5500\,\AA) with no target pre-selection based on magnitudes, colors or morphologies, enabling us to select AGN based solely on their spectral features. Both luminous quasars and low-luminosity Seyferts are found in our catalog. AGN candidates are selected with at least two significant AGN emission lines, such as the $\rm Ly\alpha$ and \ion{C}{4}
$\lambda1549$ line pair, or with a single broad emission line with FWHM$>$1000~km\,s$^{-1}$. Each source is further confirmed by visual inspections. This catalog contains 5,322 AGN, covering an effective sky coverage of 30.61 deg$^2$. A total of 3,733 of these AGN have secure redshifts, and we provide redshift estimates for the remaining 1,589 single broad-line AGN with no cross matched spectral redshifts from SDSS DR14Q\null. The redshift range of the AGN catalog is $0.25<z<4.32$, with a median of $z=2.1$. The bolometric luminosity range is $10^{9}-10^{14}\ L_\sun$ with a median of $10^{12}\ L_\sun$. The median $r$-band magnitude of our AGN catalog is 21.6 mag, with 34\% having $r>22.5$, and 2.6\% reaching the detection limit at $r\sim26$ mag of the deepest imaging surveys we searched. We also provide a composite spectrum of the AGN sample covering 700\,\AA\ - 4400\,\AA\null. 

\end{abstract}


\section{Introduction}
\label{sec_intro}

Active Galactic Nuclei (AGN) are most active at redshifts around $z \sim 2$ \citep[e.g.][and references therein]{Kormendy2013}, the so-called ``cosmic noon''. It is important to compile AGN catalogs with big sky coverage, a variety of AGN populations covering a wide range of luminosity, and big sample sizes to statistically study the AGN at cosmic noon and understand their evolution at early stages. The observations and identifications of AGN at $z>1$ has been revolutionized significantly in the past decade from space-based photometric surveys covering a few square degrees and tens of AGN gradually to ground-based spectroscopic surveys over thousands of square degrees with tens of thousands of AGN.

Observations of galaxies by space-based deep optical/near-infared (NIR) surveys, such as the Great Observatories Origins Deep Survey (\citealp[GOODS;][]{Giavalisco2004}) covering 0.09 deg$^2$, the Cosmic Evolution Survey (\citealp[COSMOS;][]{Scoville2007}) covering 2 deg$^2$, and the Cosmic Assembly Near-infrared Deep Extragalactic Legacy Survey (\citealp[CANDELS;][]{Grogin2011,Koekemoer2011}) covering 0.26 deg$^2$, help to shed light on the AGN population of the high-$z$ universe. These surveys 
usually overlap with multi-band imaging from the X-ray through to the infrared and radio, and/or with ground-based spectroscopy,  allowing us to identify and study a few hundreds of AGN down to $\sim$ 28 mag at $0<z<5$ (\citealp[e.g.][]{Giavalisco2004, Scoville2007,Grogin2011,Koekemoer2011}). 

Searching for X-ray excess with space telescopes is a straightforward way of finding AGN, especially the low-luminosity AGN, since X-rays have strong penetrability and are less affected by the host galaxies. The Chandra Deep Field-South (\citealp[CDFS;][]{Xue2011,Luo2017}, 0.135 deg$^2$) and the Chandra Deep Field-North (\citealp[CDFN;][]{Alexander2003,Xue2016}, 0.124 deg$^2$) are the deepest X-ray observations with about a thousand AGN candidate detections
between $0.1<z<5.2$. AGN are mainly identified from the X-ray sources as the ones with high X-ray luminosity, or X-ray excess compared to optical/near-infrared/radio band(s).

Space-based observations are deep enough to contain AGN spanning a large range of luminosity. The multi-band observations provide a good tool to help identify and study the SED properties of AGN\null. However, these surveys have small sky coverage ($< 1$ to a few deg$^2$) 
and overlap spatially, limiting the number of AGN identified.  The statistics based on a small number of AGN of a small sky coverage may not represent the full AGN population at high redshifts very well.


Ground-based narrow-band (NB) imaging provides a less expensive and more efficient way of searching for Ly$\alpha$ emitters (LAEs) over a large sky coverage at the cosmic noon. For example, Classifying Objects by Medium-Band Observations in 17 Filters (\citealp[COMBO-17;][]{Wolf2003}) was a pioneering ground-based medium-Band ($\sim 10\,000$~km\,s$^{-1}$) survey for high redshift galaxies. This survey identified 192 AGN over 0.78 deg$^2$ based on the 
medium-band photometry at $1<z<5$ down to $R=24$. The Javalambre Photometric Local Universe Survey (\citealp[J-PLUS;][]{Cenarro2019,Spinoso2020}) selected 14,500 L$_{Ly\alpha}>10^{43.3}$ erg\,s$^{-1}$ bright LAEs across 1000 deg$^2$ with multiple NB filters within four redshifts windows ($\Delta z<0.16$) at $2.2\lesssim z\lesssim3.3$. Their spectroscopic follow-up on a small sub-sample shows that 64\% of their bright LAEs are 
actually $z\sim 2.2$ quasars (QSOs). A major weakness of such NB quasar surveys is that their redshifts are not as reliable. A NB set of Ly$\alpha$ emitting AGN can be contaminated by many other emission lines, such as the \ion{C}{4} $\lambda1549$, \ion{C}{3}] $\lambda1909$, \ion{Mg}{2} $\lambda2799$ lines of foreground quasars, and the emission of [\ion{O}{2}] $\lambda3727$ of low-$z$ star-forming galaxies. Additionally, the redshift coverage is limited by the filters widths, making the redshifts limited to a few narrow redshift intervals. 

There are also many spectroscopic surveys studying AGN at cosmic noon. The SDSS-III Baryon Oscillation Spectroscopic Survey (\citealp[BOSS;][]{Eisenstei2011,Ross2012,Dawson2013,Paris2018}) is the largest to date. Their latest data release of the QSO catalog \citep[DR14Q;][]{Paris2018} consists 526,356 QSOs down to $g\sim22$, spanning over 9376 deg$^2$. The targeting of the spectroscopic observations of QSOs are pre-selected based on photometric observations \citep{Ross2012}. QSO candidates are first distinguished from extended nearby galaxies by their point-like morphologies, and are distinguished from stars using colors to characterize their continuum shape at different redshifts \citep{Ross2012}. The QSOs are confirmed by spectroscopic follow-up. 

The pre-selection based on imaging is a traditional method of spectroscopic surveys of QSOs, but it also brings in selection effects that could potentially miss some AGN populations. Seyfert galaxies, whose continuum shape is affected by their host galaxies, can have very different colors compared to those of bright QSOs. Some optical continuum faint AGN might have strong enough emission lines to overcome the obscuration from dust and be detected as optical emission-line AGN. The imaging pre-selection can potentially miss some very extreme objects such as the naked black holes, which are not hosted by galaxies at all \citep{Bahcall1994,Bahcall1997,Loeb2007,Haiman2009}. They could be formed as they were, or they could be ejected black holes by merger events.


The VIMOS VLT Deep Survey (\citealp[VVDS;][]{Gavignaud2006,LeFevre2013}) is a pure magnitude-limited survey without morphological or color pre-selections. The survey used the VIsible Multi-Object Spectrograph (VIMOS) installed on the European Southern Observatory Very Large Telescope. Their final data release consists 422 type-I AGN, identified with the existence of broad emission lines, covering the redshift range of $0<z<6.7$, spanning over 8.7 deg$^2$ down to $i_{\rm AB}=24.75$. It provides both broadband imaging and spectra covering 3600\,\AA$<\lambda<$10 000\,\AA. The AGN in the VVDS survey provide the faint-end of the AGN luminosity function (LF) \citep{Bongiorno2007A&A}.

The Hobby-Eberly Telescope Dark Energy Experiment Survey (HETDEX) is a wide-area untargeted spectroscopic survey using the Integral Field Unit (IFU) technique. It allows the identification of an AGN catalog free of morphological, color, and the magnitude pre-selections based on imaging. We will introduce the technical details of HETDEX in Section \ref{sec_hetdex}. In this paper, we will release the first 5k AGN catalog in HETDEX observed from January 2017 to June 2020 with an effective coverage of 30.61 deg$^2$. 

\section{The HETDEX Survey}
\label{sec_hetdex}
HETDEX \citep{Hill08,Gebhardt2021} is a spectroscopic survey without target pre-selection on the 10-m Hobby-Eberly Telescope (HET, \citealt{Ramsey98,Hill2021}). It uses the Visible Integral-field Replicable Unit Spectrograph (\citealp[VIRUS;][]{Hill2018,Hill2021}) to record spectra for all targets within its field of view (FoV)\null. At completion, VIRUS is a collection of 78 IFUs based on a common design, with $\sim$35,000 fibers feeding into 156 spectrographs. Each IFU consists of 448 fibers, covering 51 arcsec$\times$51 arcsec, with the space between fibers filled in via a three position dither set. The 78 IFUs are arranged in an array with a 100 arcsec\ spacing over the 18\arcmin\ focal surface, resulting in a 1$/$4.5 fill-factor.  Each fiber is 1.5 arcsec\ in diameter and feeds a spectrograph covering the wavelength range of 3500\,\AA\ - 5500\,\AA, and a resolution of $R\sim800$. Each pointing takes three dither exposures of 6 minutes per exposure. 

The data processing of HETDEX frames is detailed in \citet{Gebhardt2021}.  Briefly, bias frames, pixel flats, twilight sky flats, and the background on the science frames themselves are used to produce a wavelength calibrated, sky-subtracted spectrum for each fiber in the array.  Astrometric calibrations are achieved by measuring the centroid of each field star from fiber counts between 4400\,\AA\ and 5200\,\AA\ and comparing their IFU positions to the stars’ equatorial coordinates in the Sloan Digital Sky Survey \citep[SDSS;][]{York2000, Abazajian2009} and \textit{Gaia} \citep{Gaia2018} catalogs.  This process typically results in global solutions which are good to $\sim 0\farcs 2$ with the exact precision of a measurement dependent on several factors including the number of IFUs in operation.  The absolute flux calibrations for the HETDEX observations are produced by using $g < 24$ SDSS field stars as \textit{in situ} standards and using their $ugriz$ colors \citep{Padmanabhan2008}, \textit{Gaia} parallaxes \citep{Gaia2018}, and foreground reddenings \citep{Schlafly2011} to find the their most likely spectral energy distribution in a grid of model spectra \citep{Cenarro2007, Falcon-Barroso2011}.  The final system throughput curve is then derived from the most likely flux distribution of $\sim 20$ stars, and is generally good to $\sim 5\%$.

The HETDEX survey is expected to be active from 2017 to 2024, and eventually will cover 540 deg$^2$, with a filling factor of 1/4.5. The final effective sky coverage is expected to be about 90 deg$^2$. Over a million  $1.9<z<3.5$ LAEs and another million $z < 0.5$
[\ion{O}{2}] emitters are expected to be observed. As of 2020-06-26, there were 71 IFUs mounted on the telescope, and the total effective sky coverage was 30.61 deg$^2$. All AGN are selected solely based on their spectral features. Therefore, our catalog provides a census, free from pre-selection bias, of AGN including low-luminosity Seyferts, type-II AGN, and QSOs that do not satisfy the traditional photometric color selections.


We summarize the detection catalog of HETDEX in Section \ref{sec_detection}. In Section \ref{sec_selection}, we describe the selection of AGN from the detection catalog in detail. In Section \ref{sec_completeness}, we discuss the incompleteness of the HETDEX AGN catalog. We present the catalog in Section \ref{sec_catalog}. In Section \ref{sec_statistics}, we show the statistics of the HETDEX AGN catalog, include its distribution of the redshifts, photometric properties, and the composite spectrum of its AGN\null. We summarize this paper in Section \ref{sec_summary}.
We use a flat $\rm\Lambda$CDM cosmology with $H_0=\rm 70\ km\ s^{-1}\ Mpc^{-1}$, $\rm \Omega_M=0.3$, $\rm \Omega_\Lambda=0.7$ throughout this paper. 


\section{HETDEX source detection methods}
\label{sec_detection}

We refer the readers for details in the survey design, current status, and the detection algorithm to \cite{Gebhardt2021}. Here we briefly summarize the current status of the HETDEX survey and the parent sample of our AGN catalog.  

The survey began with 16 active IFUs on 2017-01-01. By 2020-06-26, there were 71 IFUs actively working on the telescope, and as of 2021-09-01, the full array 78 IFUs were actively producing data. As of 2020-06-26, the end date of this catalog, the effective sky coverage of HETDEX was 30.61 deg$^2$ (see Table \ref{t_survey}). The typical seeing is 1.8 arcsec. A full catalog of sources will be described in Mentuch Cooper et al. (in preparation). 

\begin{table}[htbp]
\centering

\begin{tabular}{c|c|c|c}
\hline\hline
Field Name          &  Field center (RA, DEC)  & Area    & N(AGN) \\
           &    deg                   & deg$^2$ &        \\\hline
DEX-spring &  (199.305607, 51.627824) & 20.69   & 3,763  \\
 DEX-fall  &  (22.499361,   0.001176) &  7.85   & 1,292  \\\hline
   COSMOS  &  (150.186756,  2.287567) &  0.33   &    25  \\
       EGS &  (214.855923, 52.849597) &  0.37   &    46  \\
   GOODS-N &  (189.052168, 62.204371) &  0.21   &    18  \\\hline
       NEP &  (271.404386, 65.014922) &  1.09   &   177  \\
     other &  (187.276667, 2.052322)  &  0.07   &     1  \\ \hline
     total &        --                & 30.61   & 5,322  \\
\hline\hline
\end{tabular}
\caption{Summary of the survey areas of the AGN catalog}
\begin{tablenotes}[flushleft]
\scriptsize
\item Note: DEX-spring and DEX-fall fields are HETDEX survey fields.
COSMOS, EGS, and GOODS-N fields are HETDEX science verification fields.
The NEP field (Chavez Ortiz et al. in preparation) and the other field are taken for our collaborators with their own scientific purposes.
\end{tablenotes}
\label{t_survey}
\end{table}
Objects which fall onto the VIRUS IFUs are detected automatically without any pre-selection via the HETDEX detection pipeline \citep{Gebhardt2021}. This pipeline  consists of two detection algorithms, one for identification of emission lines and the other for the detection of continuum sources. Emission line detection is optimized to find low signal-to-noise single-wavelength features, while the continuum source detection is optimized for the identification of broad-band spectral emission. Both methods are performed on 1D point-source extracted spectra and do not consider whether the source is extended or point-source like. Further details on each method is found in Sections \ref{sec_lines} and \ref{sec_conts} below.



\subsection{The emission line detection database}
\label{sec_lines}

The data reduction pipeline of HETDEX produces the direct outputs as 3-dimensional data cubes: 2 dimensions in the spatial direction and 1 dimension in the wavelength direction. The detection algorithm of the emission-line catalog searches emission-line signals in the 3 dimensional data base with 3-D grids.
There are two steps in the grid search. The initial grid search uses 3$\times$3 spatial grids with the step of $0\farcs 5$, and the wavelength grids are in steps of $\rm 8$ \AA\ with windows size of $\rm \pm 50$ \AA\ from 3500\,\AA\ to 5500\,\AA\null. The local 1-D spectrum of each grid element is then fitted with a Gaussian profile with an initial $\sigma$ of 2.2\,\AA\null. Emission-line candidates are the elements with $\rm S/N>4$ in the initial grid search. A second fine grid search, using a $5\times 5$ raster with $0\farcs 15$ spatial steps, is then executed around the emission-line candidates selected in the initial search. This allows the coarse $0\farcs 75 \times 0\farcs 75$ positions of the detections found in the initial search to be refined, centering on the highest signal-to-noise ratio. The final raw emission-line catalog provides 1.5 million emission-line candidate detections down to S/N$>$4.5 with their RA, DEC, wavelengths reported. Each emission-line detection is also provided with the spectrum from 3500\,\AA\ to 5500\,\AA\ \citep{Gebhardt2021}. 
The threshold of S/N$>$4.5 is chosen as a balance between the false positive rate and the completeness of detections. For HETDEX, we currently use S/N>4.8, as that allows to reach specification of an integrated false positive rate of 10\%. The initital catalog for detections include sources down to S/N=4.5, and we are in the process of pushing down in our cut of 4.8 while maintaining the specification on false positive rate. For this paper, we push to lower S/N by incorporating a visual classification. We carried out simulations to estimate the completeness and the false positive rate with different S/N cuts. The simulations are first introduced in Section 8 of \cite{Gebhardt2021}, and will be further presented with more details  in Farrow et al. in preparation.

\subsection{The continuum source detection catalog}
\label{sec_conts}

The emission line detection pipeline excludes source spectra above a counts threshold of 50 electron counts. These sources are placed in the separate, continuum source detection catalog. For each spectrum in a field, the detector counts are measured in two windows -- a blue window (3700\,\AA\ - 3900\,\AA) and a red window (5100\,\AA\ - 5300\,\AA\null). Continuum candidates are then selected as detections with either window containing more than 50 electron counts. This cut roughly corresponds to the level of $g\sim22.5$. Similar to the position determinations of the emission line catalog, the locations of continuum source candidates are then found using a grid search within a $15\times 15$ element raster and $0\farcs 1$ spatial bins. The spatial location reported in the catalog is the location of the element that achieves the lowest $\chi^2$ fit to the point spread function (PSF) model of the center of each source, the accuracy of which is about 0.5 arcsec. All candidates in the continuum catalog are reported with their RA, DEC coordinates and their spectra from 3500 \AA\ to 5500 \AA. The raw continuum catalog consists 80,000 source candidates, most of which are stars, though there are other types of luminous objects found in the raw continuum catalog including bright nearby galaxies, meteors, QSOs, etc.

\begin{figure}[htbp]
\centering
\includegraphics[width=\textwidth]{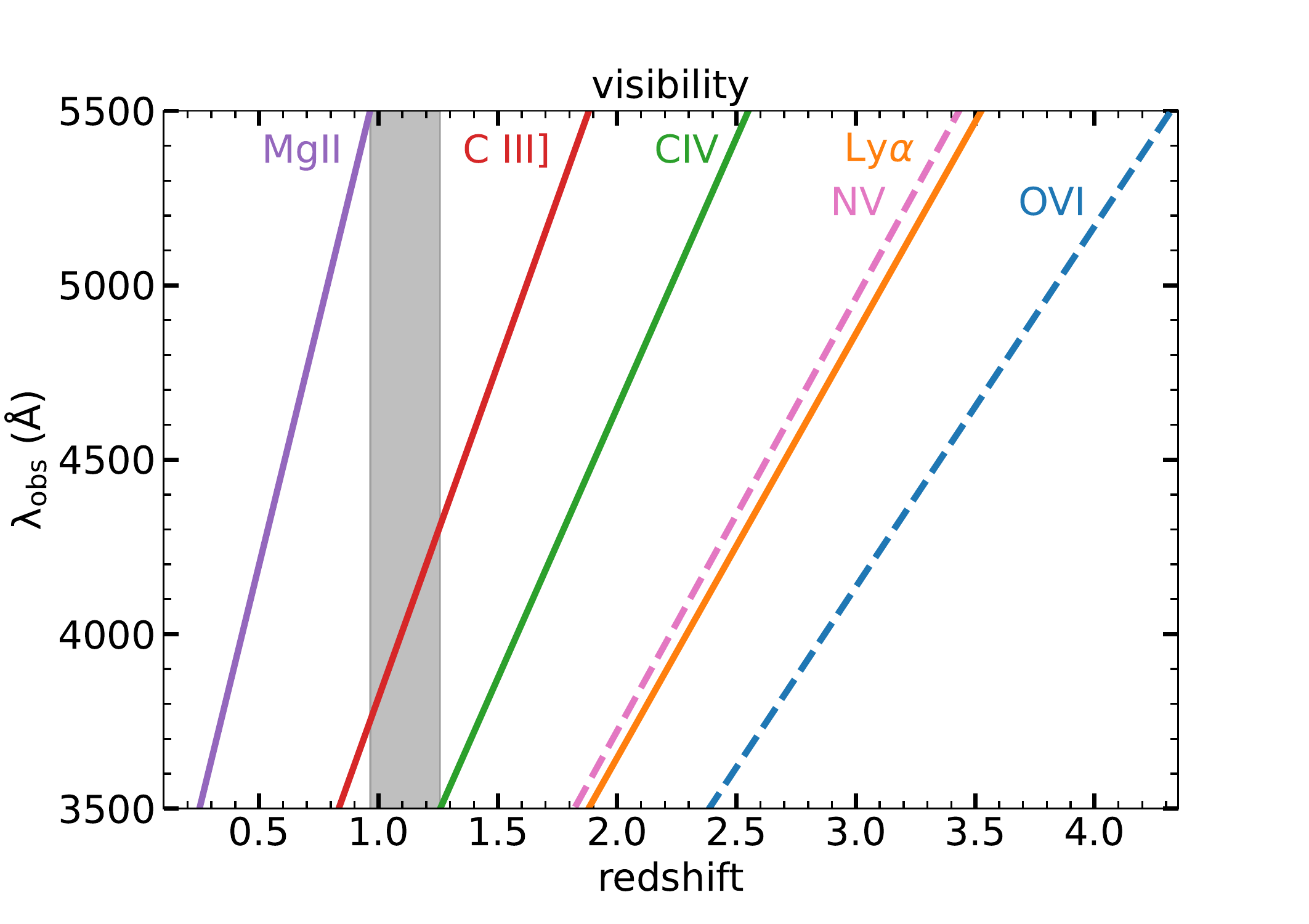}\\
\caption{The visibility of strong AGN emission lines as a function of redshift in the wavelength range of the HETDEX survey. The \ion{O}{6} $\lambda1034$ emission is usually highly noisy due to strong foreground absorbers, and \ion{N}{5} $\lambda1241$ is often  blended with Ly$\alpha$.  These two lines are thus plotted as dashed lines. 
The grey shaded region at $0.96<z<1.26$ shows one representative redshift range where only one emission 
is visible. 
We note here that $\rm Mg_{\ II}\ \lambda2799$ is the reddest emission we searched in this paper. We leave the search for emission lines of AGN at lower redshifts, such as the $\rm [Ne_{\ V}]\ \lambda3426$ emission and the Balmer emissions, for the next data release.}
\label{f_visibility}
\end{figure}

\section{AGN selection}
\label{sec_selection}

AGN are selected from the raw emission line catalog and the raw continuum catalogs in two ways: (1) the detection of two emission lines characteristic of AGN or (2) the presence of one or more broad emission lines. 

Figure~\ref{f_visibility} shows the visibility of the six strong emission lines used for the identification of AGN as a function of redshift within the HETDEX wavelength range.  If at least two of these lines are significantly detected ($5\sigma$ and $4\sigma$ for the strongest and the secondary lines) in the spectrum of an object, the source is classified as an AGN, regardless of line-width.  These sources have secure redshifts and are identified in this catalog with \texttt{zflag=1}. They include type-I and type-II AGN\null.  



As Figure~\ref{f_visibility} illustrates, there are some redshifts where we can expect to see only one emission line.  Moreover, for faint AGN, it is possible that only one emission-line will have a signal-to-noise above our threshold.  We therefore also searched for objects with only a single broad (FWHM $>1000$~km\,s$^{-1}$) emission feature. Obviously, this method can only identify  type-I AGN, and with only a single line identification, there may be some ambiguity in the redshift determination.  These objects are flagged with \texttt{zflag=0} in the AGN catalog. 


We introduce the two independent selections of AGN with more details in Section \ref{sec_2em}, and Section \ref{sec_mg}.

\subsection{AGN selected with two emission lines -- the 2em selection}
\label{sec_2em}


\subsubsection{Search in the emission line catalog}
\label{sec_2em_line}

\begin{figure*}[!htbp]
\centering
\includegraphics[width=0.275\textwidth]{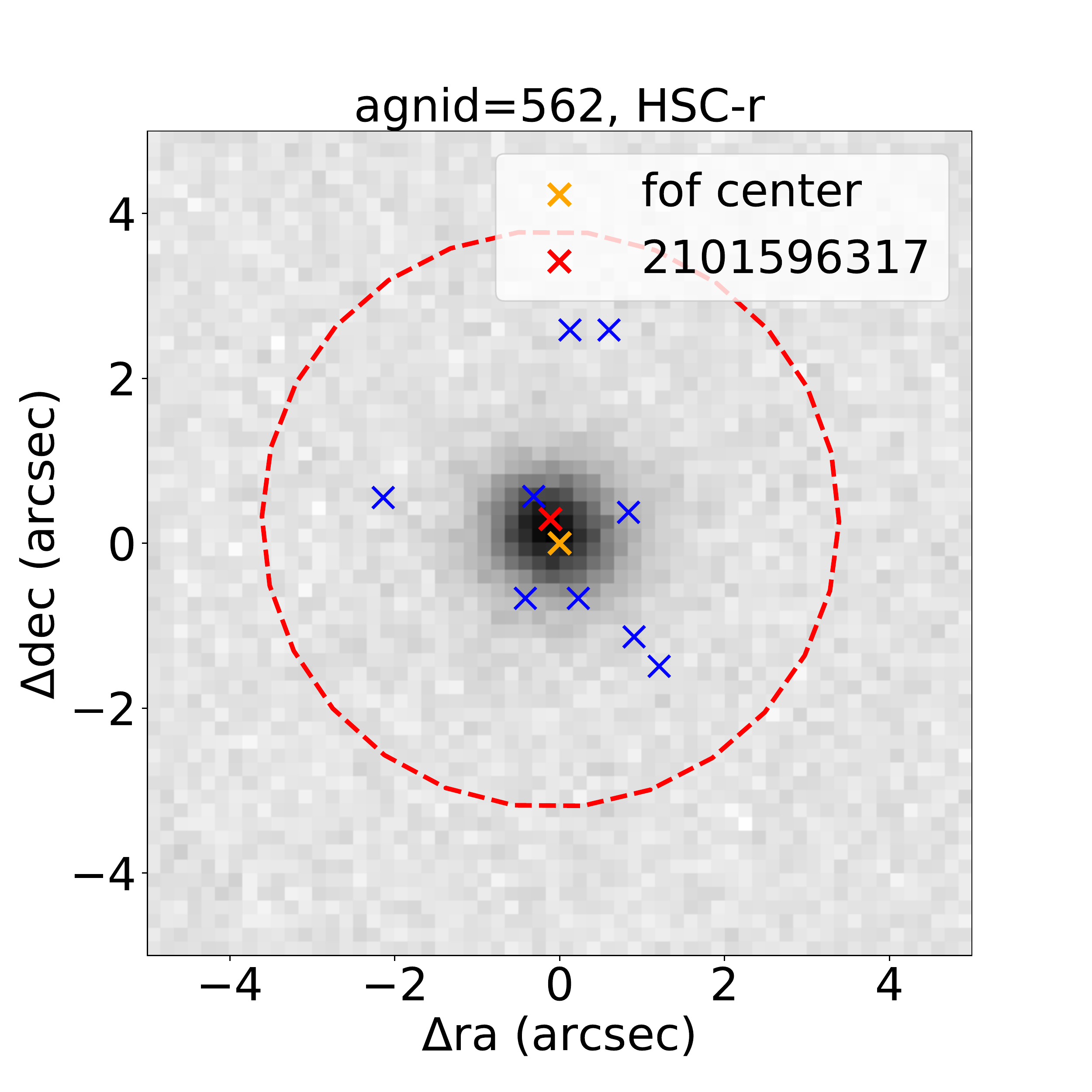}
\includegraphics[width=0.62\textwidth]{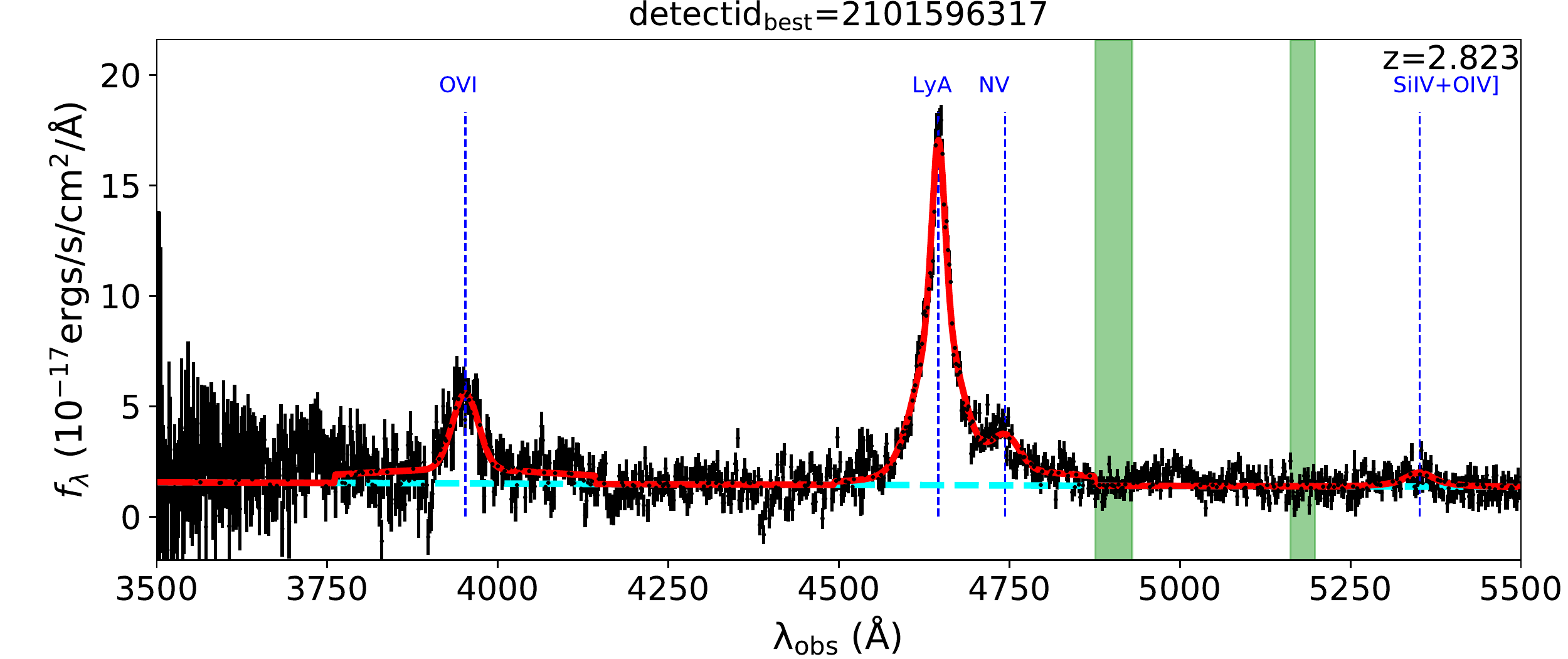}\\
\includegraphics[width=0.275\textwidth]{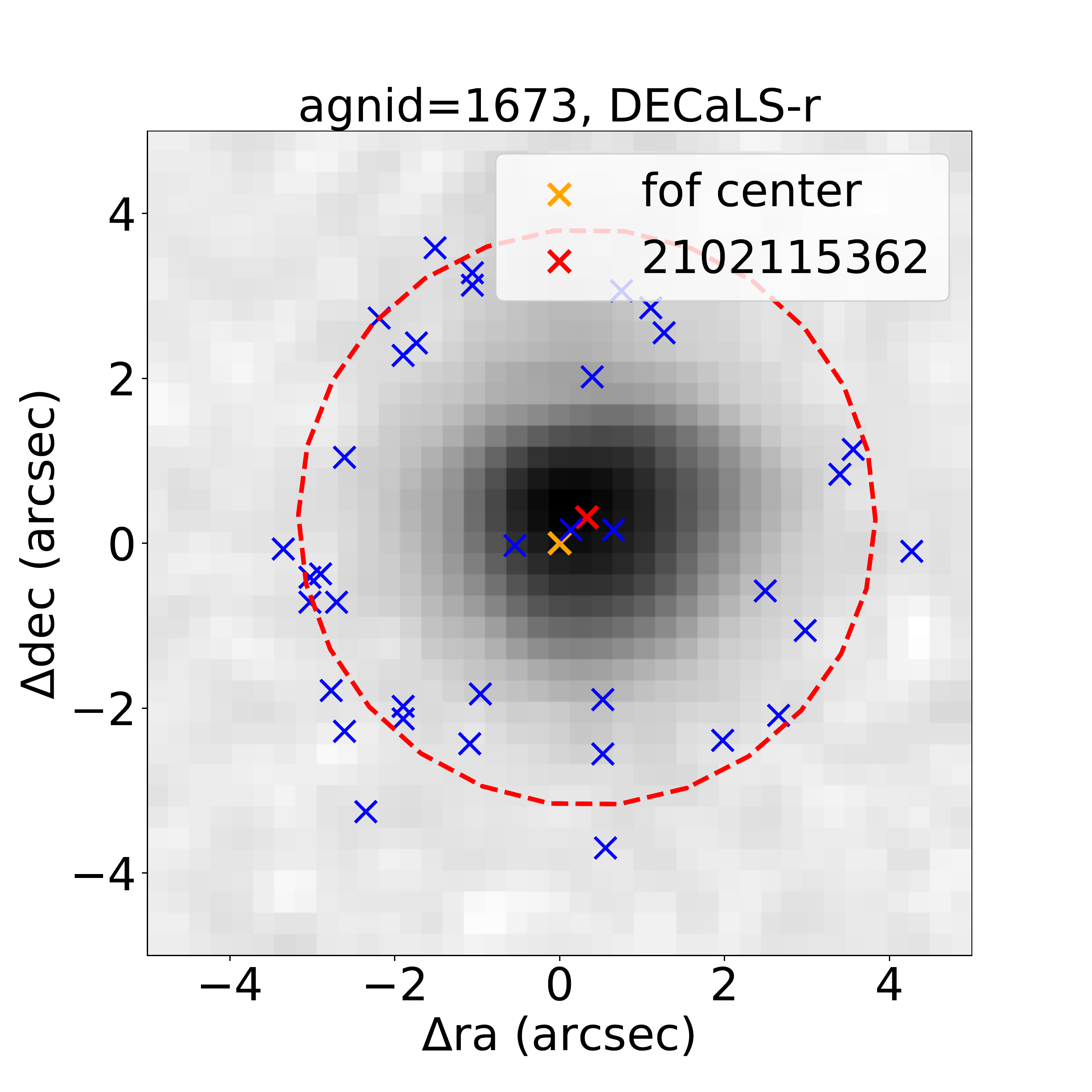}
\includegraphics[width=0.62\textwidth]{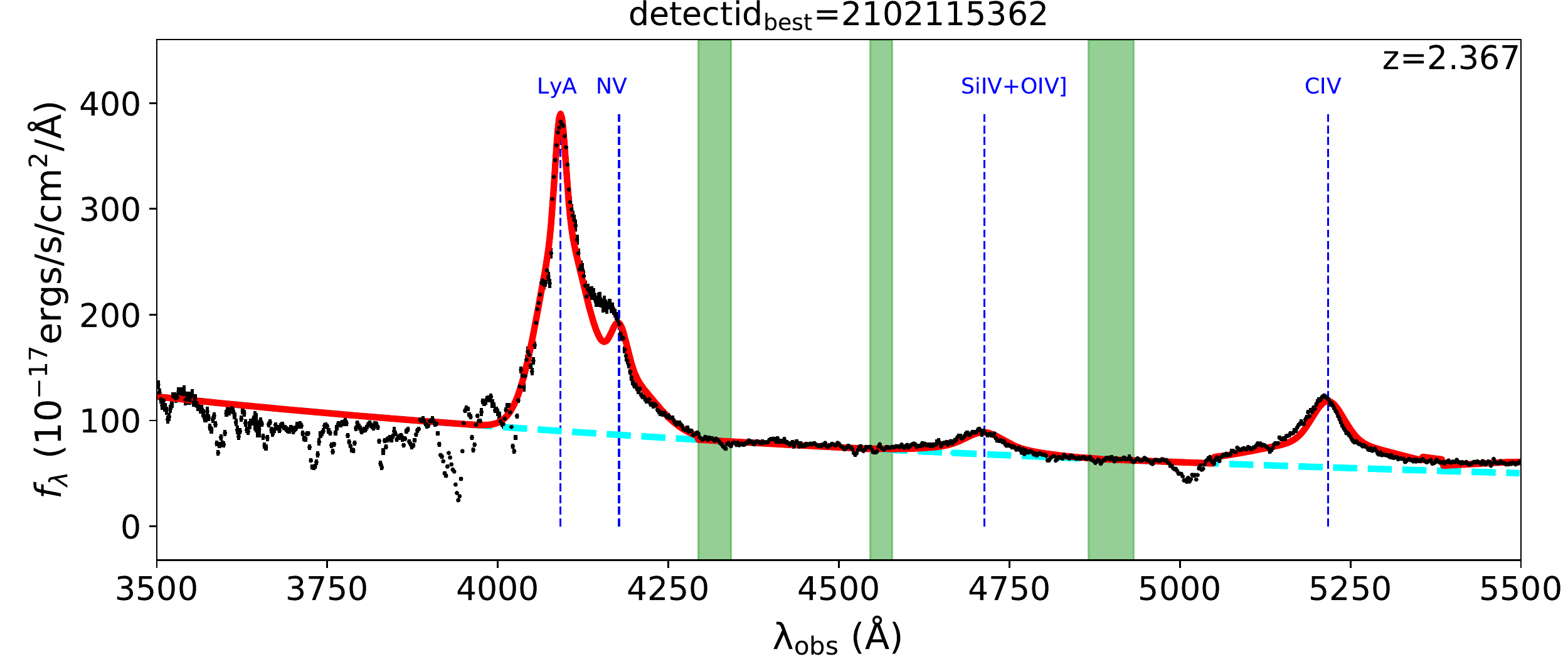}\\
\includegraphics[width=0.275\textwidth]{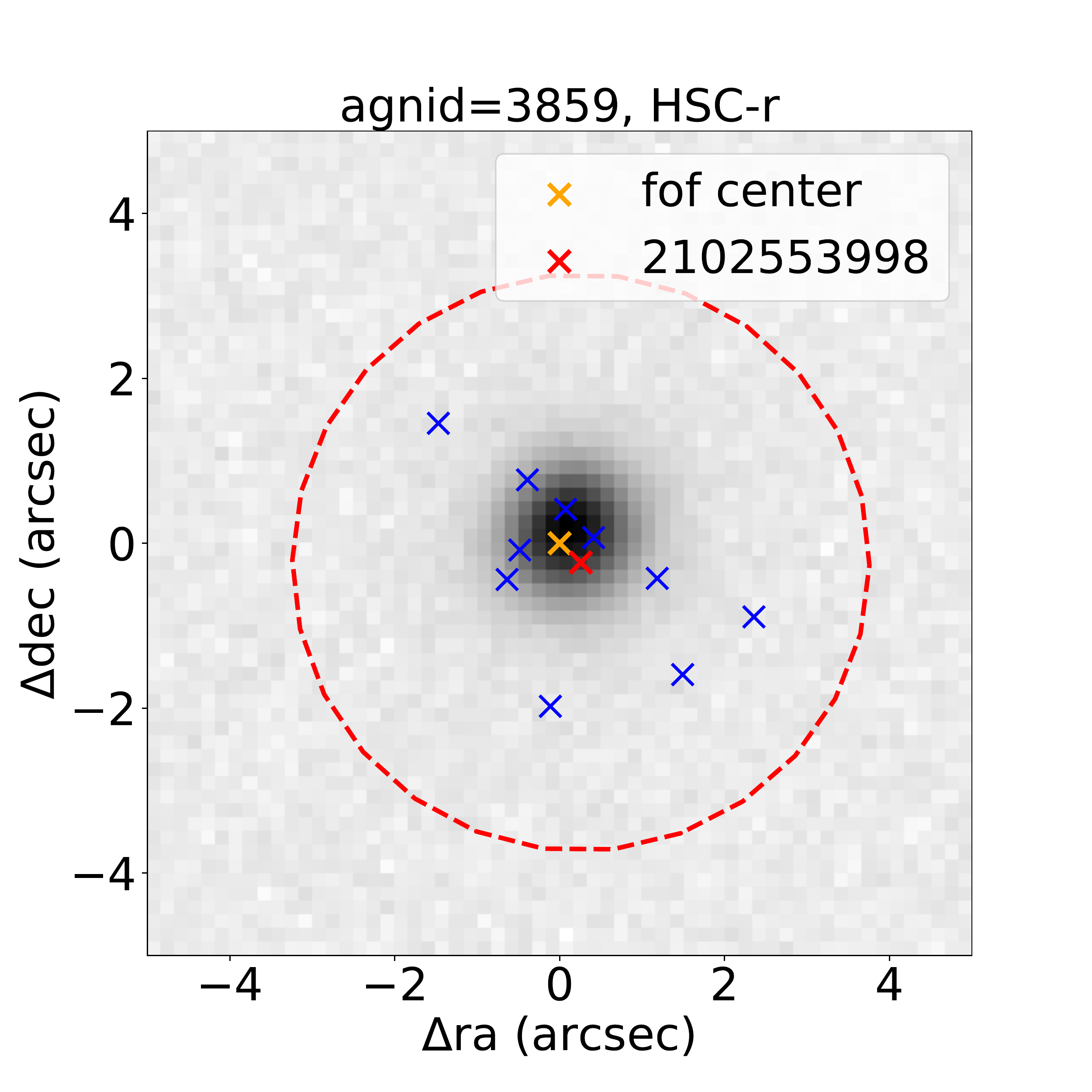}
\includegraphics[width=0.62\textwidth]{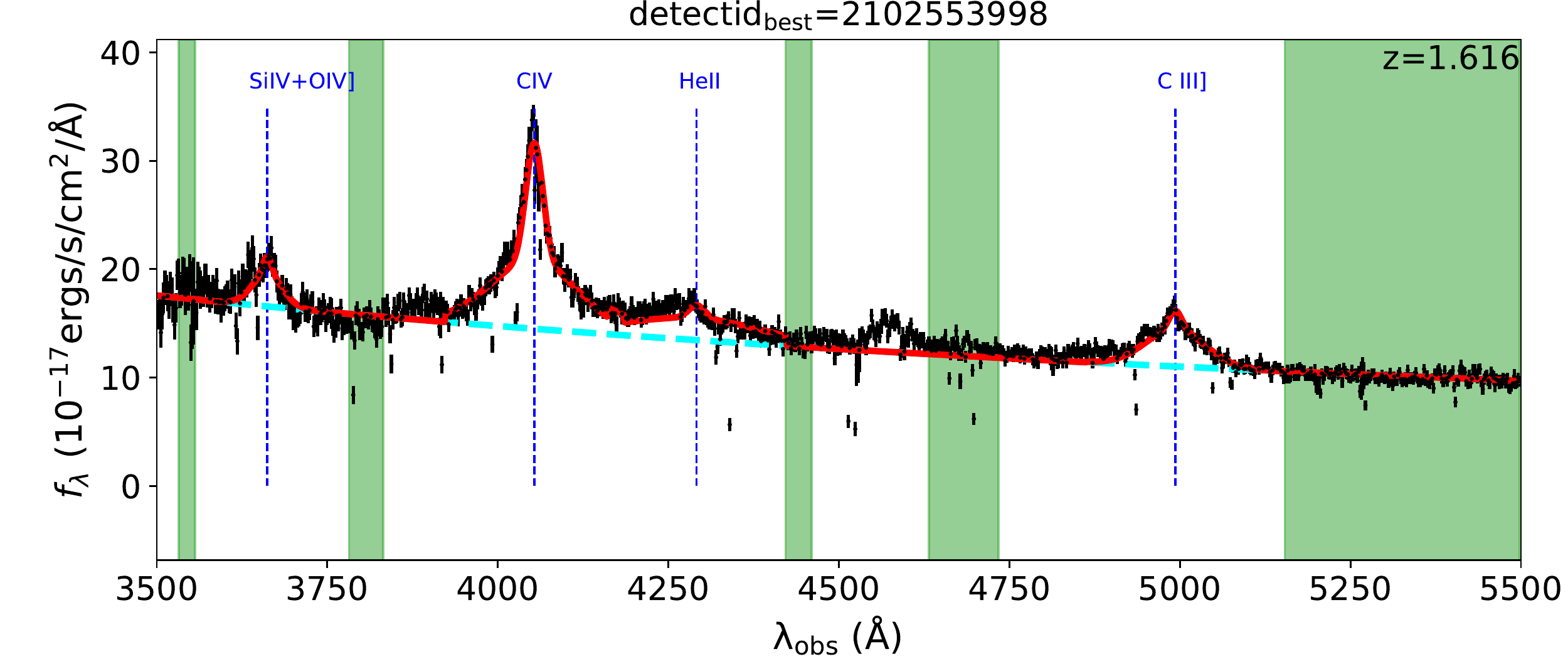}\\
\includegraphics[width=0.275\textwidth]{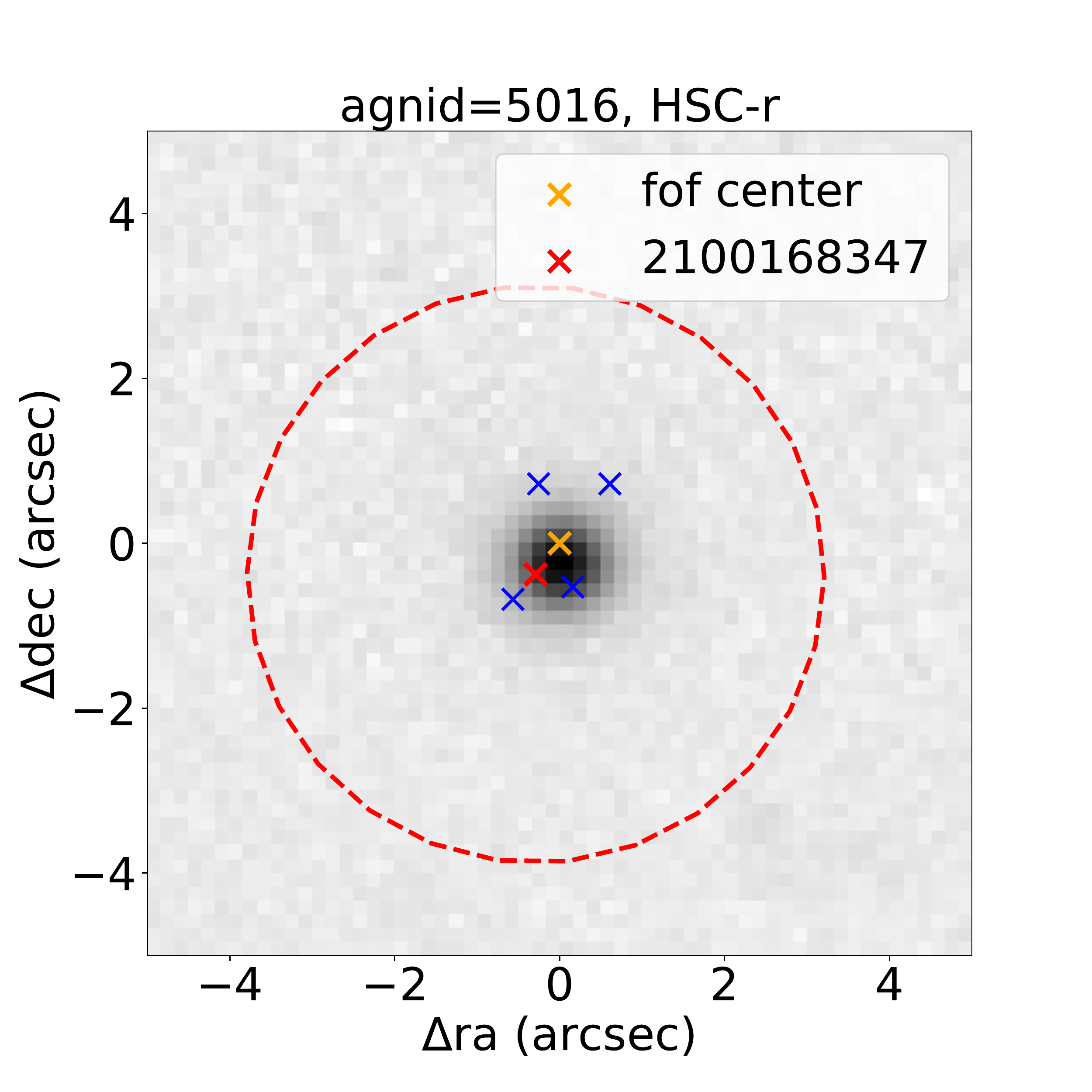}
\includegraphics[width=0.62\textwidth]{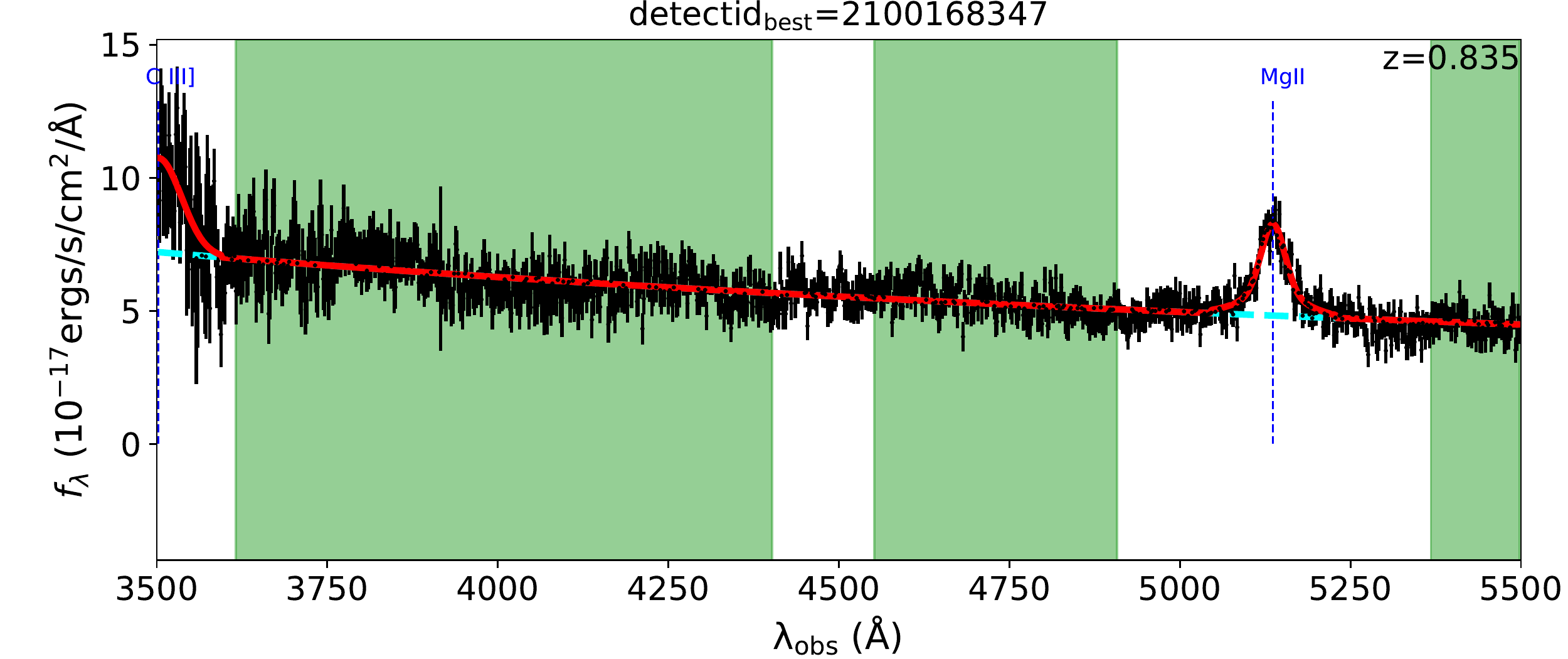}\\
\caption{Example AGN at different redshifts selected by the 2em search. Left: $r$-band images collected from various surveys. More information of the $r$-band imaging surveys searched can be found in Appendix \ref{sec_appendix_photo}\null. Blue crosses show the positions of all detections in the raw emission line and continuum catalogs within the field. The orange cross shows the emission-line-flux-weighted friend-of-friend (fof) center of these detections (see Section \ref{sec_fof} for more details), while the red cross indicates the position of the best detection ($\rm detectid_{best}$), i.e., the detection closest to the emission-line flux weighted fof center among all member detections. The red dashed circle shows the aperture of the PSF-extracted spectra of $\rm detectid_{best}$. The PSF-extraction collects all fibers within a $3\farcs 5$ radius aperture and then applies flux weights to each fiber spectrum that consider both the PSF and the differential atmospheric refraction (DAR) of light across the aperture. Right: The spectrum of the $\rm detectid_{best}$ of each AGN extracted with the aperture shown by the red circle in the left panels. Black data points with error bars are the observed spectra. The red curve shows our best fitted spectrum. The cyan dashed line is our best fitted model to the continuum with the green shaded windows. The best-fit redshift of each AGN is labeled in the upper right corner.}
\label{f_specs_2em}
\end{figure*}

The raw emission line catalog (Section \ref{sec_lines}) reports all emission line candidates with their observed wavelengths, and the extracted spectra in the full HETDEX wavelength range (3500\,\AA\ - 5500\,\AA). We consider six possible redshifts
for each emission line candidate, corresponding to the rest-frame wavelengths of six strong emission lines of high redshift AGN: \ion{O}{6} $\lambda1034$, Ly$\alpha$, \ion{N}{5} $\lambda1241$, \ion{C}{4} $\lambda1549$, \ion{C}{3}] $\lambda1909$, and \ion{Mg}{2} $\lambda2799$. For each possible redshift, we convert the observed 3500~\AA\ - 5500~\AA\ data to the rest frame, and fit the spectrum with a power-law continuum with multiple Gaussian emission-line profiles at the expected the wavelength range of the various lines. We define 
a parameter, $\rm sn_{global}$, as the peak of the emission line divided by the rms between the model and the spectrum from 3500\,\AA\ to 5500\,\AA\null. 
The spectrum that has the smallest fitted $\chi^2$ from our six modeled redshifts is defined as the best-fit solution, and we run this algorithm on every object in the 1.5 million raw emission-line catalog.  AGN candidates are then selected as those emission line candidates where the continuum subtracted local signal-to-noise ratio of their most significant emission line ($\rm S/N_{em,1st}$) is greater than 5, and that of the second most significant emission ($\rm S/N_{em,2nd}$) is greater than 4. These are then subjected to visual inspections before inclusion in the final AGN catalog. We discuss this process in greater detail in Section \ref{sec_visual}.

Continuum windows chosen for the fit of the power-law continua of sources whose redshifts are lower than 3.04 are rest frame [1275\,\AA, 1290\,\AA], [1350\,\AA, 1360\,\AA], [1445 \AA, 1465 \AA], [1690\,\AA, 1705\,\AA], [1770\,\AA, 1810\,\AA], [1970\,\AA,\,2400\,\AA], [2480\,\AA, 2675\,\AA], [2925\,\AA, 3400\,\AA], [3775\,\AA, 3832\,\AA], [4000\,\AA, 4050\,\AA], [4200\,\AA, 4230\,\AA\null]. For sources at $z>3.04$, only the first one of the above continuum windows remains within our wavelength coverage, so we have to include an additional window at [1140\,\AA, 1160\,\AA] which is blueward of the $\rm Ly\alpha$ emission, and usually heavily affected by absorption from the inter-galactic medium (IGM) or foreground galaxies. For these $z>3.04$ objects, we can only derive a rough estimate for the continuum subtracted fluxes for \ion{O}{6} $\lambda1034$ and Ly$\alpha$. For the seven AGN in our catalog with $z>3.74$, we further include the continuum window of [830\,\AA, 880\,\AA], which is bluer than the Lyman limit, since [1140\,\AA, 1160\,\AA] is also out of range at such high redshifts.

Figure \ref{f_specs_2em} presents four AGN at different redshifts selected by the 2em search. From top to bottom, the AGN are arranged by decreasing redshift.  In order, the major emission lines used in the 2em selection go from \ion{O}{6} $\lambda 1034$+Ly$\alpha$ in AGN 562, Ly$\alpha$+\ion{C}{4} $\lambda1549$ in AGN 1673, \ion{C}{4} $\lambda 1549$+\ion{C}{3}] $\lambda1909$ in AGN 3859, and \ion{C}{3}] $\lambda 1909$+\ion{Mg}{2} $\lambda2799$ in AGN 5016. In some cases, other weak emission features, such as \ion{Si}{4}+\ion{O}{4}] $\lambda 1400$ in AGN 562, 1673, and 3859, are also detected, but usually at a level that is not helpful for the identifications of the AGN\null. We re-visit the strong emissions and the weak emissions and their relative strengths in Section \ref{sec_compspec}.
Note that in the case of AGN 5016, we were able to confirm the identity of the \ion{Mg}{2} $\lambda 2799$ line because \ion{C}{3}] $\lambda 1909$ lies at the extreme blue edge of our wavelength range.

\subsubsection{Search in the continuum catalog}
\label{sec_2em_cont}
The continuum catalogue (Section \ref{sec_conts}) does not include any information about emission lines, as most of its sources are stars without emission lines.
We examined this catalog because the 50 electron counts threshold roughly corresponds to a $g<22.5$ magnitude cut. At these magnitudes, bright QSOs are still very common. To find the QSOs in the continuum catalog with no guidance from the detected emission lines, we identify the wavelengths of the highest two peaks in the 3500\,\AA - 5500\,\AA\  spectrum, and then assume each of the two peaks to be from \ion{O}{6} $\lambda 1034$, Ly$\alpha$, \ion{N}{5} $\lambda1241$, \ion{C}{4} $\lambda1549$, \ion{C}{3}] $\lambda1909$, or \ion{Mg}{2} $\lambda2799$. Therefore, for each detection in the continuum catalog, we have twelve redshift guesses based on observed wavelengths of the two highest peaks and six rest-frame wavelengths. We then fit each detection twelve times, and choose the redshift that has the lowest fitted $\chi^2$ to represent the best fit to each detection. The candidate selection is the same as the criteria in the search in the emission-line catalog: $\rm S/N_{em,1st}>5$, and $\rm S/N_{em,2nd}>4$.


\subsection{AGN selected with single broad emission lines -- the sBL selection}
\label{sec_mg}

Besides the AGN that have at least two significant emission lines, there are many AGN that have only one emission line significantly detected in the HETDEX wavelength range, either because the second most significant emission is not observed with high enough signal or because because no other emission line falls within 3500\,\AA\ - 5500\,\AA\ (Figure \ref{f_visibility}). 

For the single emission lines in the raw emission line catalog (Section \ref{sec_lines}), we cannot be sure which emission is being detected, but if the line is significantly broad, it is likely that the candidate is an AGN. We can not rule out the possibility that some single broad-line (sBL-) AGN candidates are only star-forming galaxies with strong outflows caused by the stellar winds or some other physical processes. However, most of which are proved to be real type-I AGN when cross matched with the SDSS DR14Q QSO catalog \citep{Paris2018} (more details can be found in Section \ref{sec_boss}).

The detection algorithm for the emission line catalog is primarily designed to search for the more common narrow emission lines and map the $z\sim3$ universe with LAEs without pre-selection biases. The line fitting algorithm in the search of the emission lines uses a single Gaussian component with an initial guess of $\sigma$ of 2.2 \AA\ (Section \ref{sec_lines}), which is clearly not appropriate for broad line AGN, especially those whose spectra are complicated by strong absorptions. If we select the objects using a simple $\chi^2$ cut from this single-Gaussian fit, many real broad-line sources with complicated profiles would be rejected. 
We therefore perform multi-Gaussian fits on all sources in the 1.5 million raw emission line catalog that were not previously identified as AGN by the 2em selection (Section \ref{sec_2em}). 

A multi-Gaussian fit for each line detection in the emission line catalog is performed in the observed-frame.  We consider a region centered at the wavelength recorded for the line, and adopt a window size of $\pm 200$\,\AA\null.  We approximate the continuum between 3500\,\AA\ and 5500\,\AA\ using a linear fit to all data points outside the emission-line window. We then subtract this linear continuum from the full spectrum, and fit the emission-line window with a set of Gaussian profiles. We note here that this simple continuum subtraction is only used for the sBL search and observed wavelengths determination. All emission line measurements reported in our catalog (Section \ref{sec_catalog}), regardless of the detection method, are subtracted for the power-law fitted continuum with the rest-frame continuum windows introduced in Section \ref{sec_2em}.

We allow up to eight Gaussian components (four emission lines $+$ four absorption lines) for the most complicated line profile. We begin our modeling with a simple two Gaussian component template, using a broad emission component and a narrow emission component at the same wavelength. If this two-Gaussian fit is insufficient (i.e. $\chi^2_{2e} >1.21$), two more absorption lines are then added gradually. The first (broad $+$ narrow) emission set is required to be centered at the same wavelength, while the wavelengths of the absorption lines can vary within the fitted wavelength window. If needed ($\chi^2_{2e2a} >4.9$, where $\chi^2_{2e2a}$ is the reduced $\chi^2$ of the two emission lines $+$ two absorption lines fit), the second (broad $+$ narrow) emission set is required to be centered at $\rm \lambda_{obs} \times (\lambda_{N\,V,rest}/\lambda_{Ly\alpha,rest})$, where $\rm \lambda_{obs}$ is the wavelength of the first (broad $+$ narrow) emission set. This addresses the Ly$\alpha$+\ion{N}{5} emissions, which are often blended together and are difficult to fit separately. The threshold cut on the reduced $\chi^2_{th}$ in each step is different and set empirically based on our visual inspection experiences. Our current set of $\chi^2_{th}$ values are our best estimates for this catalog, in future work we may update the exact values used. 

Broad-line candidates are selected when the emission-line multi-Gaussian fitted signal-to-noise ratio $\rm S/N_{mg} > 3.8$, $\chi^2_{\rm mg}<2$, and $\rm FWHM_{mg} > 1000\ km/s$. 
All candidates are then all followed up with careful visual inspections before being included in the AGN catalog (see Section \ref{sec_visual} for more details).

\begin{figure*}[htbp]
\centering

\includegraphics[width=0.65\textwidth]{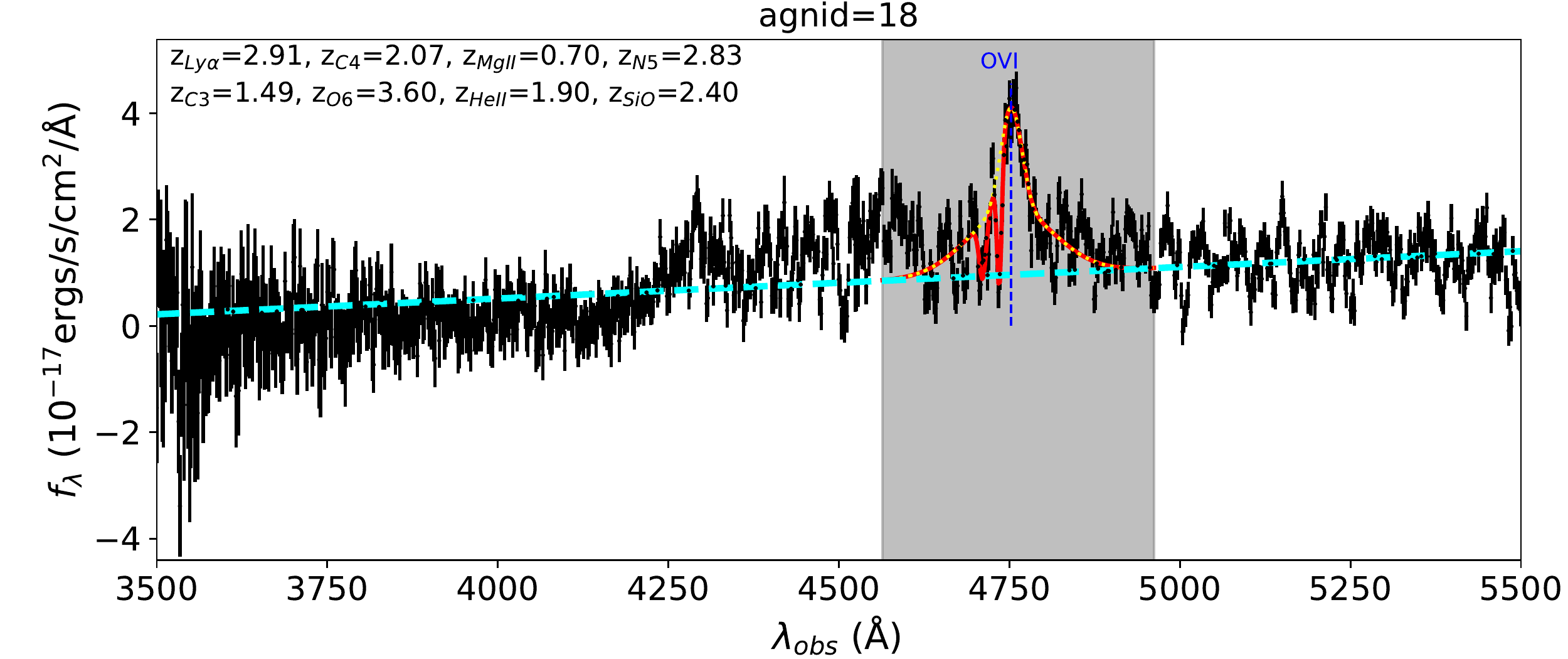}\\
\includegraphics[width=0.65\textwidth]{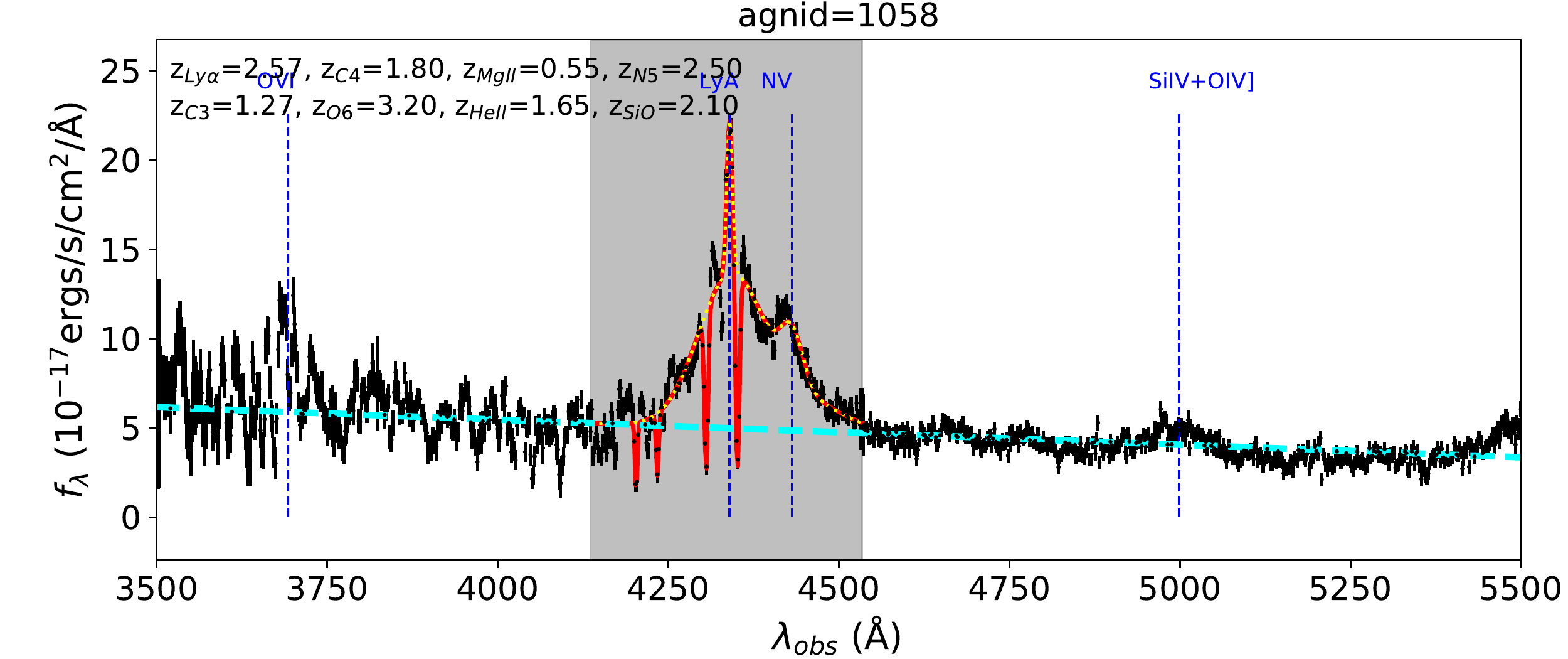}\\
\includegraphics[width=0.65\textwidth]{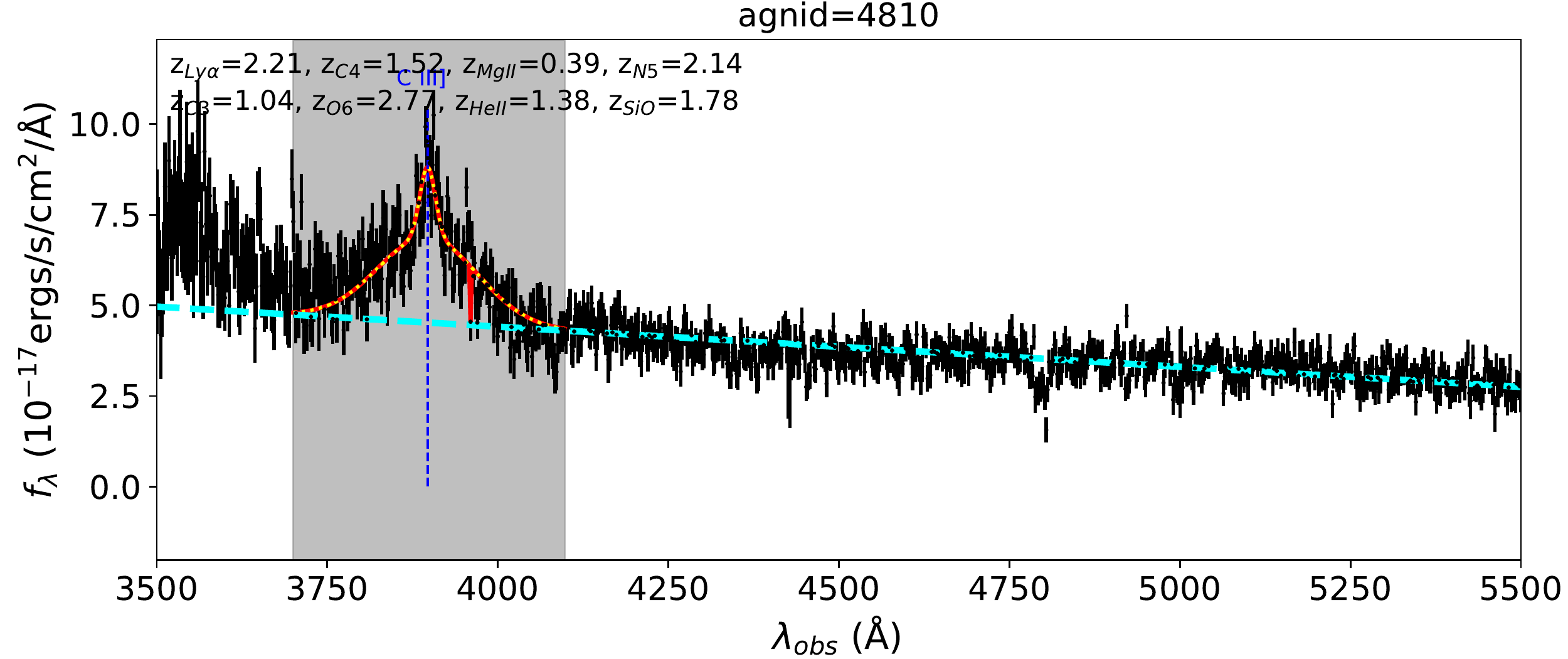}\\
\includegraphics[width=0.65\textwidth]{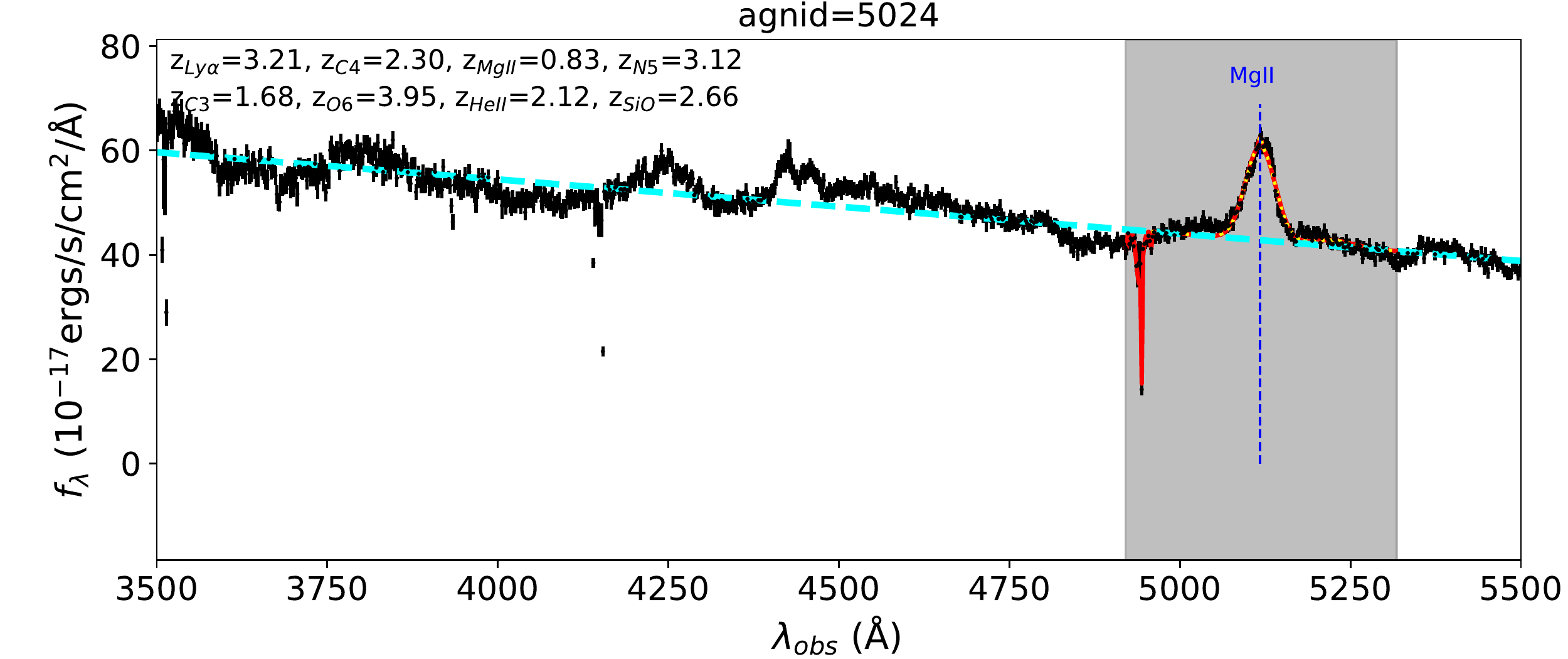}\\
\caption{Example AGN selected by the sBL method. The black data points with error bars are the observed spectrum, the red curve is the best fitted model of the emission-line window, and the grey shaded area is the window for the fit of the emission line. The cyan dashed line is a simple linear fit to the data points outside the grey shaded area. The yellow dotted line shows the sum of all emission components in the multi-Gaussian fit. The title of each panel is the catalog AGN ID  (Appendix \ref{sec_append}). We give eight possible redshifts in the upper left corner of each panel, corresponding to the detected emission line being Ly$\alpha$, \ion{C}{4} $\lambda 1549$, \ion{Mg}{2} $\lambda 2799$, \ion{N}{5} $\lambda 1241$, \ion{C}{3}] $\lambda 1909$, \ion{O}{6} $\lambda1034$, \ion{He}{2} $\lambda 1640$, and \ion{Si}{4}+\ion{O}{4}] $\lambda 1400$.  The blue dashed vertical lines accompanied with line names above show the positions of common emission lines, assuming the four single broad emissions of AGN 18, 1058, 4810, 5024 are \ion{O}{6} $\lambda 1034$, Ly$\alpha$, \ion{C}{3}] $\lambda 1909$, and \ion{Mg}{2} $\lambda 2799$, respectively.}
\label{f_specs_mg}
\end{figure*}

Figure \ref{f_specs_mg} shows some examples of AGN selected by the single broad-line (sBL) selection. The blue dashed lines accompanied by line names show the expected locations of other lines common to AGN\null.  These four single broad-line emission objects all have confirmed redshifts from SDSS DR14Q at their RA and DEC, and the matched redshifts suggest the emissions highlighted by the grey shaded areas of AGN with \texttt{agnid} $=18$, 1058, 4810, 5024 to be \ion{O}{6} $\lambda 1034$, Ly$\alpha$, \ion{C}{3}] $\lambda 1909$, and \ion{Mg}{2} $\lambda 2799$, respectively.

We note here that $\rm Ly\alpha$ is not the only emission line in the spectrum of AGN 1058; emission from \ion{O}{6} $\lambda 1034$, \ion{N}{5} $\lambda 1241$, and \ion{Si}{4}+\ion{O}{4}] $\lambda 1400$ is also detected. The spectrum to the blue side of  Ly$\alpha$ is usually very noisy with many foreground narrow-line absorption features, making it hard to fit \ion{O}{6} $\lambda1034$.  Similarly \ion{Si}{4}+\ion{O}{4}] $\lambda 1400$ is usually too weak to be identified by the code.  The \ion{N}{5} $\lambda1241$ line of AGN 1058 is quite well separated from Ly$\alpha$ and easy to measure. However, there are many cases where both Ly$\alpha$ and \ion{N}{5} $\lambda1241$ are too broad to too blended to identify as separate emission features. As a result \ion{N}{5} $\lambda1241$ is not usually used as evidence to the confirmation of Ly$\alpha$. The continuum subtracted grey shaded area of AGN 1058 is fitted with four emission Gaussian components $+$ four absorption Gaussian components, including a broad Gaussian emission and a narrow Gaussian emission both centered on $\rm Ly\alpha$ emission, a broad Gaussian emission and a narrow Gaussian emission both centered on \ion{N}{5} $\lambda1241$, and four narrow Gaussian absorption components centered at different appropriate wavelengths. 


\subsection{AGN selected by the cross match with the SDSS DR14Q QSO catalog}
\label{sec_boss}
In the above section, we described the procedures used to select AGN via the identification of two emission-lines (Section \ref{sec_2em}) and from the presence of a single single broad emission line (Section \ref{sec_mg}).  AGN selected by the 2em technique have secure redshifts and are flagged by \texttt{zflag=1} in our AGN catalog (Appendix \ref{sec_append}), while those found by the sBL method only have secure observed wavelengths, since there is no confirming feature in the limited HETDEX wavelength range of 3500\,\AA\ - 5500\,\AA\null.  SDSS, however, has a much larger spectral coverage of 3500 \AA\ - 10000 \AA, which can be used to confirm our identifications. We therefore cross match our sBL selected broad-line AGN candidates with the SDSS DR14Q \citep{Paris2018}; this catalog is, by far, the largest optical QSO catalog in existence, and its redshift range overlaps with our redshift range quite well.  About 60\% of the sBL identified AGN in our catalog are matched 
to an SDSS DR14Q source with a secure redshift. We flag these with \texttt{zflag=1}. The remaining sBL identified AGN candidates are only provided with our best redshift estimates and are flagged with \texttt{zflag=0}. 

The best redshift estimates for the sBL identified AGN flagged with \texttt{zflag=0} are predicated on the emission line identification in the visual inspections (Section \ref{sec_visual}) which is 
an empirical decision with the combination of the following properties. 
(i) Asymmetric redshifted line profile \citep[e.g.][]{Hashimoto2013}, heavy absorption on the blue wing, and/or a second peak at expected \ion{N}{5} $\lambda1241$ are usually strong indicators of the $\rm Ly\alpha$ line (see \texttt{agnid=1058} in Figure \ref{f_specs_mg} for a typical example). (ii) Steeper slope of the continuum in the observed frame is usually an indicator of emission lines of lower redshifts, such as \ion{C}{3}] $\lambda 1909$ and \ion{Mg}{2} $\lambda 2799$, because given observed wavelength range corresponds to longer rest-frame wavelength range at lower redshifts. For example, \texttt{agnid=5024} has deeper observed continuum slope compared to \texttt{agnid=4810} and \texttt{agnid=1058} in Figure \ref{f_specs_mg}. (iii) The increment of noisy fluxes at expected weak emissions can also help confirm the observed single emission line. For example, the spectrum of \texttt{agnid=1058} in Figure \ref{f_specs_mg} shows increment of noisy fluxes at expected \ion{O}{6} $\lambda1034$. The signal at \ion{O}{6} $\lambda1034$ is too noisy to be identified as a significant emission line by the code, but it can help confirm the $\rm Ly\alpha$ emission line. (iv) The observed EW can further help to distinguish different emission lines, as the observed EW decreases from Ly$\alpha$ through \ion{C}{4} $\lambda 1549$ and \ion{C}{3}] $\lambda 1909$ to \ion{Mg}{2} $\lambda 2799$ (see \texttt{agnid=1058}, \texttt{agnid=4810}, and \texttt{agnid=1058} in Figure \ref{f_specs_mg} for example, and see Section \ref{sec_ew} for more information). (v) The absence of strong emission lines within the wavelength range is an additional crucial information in the emission line identification. For example, the combination of (i)-(iv) sometimes suggests a single broad emission line might either be \ion{C}{4} $\lambda 1549$ or \ion{C}{3}] $\lambda 1909$. When assuming the detected line is \ion{C}{4} $\lambda 1549$, the $\rm Ly\alpha$ is absent at expected wavelength. This is then a strong rejection of the detected line to be \ion{C}{4} $\lambda 1549$. 

Besides helping confirm the redshifts of the sBL selected AGN, the cross match between the HETDEX raw emission line (Section \ref{sec_lines}) and  raw continuum (Section \ref{sec_conts}) catalogs and the SDSS DR14Q catalog can allow us to explore the incompleteness in our catalog.  Together, the 2em selection and the sBL selection should have found most of the broad-line AGN contained with the HETDEX survey to date.  A comparison 
can identify most AGN, especially the broad-line AGN in our catalog. However, the code may have missed a number of AGN in the parent sample that are difficult to identify, and we wish to recover them. In our 5,322 AGN catalog, 2,248 are covered by SDSS DR14Q, and 91 out of the 2,248 (4\%) are not initially identified as AGN by our automated search and are found only because of the SDSS cross match. 

\begin{figure}[htbp]
\centering
 \includegraphics[width=\textwidth]{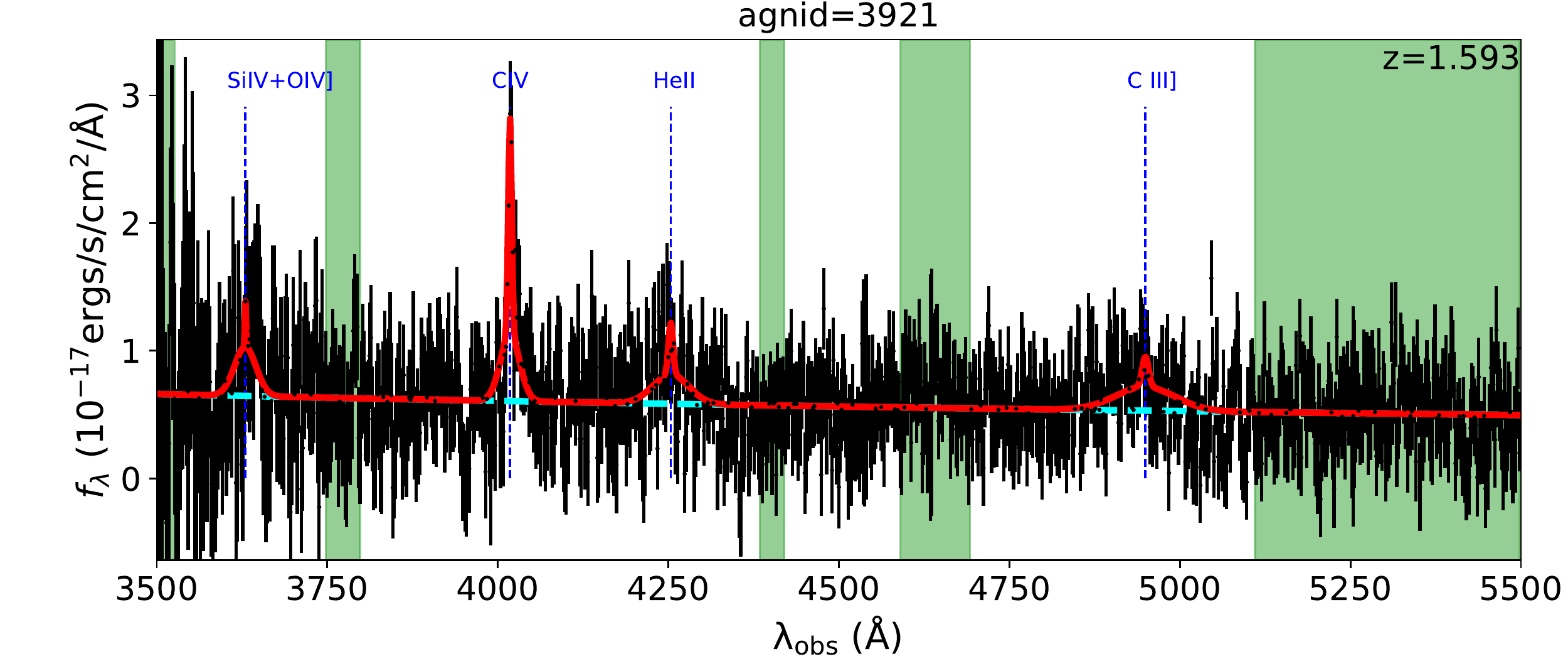} \\
\caption{An example AGN that is missed by both the 2em selection and the sBL selection, but caught by the cross match with the SDSS DR14Q catalog and added back into our HETDEX AGN catalog.
}
\label{f_specs_boss}
\end{figure}

Figure \ref{f_specs_boss} shows an AGN (3921) rejected by both the 2em selection and the sBL selection, but matched with a QSO in SDSS DR14Q\null. This AGN is rejected by the 2em selection because the \ion{C}{3}] $\lambda1909$ emission is not strong enough ($\rm S/N_{C_{\ III]}}=3.5<4$) to be linked with the \ion{C}{4} $\lambda1549$ emission and identified as a line pair.  Moreover, the \ion{C}{4} $\lambda1549$  line is highly affected by absorption on the blue wing, which causes the feature to be identified as a narrow emission line with $\rm FWHM_{C_{\ IV}}=528$~km\,s$^{-1}$ and rejected by the sBL search. AGN 3921 is caught by the SDSS catalog as an AGN, because its \ion{Mg}{2} $\lambda2799$ is  detected as broad emission at $\lambda_{\rm obs} = 7258$\,\AA\ in the SDSS spectrum, and thus confirms the weak emission features of \ion{C}{4} $\lambda1549$ and \ion{C}{3} $\lambda1909$.  This is a prime example of why this cross match follow-up is useful for finding AGN.

\subsection{Visual inspections}
\label{sec_visual}

The two-emission-line selection code (Section \ref{sec_2em}), the multi-Gaussian single broad-line selection code (Section \ref{sec_mg}), and the cross match with SDSS DR14Q with $r<5$ arcsec (Section \ref{sec_boss}), together provide an AGN candidate list of 33,000 AGN detections. We then visually check each candidate AGN to remove false positives (hot pixels, cosmic rays, nearby fibers affected by strong scattered light of a star, and other detector artifacts) and low redshift interlopers (meteors, low redshift galaxies, and stars) that happen to pass the selection criteria. 


Low redshift sources are an important source of contamination. Stellar features, such as a peak in continuum between the higher-order Balmer absorptions of white dwarfs, can be mistaken for broad line emission by the sBL selection code. In addition, some features of low-$z$ star-forming galaxies, such as the bump red-ward of the 4000\,\AA\ break, can be picked up by the multi-Gaussian fits. Moreover, a line can sometimes be too broad, making its centroiding difficult.  When this happens, line pairs may be mis-identified.  For example, the [\ion{O}{2}] $\lambda3727$+[\ion{O}{3}] $\lambda 5007$ line pair can sometimes be mistaken for \ion{Mg}{2} $\lambda 2799$ + [\ion{O}{2}] $\lambda 3727$ by the 2em code. Finally, the adjacent narrow emission lines of meteors can be mistaken as a single broad emission line absorbed by a few narrow features.

Visual inspection of a combination of information, such as matched photometric imaging, the flux distribution in the 2-D spectrum, the spatial separation between our detection and the associated SDSS QSO, the continuum shape of the 1-D spectrum, and the appearance of the raw CCD frame at the position of the candidate, can allow us to classify candidates as AGN, stars, low-$z$ galaxies, and false positives with higher confidence than what is possible with only the automated code.

Our three selection processes (Section \ref{sec_2em}, \ref{sec_mg}, and \ref{sec_boss}) collect a combined 33,000 AGN candidates from the emission-line catalog (1.5 million detections, Section \ref{sec_lines}) and the continuum catalog (80,000 detections, Section \ref{sec_conts}). Visual inspections confirm 17,168 of these candidates. Within this sample are 14,896 AGN detections with secure redshifts (\texttt{zflag=1}), while the remaining 2,272 sBL AGN detections only have estimated redshifts, and are labeled with \texttt{zflag=0}. We note that the automated identification codes for bad amplifiers, meteors, big nearby galaxies, and stars are built simultaneously with the AGN selection code since this is our first data release. In our next data release, we can first remove bad amplifiers, meteors, big nearby galaxies, and stars, then apply the AGN selection code. The amount of work of visual inspections would be significantly reduced.

\subsection{Friend-of-friend grouping (fof): the unique AGN catalog}
\label{sec_fof}
 
The HETDEX survey is a blind spectroscopic survey taken by an array of IFUs. Both spectral and spatial information are collected in a single exposure. 
Due to the spatial extent of AGN and the way the detection algorithm works, the same AGN can be detected multiple times spatially (see the left column of Figure \ref{f_specs_2em} for some examples) and spectrally (e.g. the $\rm Ly\alpha$ emission and the \ion{C}{4} $\lambda1549$ emission of the same source are identified as two detections by the HETDEX pipeline, Section \ref{sec_lines}). The diffuse 
emission of AGN can be caused by poor seeing, ionized inter-galactic medium around the AGN, strong outflows blowing into the outskirts of the host galaxies, etc. 
Therefore, there are many duplicates in the 17,168 AGN detection catalog visually confirmed in Section \ref{sec_visual}.  



To remove the duplicates, we apply a 3-dimensional emission-line flux-weighted friends-of-friends (fof) algorithm with linking lengths of $\Delta r=5$ arcsec\ on sky and $\Delta z=0.1$ in redshift to our 17k AGN (candidate) detection list. 
The chosen separation of $\Delta r=5$ arcsec\ works best for most of the AGN in our sample based on our experience: a larger $\Delta r$ can erroneously combine nearby AGN into a single object, and a smaller $\Delta r$ can fail to link all detections of the same source.
For normal narrow emission-line galaxies, $\Delta z=0.1$ is too broad a choice in the redshift direction. However for AGN, $\Delta z=0.1$ works well, considering their strong, broad emissions can have dispersions of several thousands of $km\ s^{-1}$.  Our tests confirm that all statistical results in Section \ref{sec_statistics} remain with slightly different $\Delta r$ and $\Delta z$ sets.

After removing duplications, the final AGN catalog is reduced from 17,168 AGN detections to 5,322 unique objects. This includes 3,733 AGN with secure redshifts (\texttt{zflag=1}, and 1,589 AGN single broad-line emitters reported with our best redshift estimation (\texttt{zflag=0}).  We note that although broad-line emission could be produced by other mechanisms, such as strong stellar outflows, all sBL selected objects are highly suspected to be type-I AGN\null. Future spectroscopic observations extending to longer wavelengths are needed for the confirmation of these \texttt{zflag=0} type-I AGN candidates.

The coordinates of each AGN reported in our catalog (the \texttt{ra} column and the \texttt{dec} column in Table \ref{t_catalog}) are determined from the emission-line flux-weighted center of the fof grouping among all member detections. In this paper, each unique AGN is represented by its best member detection ID (the $\rm detectid_{best}$ column in Table \ref{t_catalog}), which is the member detection closest to the fof center. All measurements of each AGN are made with the spectrum of $\rm detectid_{best}$, which is centered on the $\rm ra_{best}$ column and the $\rm dec_{best}$ column in Table \ref{t_catalog}. Figure \ref{f_specs_2em} shows the relative positions of the fof center (the orange cross), $\rm detectid_{best}$ (the red cross), and all member detections (the blue crosses) in the $r$-band images of four example AGN\null. The red circle shows the aperture used for the PSF-extracted spectrum of $\rm detectid_{best}$. 

\begin{figure*}[!htbp]
\centering
\includegraphics[width=0.8\textwidth]{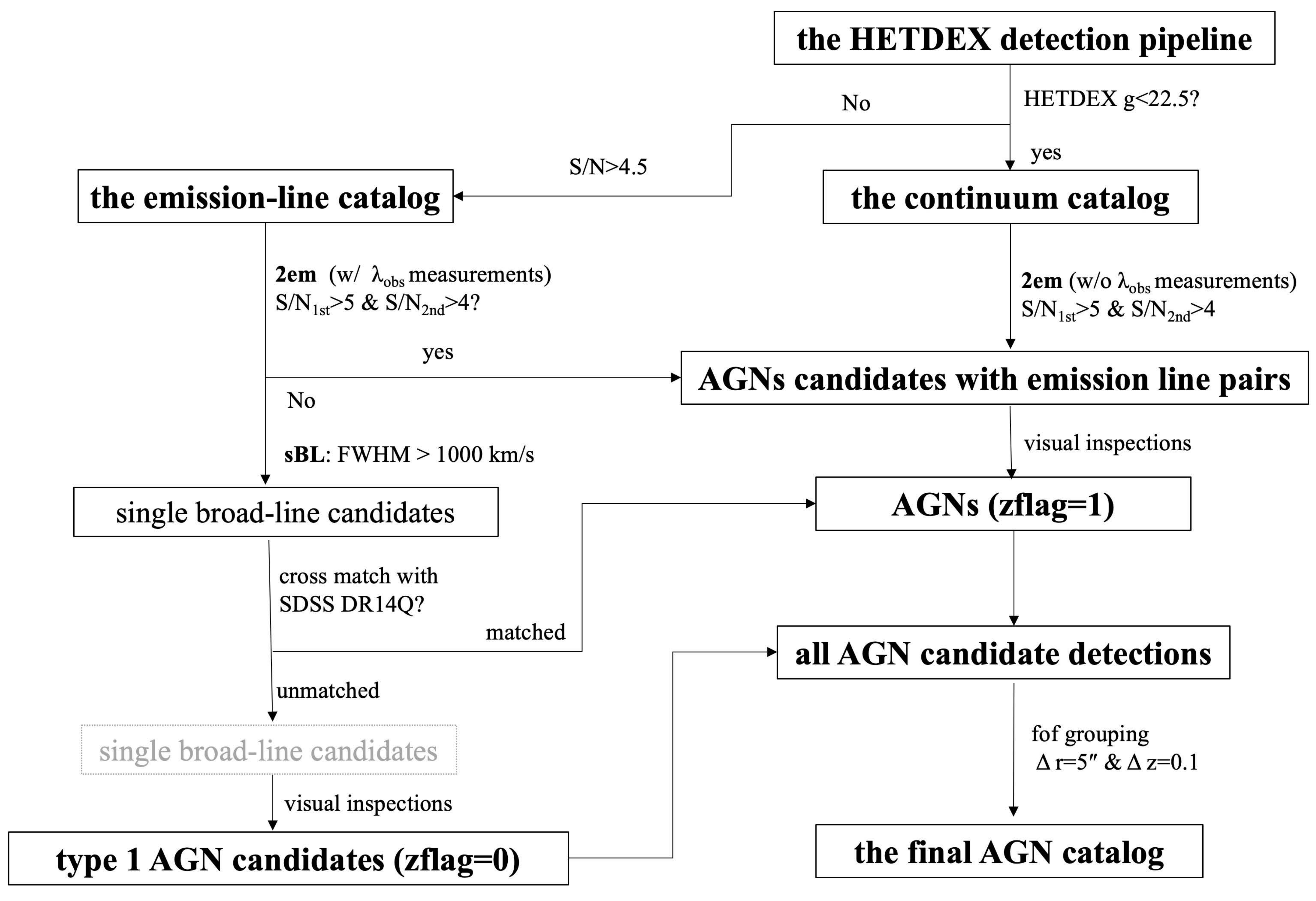}\\
\caption{The flowchart that summarizes the selection of the AGN catalog from the HETDEX data base. For simplicity, only the major selection criteria are included. More details can be found in Section \ref{sec_selection}.}
\label{f_flowchart}
\end{figure*}

We choose to represent each AGN with $\rm detectid_{best}$ because we have better control of the completeness and aperture corrections, as these have already been measured for the point sources in the original raw emission-line catalog and raw continuum catalog. 
Figure \ref{f_specs_2em} shows that the aperture of the PSF-extraction of $\rm detectid_{best}$ can roughly cover each AGN, though not perfectly. 
We compared the spectrum extracted at the fof center with that extracted at $\rm detectid_{best}$ with a small sub-sample.
We did not find significant difference between the two. 

\subsection{Summary of the selection of the AGN catalog}

Figure \ref{f_flowchart} compactly summarizes the selection of the AGN catalog (Section \ref{sec_selection}).
The sample is purely based on emission features, free of any pre-selection on morphologies, magnitudes, colors, etc.
Both bright QSOs and low-luminosity Seyferts are included in the catalog. The 2em selection method also allows for the detection of type-II AGN in our sample.

\section{Completeness}
\label{sec_completeness}
In this section, we discuss the completeness of our HETDEX AGN catalog by comparing with samples of known AGN in the literature. 

There are two major steps in our selection: the HETDEX selection pipeline (Section \ref{sec_detection}) and the AGN selection code (Section \ref{sec_selection}). The HETDEX pipeline used a 3-D grid search (spacial $+$ spectral) to find emission-line candidates (Section \ref{sec_lines}), and a 2-D grid search (spatial) to find continuum candidates (Section \ref{sec_conts}). The main purpose of the HETDEX survey is to map $z\sim3$ LAEs with no pre-selection biases. Therefore, the HETDEX pipeline is optimized to search for the more common narrow emission lines, as they represent the majority of the high-z star-forming galaxies being surveyed. Broad emitters are the minorities of the emission-line galaxies, such as the type-I AGN and extreme star-forming galaxies with strong stellar winds. The completeness of the HETDEX pipeline in the identification of the broad emission lines is then lower than that of the narrow emission lines. The wavelength grid search for the emission-line candidates in the HETDEX pipeline uses a single Gaussian profile to scan signals between 3500\,\AA\ - 5500\,\AA\ with $\pm$ 50\,\AA\ windows and 8\,\AA\ steps. The success rate of a model fit relies heavily on the initial guess of the parameter(s) built in the code, though the best fitted parameters are allowed to vary. The initial guess of the line width of the single Gaussian model in the HETDEX pipeline is $\sigma_{\rm initial}=2.2$\,\AA, slightly larger than the resolution of the spectrum 2\,\AA\ and suitable for most narrow emissions. This narrow initial guess of the line width can result in the lower completeness for the very broad line associated with bright AGN\null. 

We explore two sets of simulations with the $\sigma_{\rm initial}=2.2$\,\AA\ search and a broader single Gaussian search ($\sigma_{\rm initial}$=8.0\,\AA) by inputting simulated emissions into the database of the month of 2020-05, and run the pipeline with different initial guess of line widths. The wavelengths grid in the $\sigma_{\rm initial}$=8.0\,\AA\ search has a larger window size ($\pm$ 80\,\AA), while the step remains 8\,\AA. We find that the total loss rate of broad emissions of the current HETDEX pipeline compared to the broad search at $\rm S/N>4.5$ is $\sim$ 10\%. This is a good result considering that the false positive rate is lowered from about 80\% for the $\sigma_{\rm initial}=8.0$ \AA\ search to about 50\% for the the $\sigma_{\rm initial}=2.2$ \AA\ search. The efficiency of the visual inspections on the candidates selected by the code is significantly improved with the reduced false positive rate in the $\sigma_{\rm initial}=2.2$ \AA\ search.

\cite{Zhang2021} independently explored the HETDEX database, without using the HETDEX pipeline, in one of their samples of AGN, providing a good opportunity to check the completeness of both the HETDEX pipeline and the AGN selection. They extracted objects at the positions of the $>5\sigma$ detections in the Hyper Suprime-Cam-HETDEX joint survey (HSC-DEX, see Appendix \ref{sec_appendix_photo} for more details) and the Hyper Suprime-Cam Subaru Strategic Program (HSC-SSP; \cite{Aihara2018,Aihara2019}). The 5$\sigma$ depth in $r$-band is about 26 mag in the wide fields and about 28 mag in the deep fields. They scanned the extracted spectra repeatedly with 6, 10, 18, 34, and 110\,\AA\ bins. Real broad emission lines are selected with $\rm S/N>5.5$ and $\rm FWHM>1000$~km\,s$^{-1}$ in their single Gaussian fit. Their broad line selection process is almost a pure magnitude-limited sample based on the deep HSC survey. 

\begin{figure}[htbp]
\centering
\includegraphics[width=0.8\textwidth]{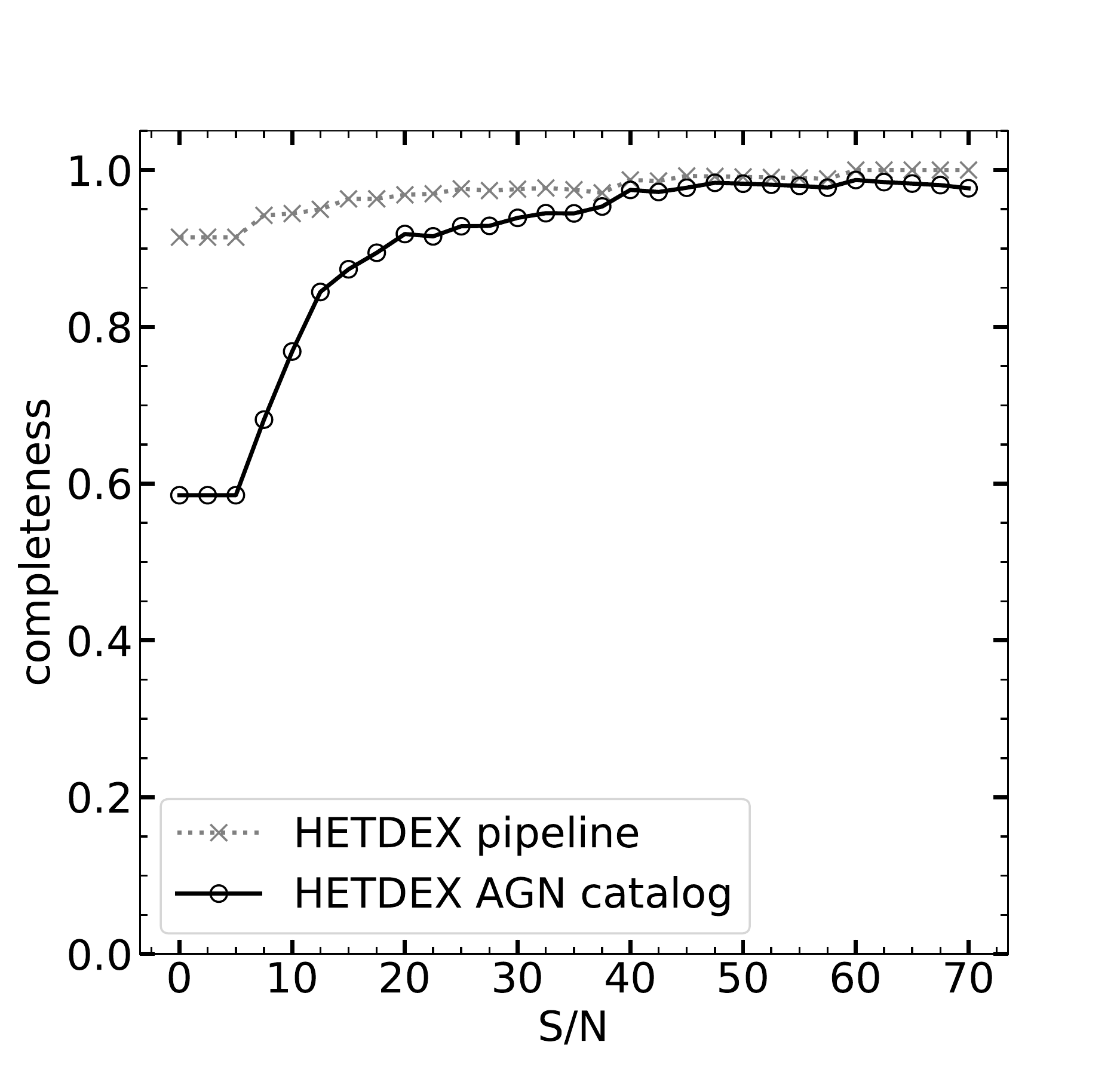}
\includegraphics[width=0.85\textwidth]{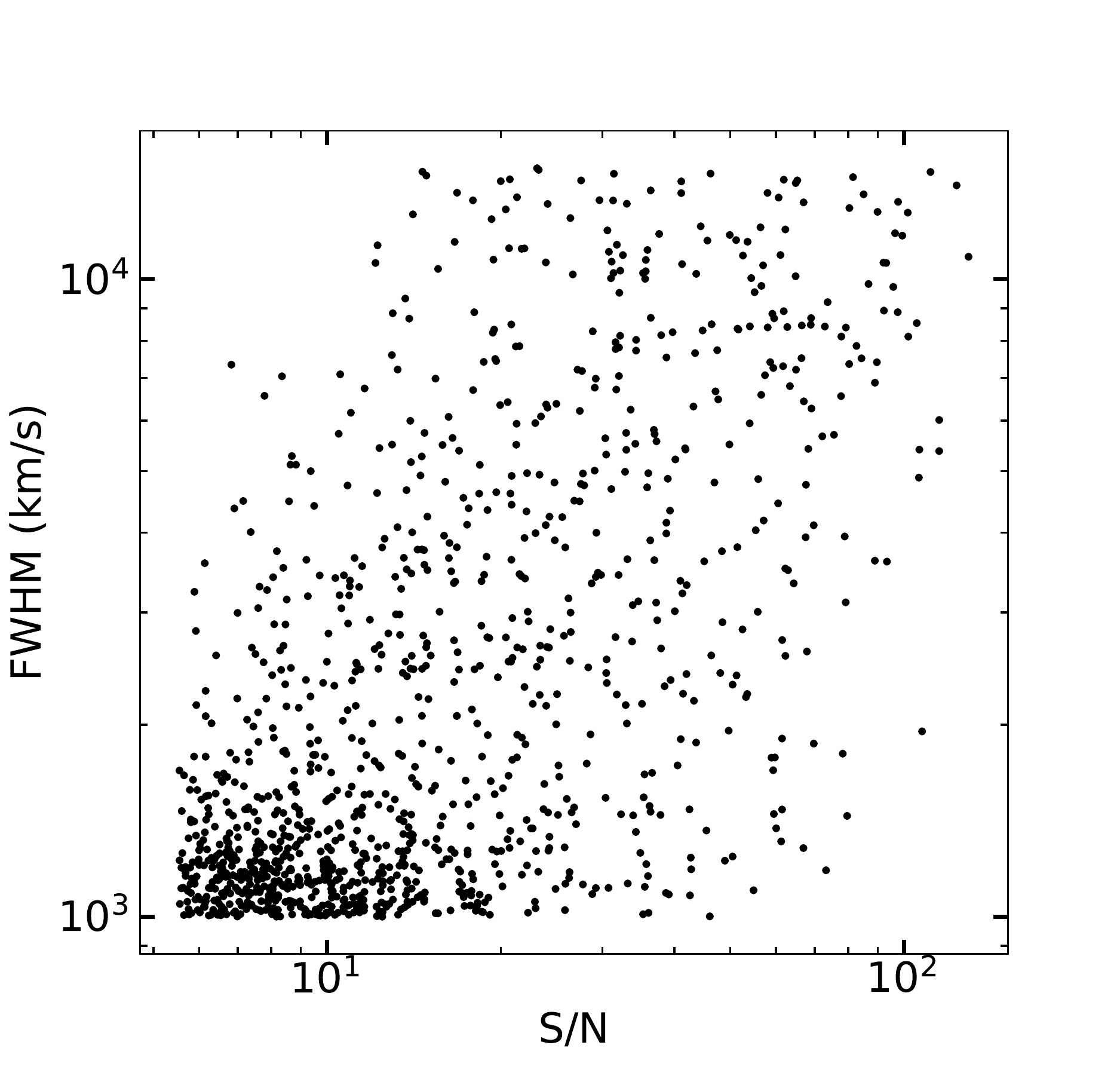}
\caption{Upper: The completeness of the HETDEX pipeline and that of the HETDEX AGN catalog out of the 1155 broad-line AGN identified in \cite{Zhang2021} as a function of the signal-to-noise ratio of their broad emissions. Bottom: The line widths versus the signal-to-noise ratio of the 1155 broad emissions identified in \cite{Zhang2021}. 
}
\label{f_yechi}
\end{figure}

\cite{Zhang2021} identified 1155 broad-line AGN in the HETDEX database. We explored the completeness of the HETDEX pipeline and that of the HETDEX AGN catalog as a function of the S/N of their broad emissions in the upper panel of Figure \ref{f_yechi}. The incompleteness of the HETDEX pipeline and the HETDEX AGN catalog at S/N\,$>$\,20, 15, 10, 5.5 are 8.6\%, 5.5\%, 3.7\%, 3.2\% and 41.5\%, 23.1\%, 12.7\%, 8.2\%, respectively.

The loss of AGN from the HETDEX pipeline is less than 10\%, even at very low signal-to-noise ratios. The loss of AGN of the HETDEX AGN catalog from the HETDEX pipeline is most severe at $\rm S/N<20$. Figure \ref{f_yechi} shows the degeneracy between the S/N and the FWHM in the bottom panel. The low detection rate of the HETDEX AGN catalog at low S/N is due to the low detection rate for line-widths less than 1500~km~s$^{-1}$.  The emission lines with 1000~km\,s$^{-1}$ $<$ FWHM $\lesssim 1500$~km\,s$^{-1}$ in the single Gaussian fit in \cite{Zhang2021} have a higher chance of falling slightly below 1000 km/s in our multi-Gaussian fit (Section \ref{sec_mg}) than broader line emission. This intermediate FWHM range is a grey region between the broad and the narrow line emission regime, and is therefore more likely to be rejected in the visual inspections at low signal-to-noise levels. For \cite{Zhang2021}, these ``intermediate broad'' emissions ($\rm 1000\ km/s < FWHM \lesssim 1500\ km/s$) were included in their catalog, because their goal was to bridge the luminosity functions of the narrow-line and broad-line emitters. Therefore, it does not matter if they call these sources as AGN or not. However, for our purpose of building an AGN catalog, it would be safe to exclude these ``intermediate broad emissions'' with low S/N. 

We also checked the completeness of the AGN selection (Section \ref{sec_selection}) by cross matching the SDSS DR14Q QSO catalog with the objects found by the HETDEX pipeline. These are relatively bright objects ($r,g\lesssim22.5$ mag), so our catalog is expected to be relatively complete at these magnitudes. There are 2,248 matched QSOs between the full SDSS DR14Q QSO catalog and the direct products of the HETDEX pipeline. Among these objects only 91 (4\%) of the matched QSOs were lost by the AGN selection code; we have added these objects back in our final AGN catalog with \texttt{sflag=0}. (see Section \ref{sec_boss} and descriptions of Column 158 in Appendix \ref{sec_append} for more details, and Figure \ref{f_specs_boss} as an example).

\begin{figure}[htbp]
\centering
\includegraphics[width=0.8\textwidth]{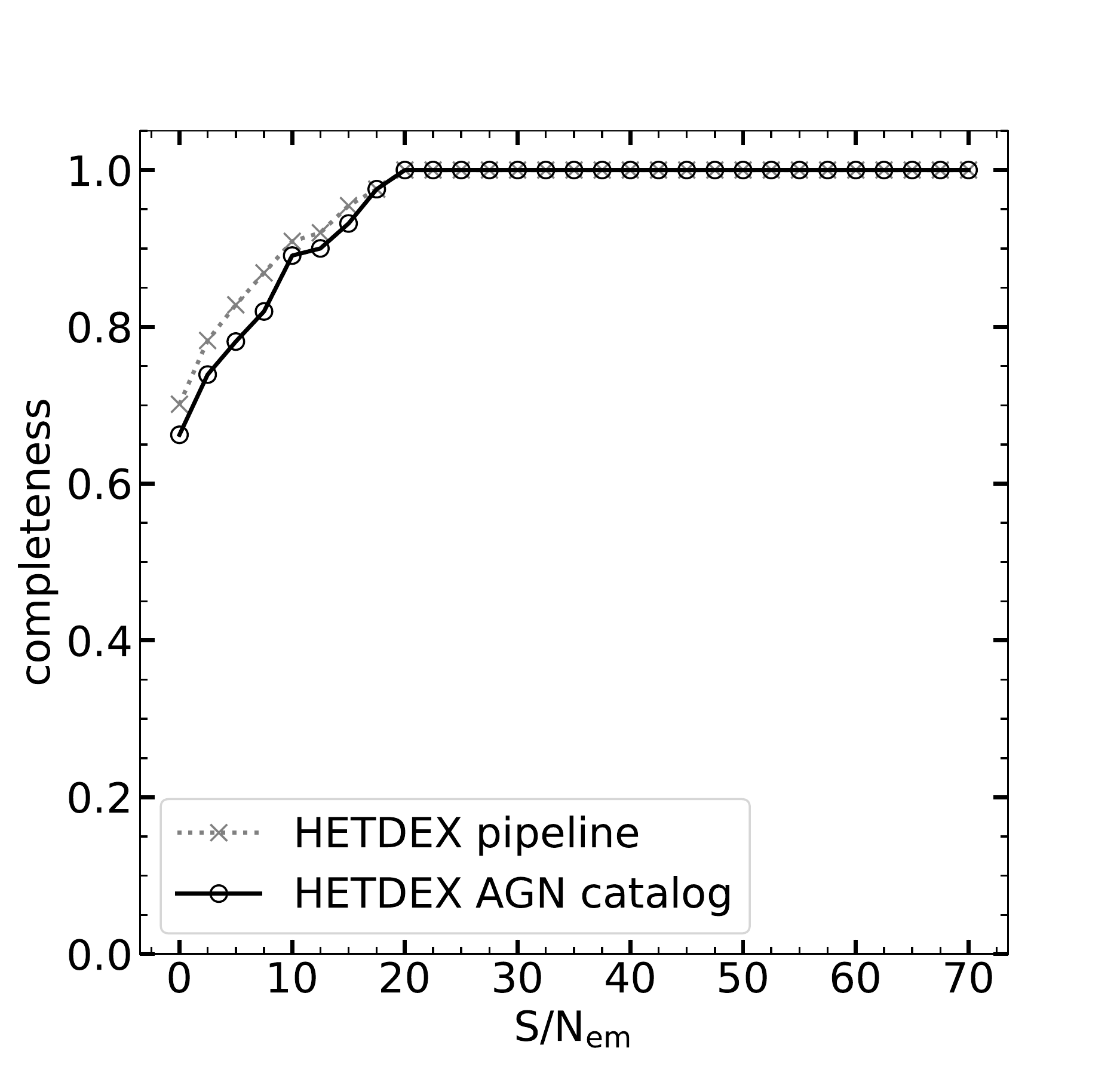}
\caption{The recovery fraction of 77 X-ray identified AGN with spectral redshifts that have HETDEX observations for the HETDEX AGN catalog (black open circles connected by solid line) and the HETDEX pipeline (grey crosses connected by dotted line) as a function of the signal-to-noise ratio of the emission lines.}
\label{f_kay}
\end{figure}

To further check the completeness of the HETDEX AGN catalog, we also performed a cross-comparison with the sample of 932 X-ray luminous AGN identified in \cite{Florez2020}.  The latter study matched galaxies in the $K_S$-selected DECam-NEWFIRM-IRAC catalog \citep{Stevans2021} with counterparts in the updated X-ray catalogs \citep{Ananna2017, LaMassa2019} from the Stripe 82X X-ray legacy survey \citep{LaMassa2016}. These catalogs have AGN with X-ray luminosity $\rm L_{2-10\ keV} \sim 10^{42}$~erg\,s$^{-1}$ in the hard (2-10 keV) X-ray band and a mix of spectroscopic and photometric redshifts.  This match leads to 932 AGN counterparts over the redshift range $0.5 < z < 3.0$ with $\rm L_{2-10\ keV} \sim 10^{42}\ erg$\,s$^{-1}$.  As noted by \cite{Florez2020},  these sources are overwhelmingly likely to be AGN since studies that use hard X-rays for selecting AGN \citep{LaMassa2019, Masoura2018, Brandt2015, Brandt2005} always classify sources with luminosities greater than $\rm L_{2-10\ keV} \sim 10^{42}\ erg$\,s$^{-1}$ as X-ray luminous AGN\null.  In contrast, galaxies without AGN typically have hard X-ray emission considerably below this threshold; only an extremely small number of starburst galaxies may exceed this threshold, but not by much \citep{Lehmer2008}.

Of these 932 X-ray AGN, 107 have HETDEX observations, and among these 107 sources, 77 have spectroscopic redshifts.  Out of this sample, we find that 51 (66.2\%) have matching counterparts in the HETDEX AGN catalog. Figure \ref{f_kay} shows a detailed detection rate of the HETDEX AGN catalog (black open circles connected by solid line) out of the 77 X-ray AGN. At S/N$>$5, 10, 15, 20, the HETDEX AGN catalog is 78.1\%, 89.1\%, 93.2\%, 100\% complete. The loss of AGN only happens at lower S/N, and is mainly rejected by the HETDEX pipeline (Section \ref{sec_detection}) before the AGN selection code (Section \ref{sec_selection}).

To sum up, the incompleteness of real broad emissions of the HETDEX pipeline is $\sim$ 10\% at $\rm S/N>4.5$, and $\sim$ 8.6\% at $\rm S/N>5.5$. The incompleteness of AGN of the HETDEX AGN catalog in the magnitude-limited sample identified in the HSC survey ($r=26$ in the wide fields, and $r=28$ in the deep fields) is 35.7\%, but the majority of the missed AGN are the ones with intermediate line widths (1000~km\,s$^{-1}$ $<$ FWHM $\lesssim 1500$~km\,s$^{-1}$) and lower S/N. When comparing with the bright QSO catalog from SDSS \citep{Paris2018}, the AGN loss rate of our AGN selection code from the HETDEX pipeline produced raw catalog is 4\%. When comparing with the X-ray luminous AGN, the HETDEX AGN catalog is 78.1\% complete at $\rm S/N>5$. We have carefully estimated the completeness corrections for every AGN in our catalog by making simulated spectra with controlled properties, such as the line flux, the line widths, etc, and run the simulated spectra through the HETDEX pipeline and the AGN selection code. We will introduce this with details in our next paper of this series that studies the luminosity function of the HETDEX AGN (Liu et al. in preparation).

\section{The catalog}
\label{sec_catalog}

The first AGN catalog of the HETDEX survey consists 5,322 unique objects covering 30.61 deg$^2$ over $0.25<z<4.32$. Slightly over 70\% of the AGN have secure redshifts \texttt{zflag=1}, while the rest are type-I AGN candidates with estimated redshifts (\texttt{zflag=0}). Among the 3,733 \texttt{zflag=1} AGN, there are 1,486 HETDEX AGN new to SDSS DR14Q \citep{Paris2018}. The raw surface density of AGN is 173.86 $\rm deg^{-2}$. The completeness corrected surface density is 339.65 $\rm deg^{-2}$ for HETDEX AGN with completeness greater than 0.05.

SDSS DR14Q has 523,638 QSOs within the redshift range of the HETDEX AGN ($0.25<z<4.32$), spanning 9376 $\rm deg^2$, giving a raw surface density of 55.8 $\rm deg^{-2}$. \cite{Paris2018} provides the completeness corrected surface density of SDSS QSOs within $0.9<z<2.2$ as 125.03 $\rm deg^{-2}$, while that of the HETDEX AGN within the same redshift interval is 172.98 $\rm deg^{-2}$ for the ones with completeness greater than 0.05.

Version 1 of the AGN catalog is made available on a public website\footnote{\url{http://web.corral.tacc.utexas.edu/hetdex/HETDEX/catalogs/agn_catalog_v1.0/}} and in the online Journal (Table \ref{t_catalog}). 
The catalog is available as a FITS file with six extensions. 
We include a simple Jupyter notebook to show how to use this FITS file in Python.
The FITS extensions are as follows.
  
\begin{enumerate}
  \item Table extension (5322 rows $\times$ 163 columns): Basic information, one row for one unique AGN, arranged in a descending order of redshifts. Column information can be found in the header, and also summarized in Appendix \ref{sec_append}. We note that if any measurement is not available, it is set to the default value of $-99$. For example, all measurements for the \ion{Mg}{2} $\lambda2799$ emission of $z > 2.5$ AGN are set to $-99$, because this emission is out of the wavelength coverage of the HETDEX survey.
      
  \item Image extension (5322 rows $\times$ 1036 columns): 1-D spectra for the $\rm detectid_{best}$ of each AGN (a universal $\rm E(B-V)=0.02$ extinction correction applied). The dust law used was \cite{Cardelli1989} assuming $\rm R_V=3.1$. Rows are arranged in the same order as extension 1. The wavelength array starts from 3470\,\AA\ and contains 1036 elements with a stepsize of 2.0\,\AA\null. The fluxes are in units of  $10^{-17}$ erg\,cm$^{-2}$\,s$^{-1}$\,\AA$^{-1}$. The wavelength solution and the flux units are recorded in the extension header.
      
  \item Image extension (5322 rows $\times$ 1036 columns): The error array for extension 2, with the same wavelength solution and flux units.

  \item Table extension (6004 rows $\times$ 3 columns): Repeat information for each AGN. Some AGN were observed multiple times. In this table, each unique observation is listed with their \texttt{shotid}. Column 1 is the \texttt{agnid}, Column 2 gives the number of repeat observations of \texttt{agnid}, and Column 3 gives the \texttt{shotid} of the observation. An AGN observed multiple times will have multiple entries in this table with the same \texttt{agnid} and \texttt{nshots}, but different \texttt{shotid}. A value of \texttt{nshot=1} means no repeats; the AGN was only observed once. The first eight digits of \texttt{shotid} is the date the observation is taken. For example, \texttt{shotid=20191006024} says the observation is the 24th exposure taken on UT Date 2019-Oct-06.
     
  \item Image extension (6004 rows $\times$ 1036 columns): 1-D spectra at the fof center, repeat observations included, no extinction correction applied. Rows are arranged in the same order as extension 4. The wavelength solution and flux units are the same as extension 2, and can be found in the header.
      
  \item Image extension (6004 rows $\times$ 1036 columns):  Error array for extension 5.
\end{enumerate}

\begin{figure}[!htbp]
\centering
\includegraphics[width=1.21\textwidth]{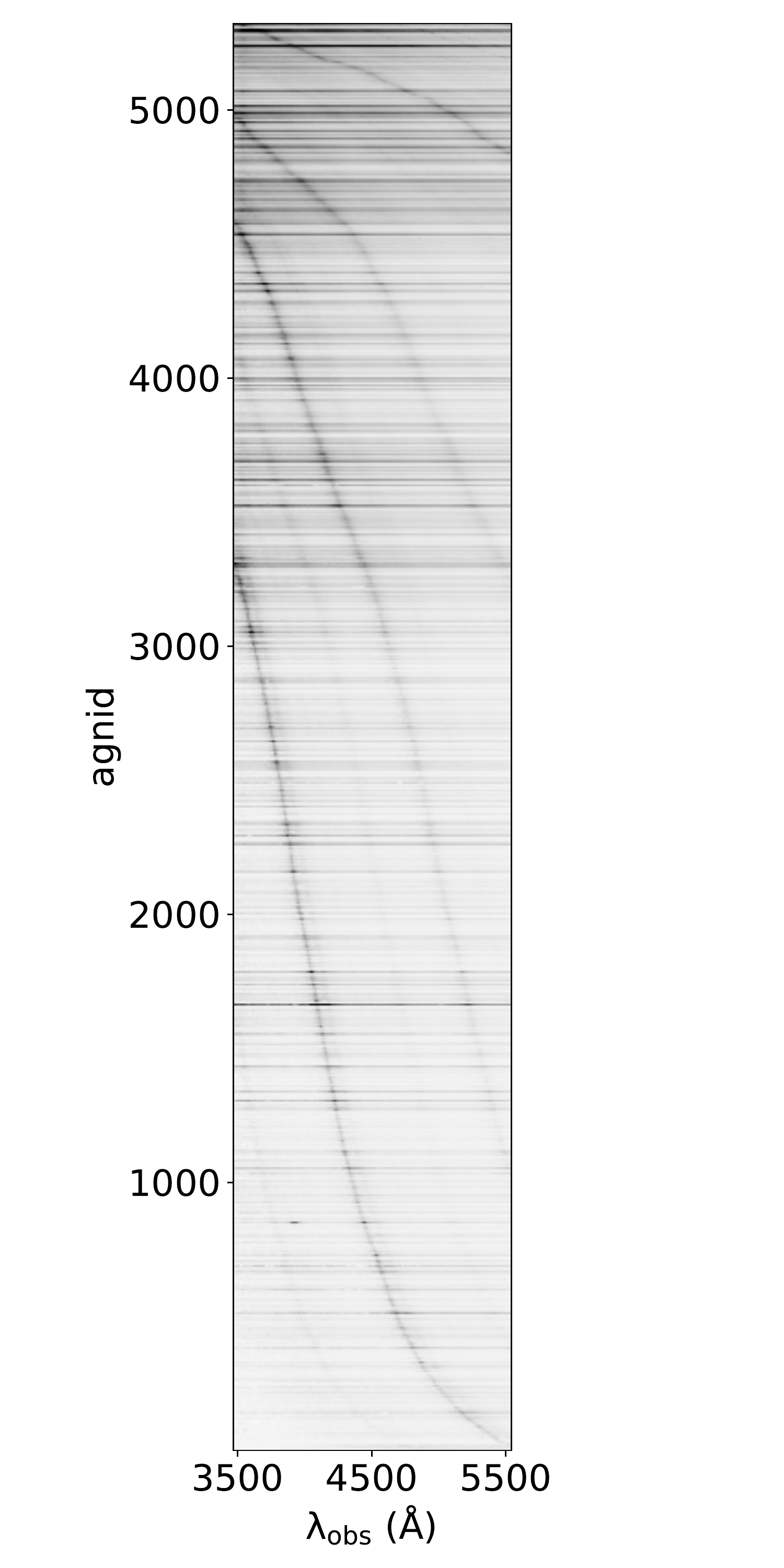}
\caption{The image in the extension 2 of the FITS file. The vertical direction has the AGN arranged with descending redshifts from bottom to top. \texttt{agnid} is the unique sequential numerical identifiers for each AGN. The horizontal direction has the 1-D spectrum with increasing observed wavelengths from left to right.}
\label{f_ext2}
\end{figure}

Figure \ref{f_ext2} shows the second extension of the FITS file. 
The four strongest emission lines in the image, from bottom to top and from left to right, are Ly$\alpha$, \ion{C}{4} $\lambda1549$, \ion{C}{3}] $\lambda1909$, and \ion{Mg}{2} $\lambda2799$. The weak emission from \ion{O}{6} $\lambda1034$ shows up to the bottom left of the Ly$\alpha$. The \ion{Si}{4}+\ion{O}{4}] blend at $\lambda 1400$ is also visible between Ly$\alpha$ and \ion{C}{4} $\lambda1549$. We will re-visit the relative line ratio between different emissions in Sections \ref{sec_ew} and \ref{sec_compspec}.

\section{Statistics}
\label{sec_statistics}

\subsection{Redshift Distribution}
\label{sec_redshifts}

\begin{figure}[htbp]
\centering
\includegraphics[width=\textwidth]{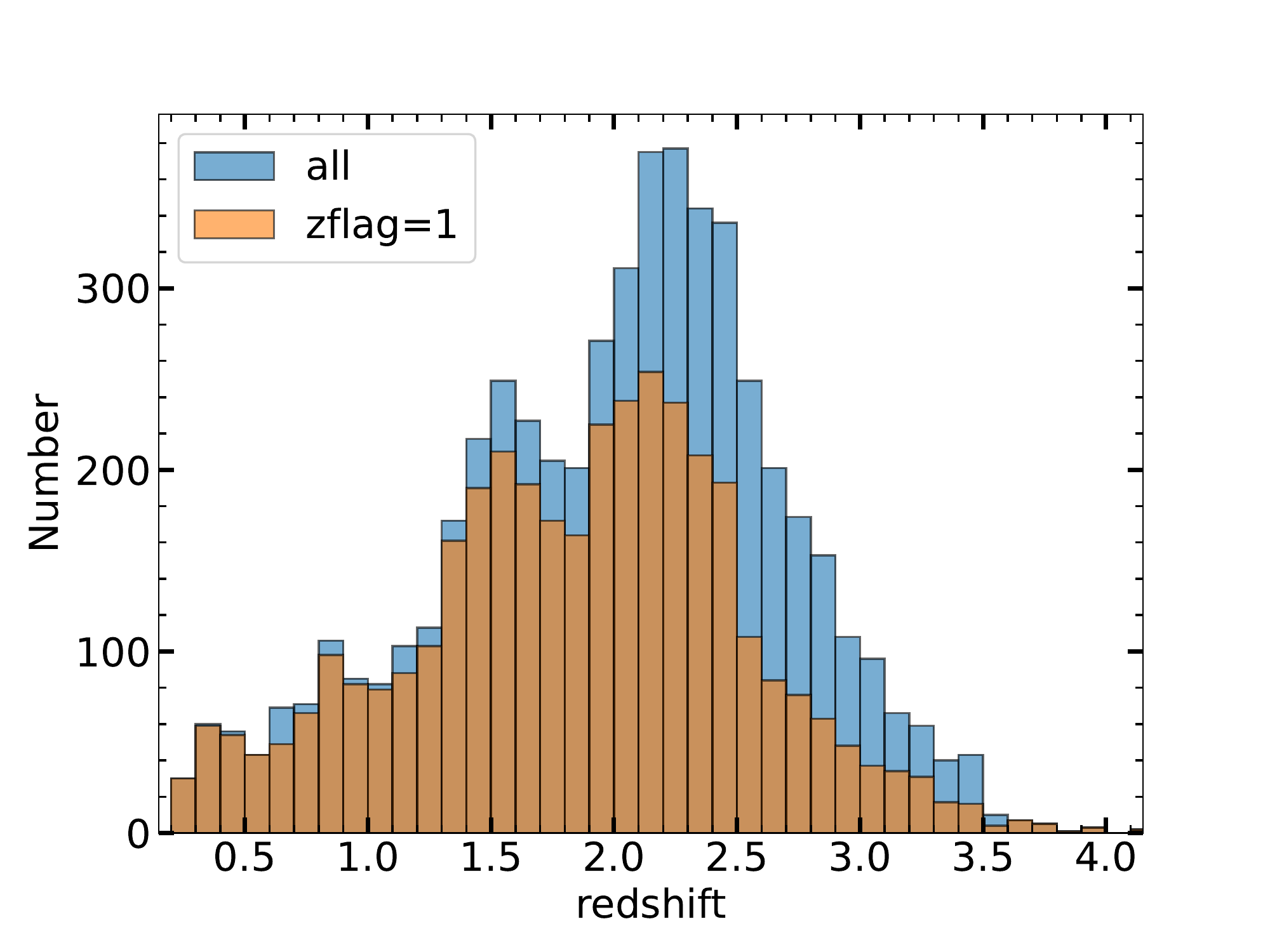}
\caption{The redshift distribution for all AGN (blue) and AGN with secure reshifts (orange).}
\label{f_z}
\end{figure}

Figure \ref{f_z} shows the distribution of the redshifts of the full AGN catalog (blue), and the AGN with secure redshifts only (orange). The former sample has a median redshift of $z=2.1$, and the latter sample has a median redshift of $z=1.9$. Neither sample is uniformly distributed, with peaks at $z\sim2.2$, $z\sim1.5$, and $z\sim0.9$. These are the redshifts where the line pairs Ly$\alpha$+\ion{C}{4} $\lambda1549$, \ion{C}{4} $\lambda1549$+\ion{C}{3}] $\lambda1909$ and \ion{C}{3}] $\lambda1909$+\ion{Mg}{2} $\lambda2799$ are simultaneously available for our 2em selection algorithm. The redshift distribution of SDSS DR14Q also show such redshift peaks due to similar reasons \citep{Paris2018}. With careful completeness corrections the impact of these peaks on the statistics of AGN can be accounted for. We will introduce the completeness correction with more details in our next paper of the luminosity function (LF) of our AGN catalog (Liu et al. in preparation). 

\subsection{Photometric Properties}
\label{sec_photo}


Traditional surveys for QSO rely heavily on the photometry for target selection due to their instrumental designs. For example, the SDSS spectroscopic survey only targets objects found of interest to the project.  The pre-selection of QSOs requires them to have point-like morphologies, continua bright enough for accurate photometry, and colors that distinguish the QSOs from stars.  In contrast, HETDEX surveys the sky with no target pre-selection via the use of 78 IFUs (34,944 fibers) fed into 156 spectrographs, which cover $\sim$ 1/5 of the HET's useable $18\arcmin$ field of view. The HETDEX software automatically searches through this database of $> 10^9$ spectra (most of which are blank sky), and identifies emission-line and continuum sources which fall onto the fiber arrays. Our selection code then reads this database and identifies AGN from their characteristic emission lines and line widths. 
In this Section, we compare our HETDEX AGN catalog with the SDSS DR14Q QSO catalog, and discuss the AGN population(s) rejected by the traditional magnitude cuts (Section \ref{sec_r}) and color selections (Section \ref{sec_color}).

\subsubsection{Magnitudes}
\label{sec_r}

\begin{figure}[!htbp]
\centering
\includegraphics[width=\textwidth]{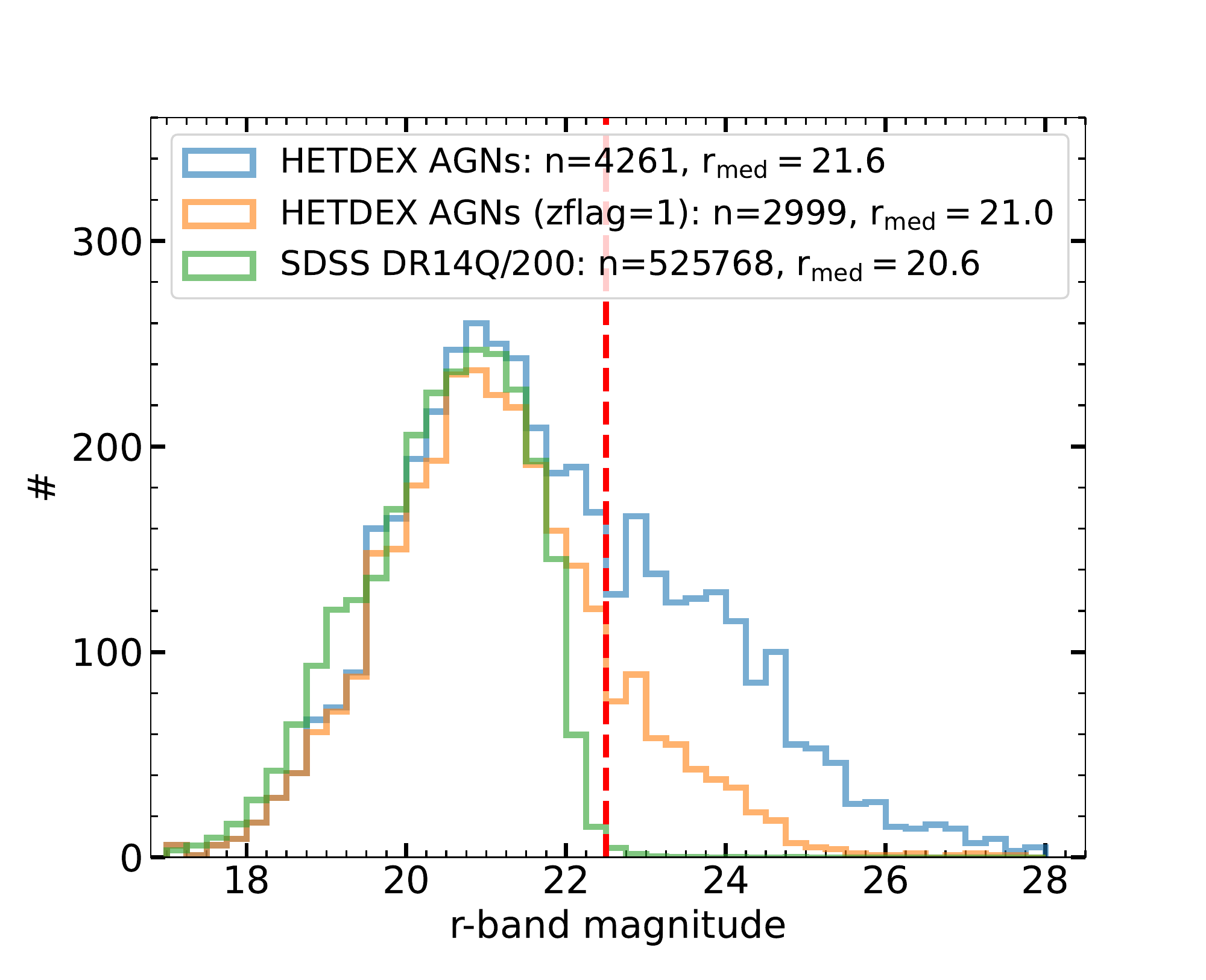}\\
\caption{The distribution of the $r$-band aperture magnitudes (see the descriptions of Column 135-139 in Appendix \ref{sec_append} and Appendix \ref{sec_appendix_photo} for more details of the measurements). The blue histogram is the distribution for all HETDEX AGN with coverage from four $r$-band imaging surveys that we have searched. The orange histogram includes only those HETDEX AGN with secure redshifts. The green histogram is the distribution for the full SDSS DR14Q catalog. The red dashed vertical line at $r=22.5$ roughly shows the flux limit of SDSS.}
\label{f_r}
\end{figure}


Figure \ref{f_r} compares the distribution of $r$-band aperture magnitudes in our HETDEX AGN catalog with that of the SDSS QSO catalog.  We note that our aperture magnitudes are not derived from a simple cross match between an existing imaging catalog with the HETDEX detections. The ELiXer software (Emission Line eXplorer; the primary automated classification and redshift identification application for HETDEX, combining multiple information sources, including emission line fitting, line and equivalent width ratios, archival photometric imaging, and others to make its determinations, more details will be found in Davis et al. in preparation) puts an aperture at the HETDEX detection on the imaging and then measures the enclosed flux with a proper aperture size, the aperture size when the enclosed flux stops growing as the aperture size increases. Details can be found in the descriptions of Column 135-138 in Appendix \ref{sec_append}\null. The aperture magnitudes allow the measurements of the AGN with very faint continuum, which can be close or even beyond the detection limits of the $r$-band imaging. 
For the continuum bright imaging sources, the aperture magnitudes agree well with the total magnitudes recorded in the imaging catalogs. The deepest $r$-band survey searched by ELiXer is the HSC survey, whose 5$\sigma$ limit is $\sim$ 26 mag in the wide fields, and $\sim$ 28 mag in the deepest regions.

\begin{figure}[htbp]
\centering
\includegraphics[width=\textwidth]{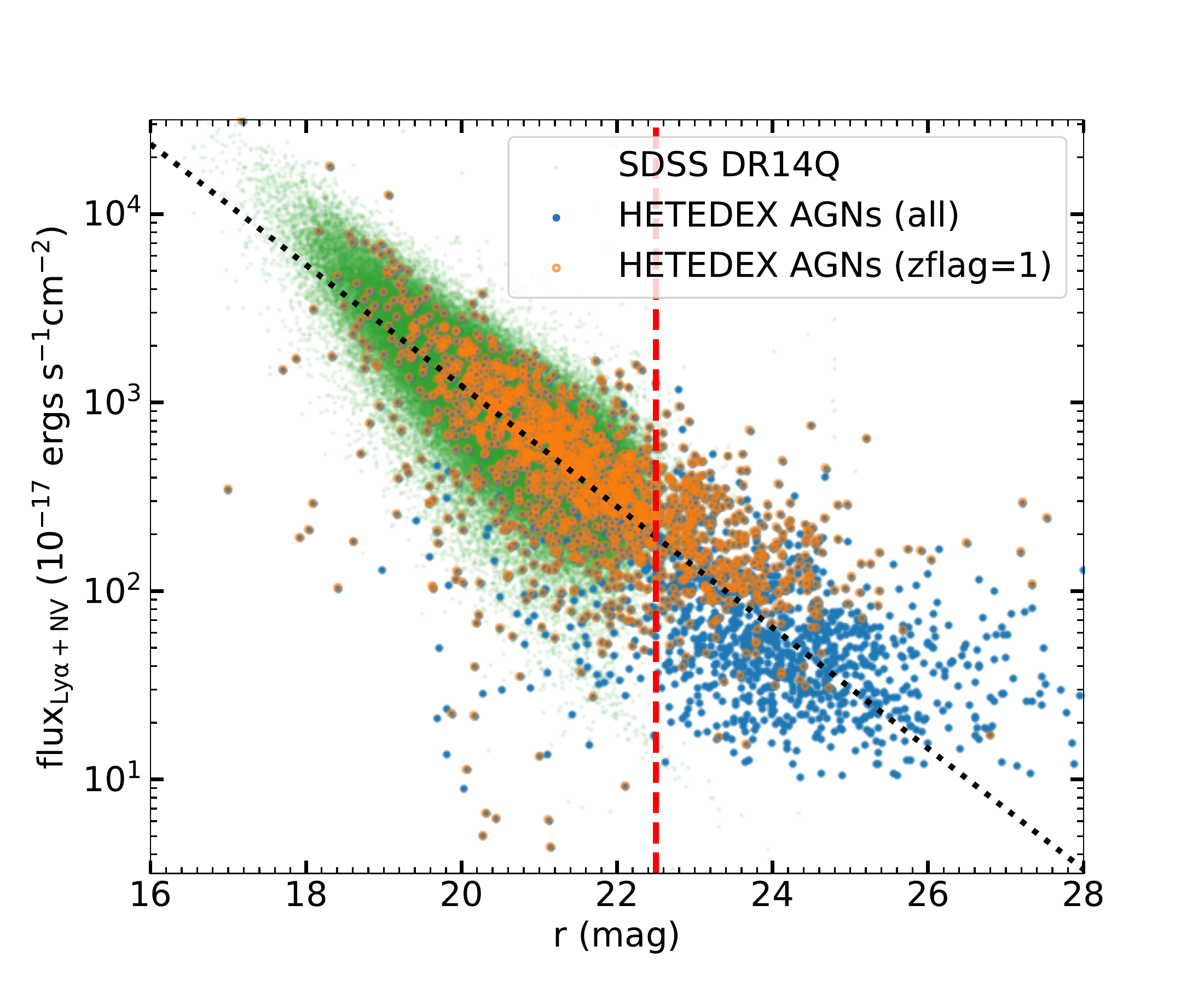}\\
\caption{The flux of the Ly$\alpha$+\ion{N}{5} $\lambda1241$ emission versus the $r$-band aperture magnitudes for the full HETDEX AGN catalog (blue), the HETDEX AGN with secure redshifts (orange), and the SDSS DR14Q catalog (green). The red dashed line again shows $r=22.5$ where the SDSS QSOs stop. The black dotted line ($\log$ flux(Ly$\alpha$+\ion{N}{5}) = 
$-0.32\,r + 9.5$) is a simple linear fit to the SDSS data points.}
\label{f_Lya_r}
\end{figure}

\begin{figure*}[!htb]
\centering
\includegraphics[width=0.3\textwidth]{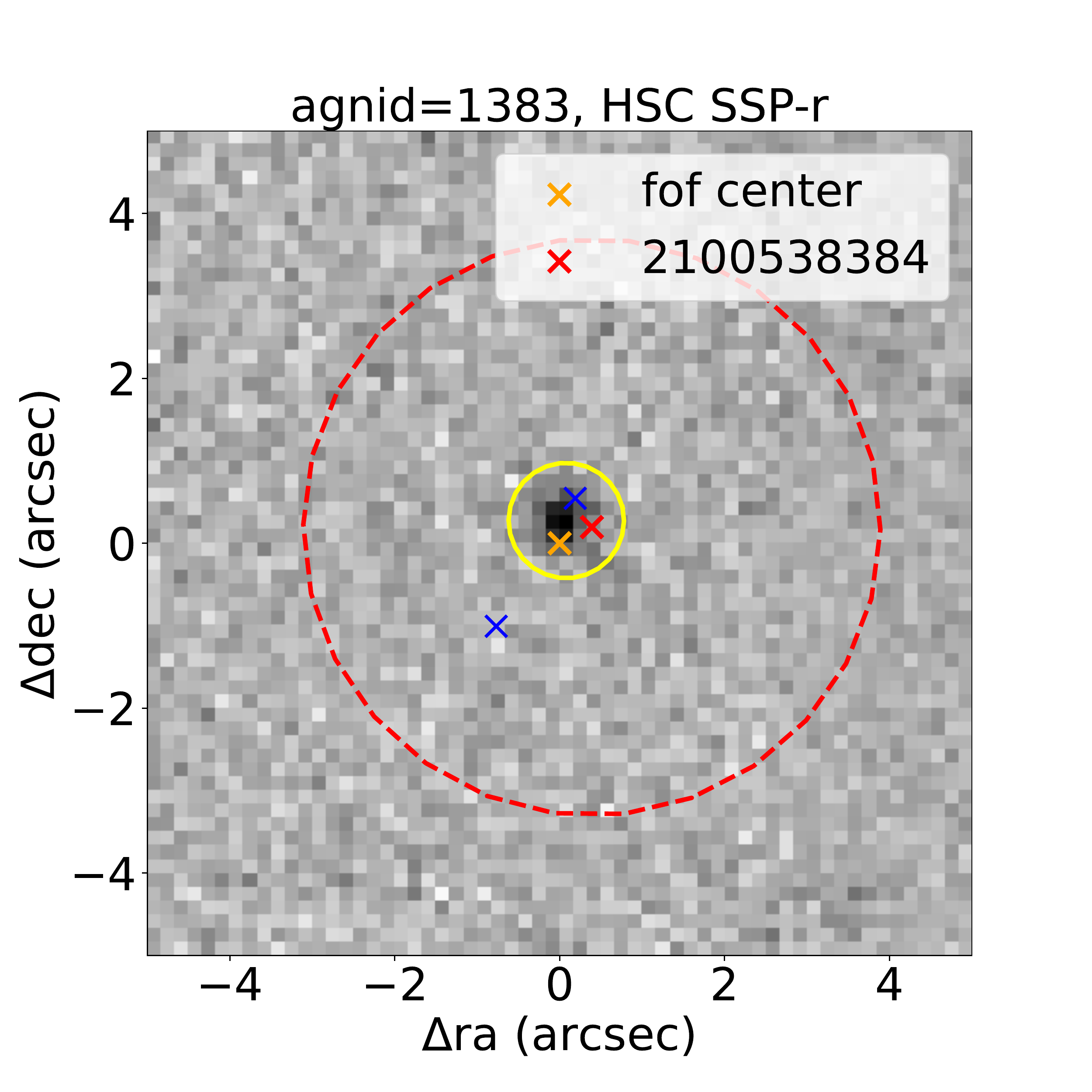}
\includegraphics[width=0.69\textwidth]{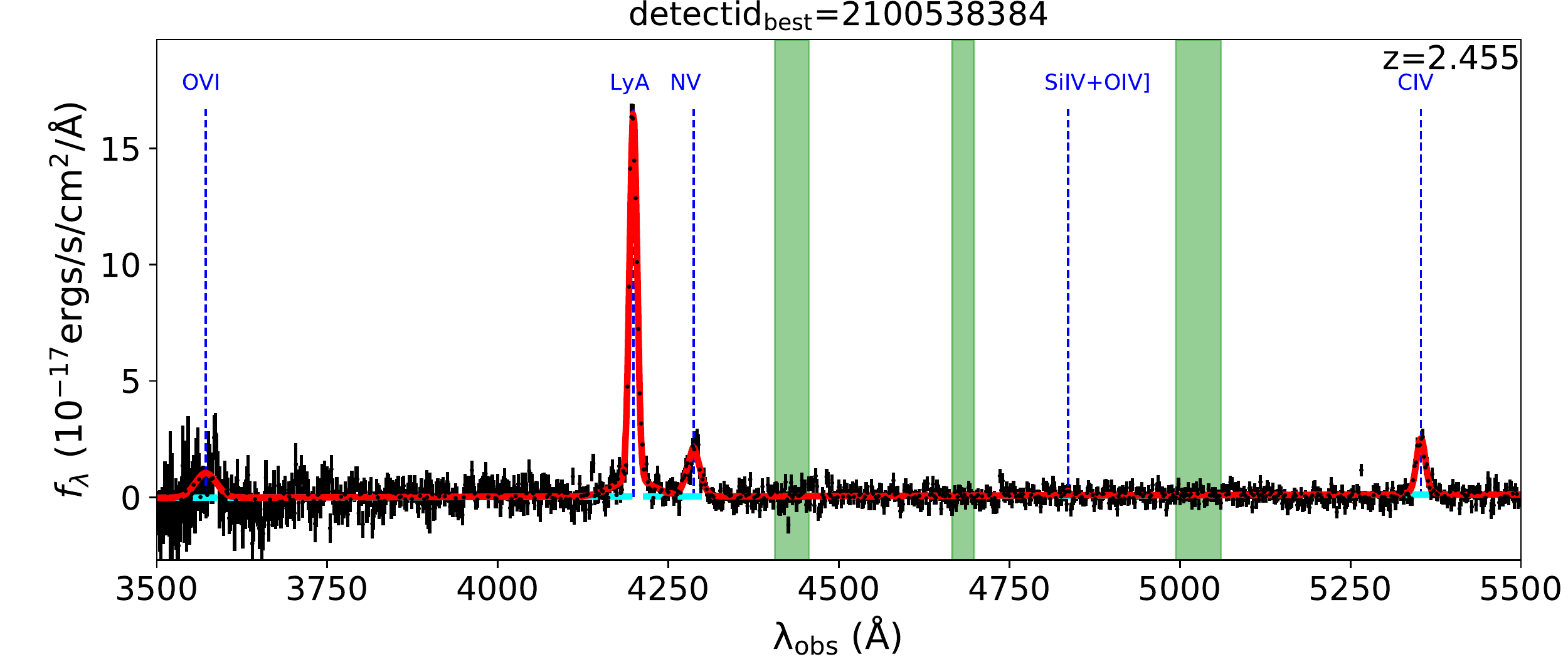}\\
\includegraphics[width=0.3\textwidth]{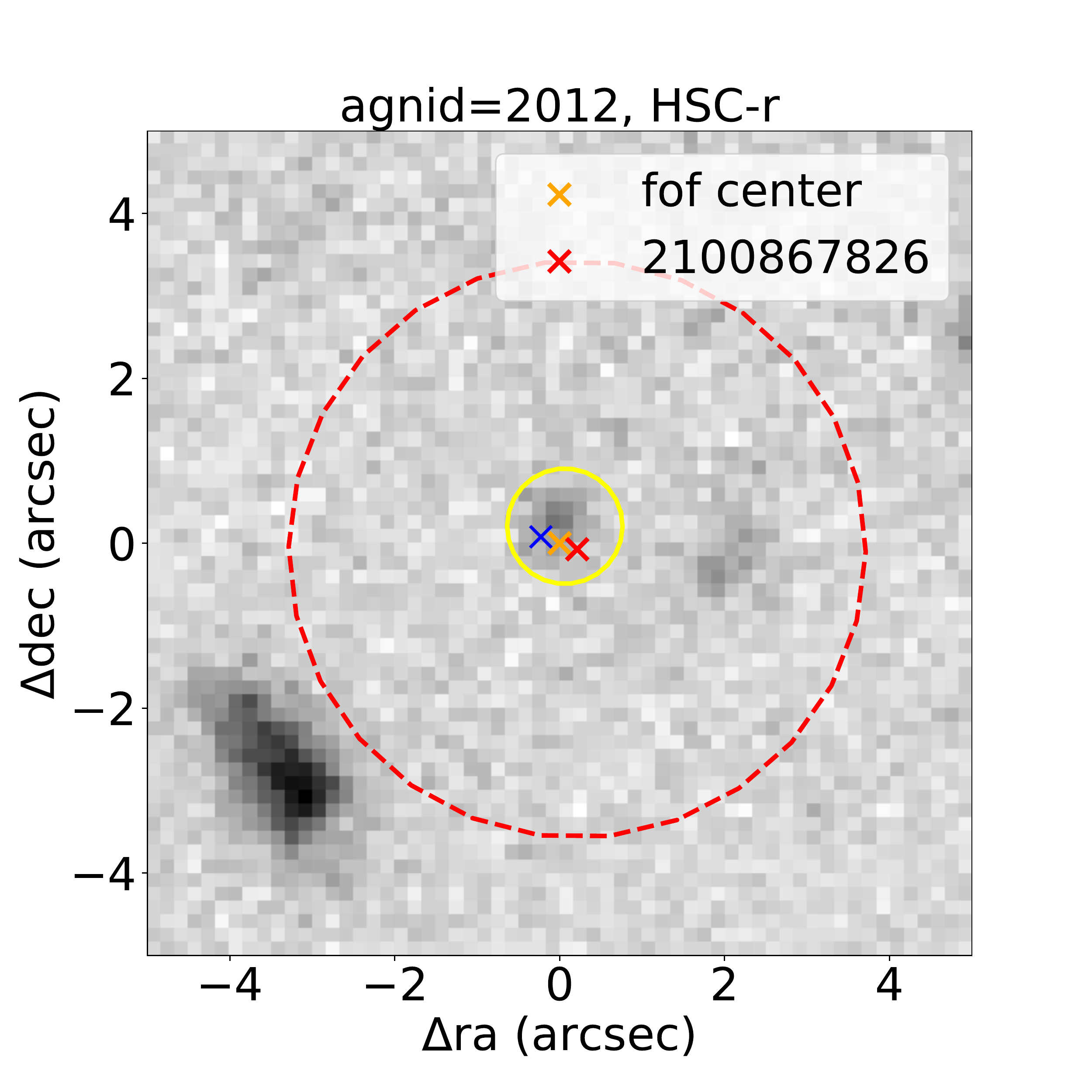}
\includegraphics[width=0.69\textwidth]{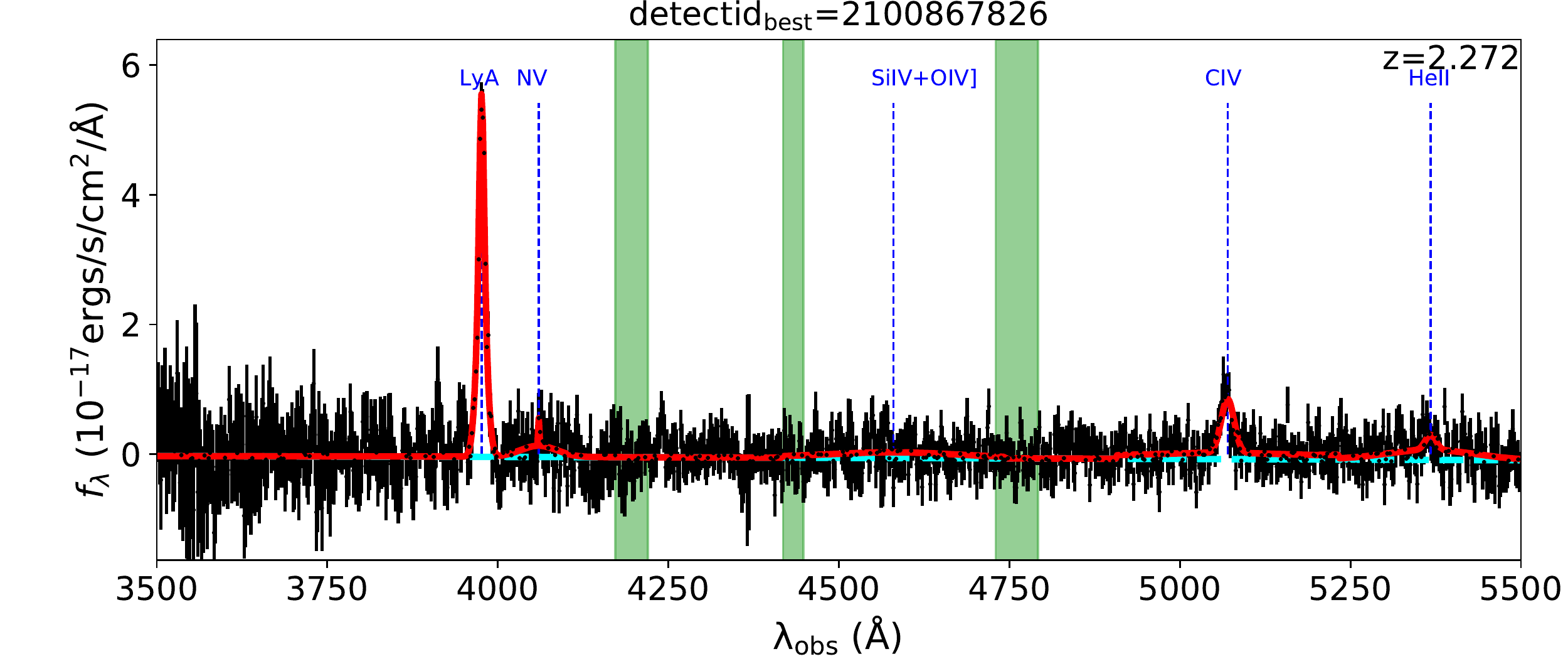}\\
\includegraphics[width=0.3\textwidth]{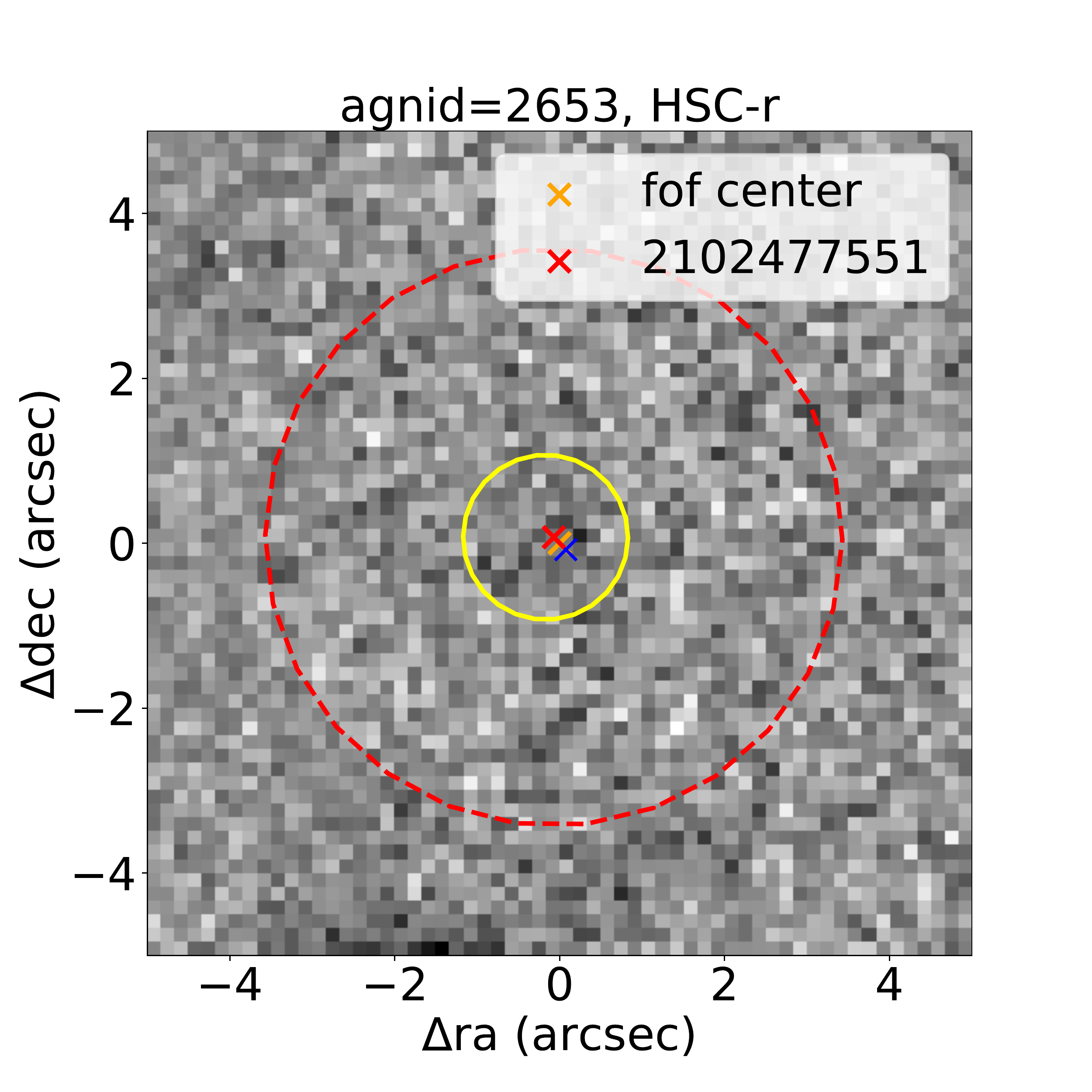}
\includegraphics[width=0.69\textwidth]{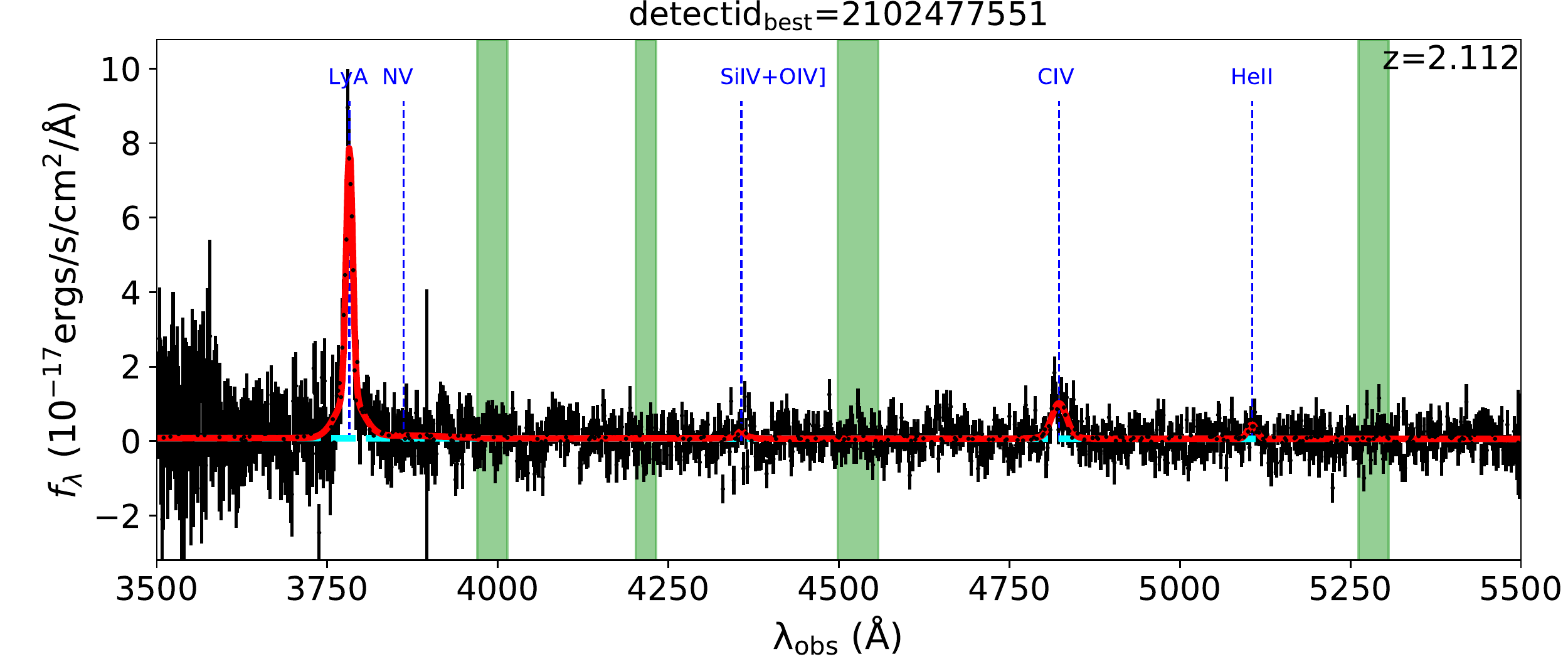}\\
\caption{The $r$-band HSC images and the HETDEX spectra of three AGN with faint continua ($r>22.5$). This figure has the same design with Figure \ref{f_specs_2em}. The yellow circles are added to show the photometric apertures used by the ELiXer software. The r-band aperture magnitudes for AGN 1383, 2012, 2653 are 24.1, 25.1, and 26.0, measured with the aperture sizes of $1\farcs 4$, $1\farcs 4$, and $2\farcs 0$ in diameter. The 5$\sigma$ depths of these three fields are 26.4, 26.2, and 26.2 respectively.} 
\label{f_faint}
\end{figure*}

When comparing the median magnitudes, the HETDEX AGN ($r_{\rm med}=21.6$) are overall one magnitude fainter than the SDSS QSOs ($r_{\rm med}=20.6$). When comparing the magnitude limits, the HETDEX AGN can be as faint as $26\sim28$ mag, which is about 4$-$5 magnitudes fainter than the effective limit of the SDSS survey ($r\sim22.5$). A simple cut at $r<22.5$ would remove 34\% of the AGN from the full HETDEX AGN catalog (the blue histogram in Figure \ref{f_r}), and 12\% of the objects with secure redshifts (the orange histogram in Figure \ref{f_r}).

Figure \ref{f_Lya_r} shows the flux of the Ly$\alpha$+\ion{N}{5} $\lambda1241$ emission versus the $r$-band aperture magnitudes for the HETDEX AGN at $1.9<z<3.5$ and the SDSS QSOs. The emission-line measurements for the SDSS DR14Q catalog are taken from \cite{Rakshit2020}. There is a strong correlation between the strengths of the emissions, indicated by the flux of Ly$\alpha$+\ion{N}{5} $\lambda 1241$ and that of the AGN's continua, as measured by its $r$-band magnitudes. We fit a simple linear model to the SDSS QSOs shown by the black dotted line. The HETDEX AGN follow the correlation of the SDSS QSOs, and extend to fainter magnitudes. This indicates that the blind spectroscopic survey is not finding a distinctive AGN population with high equivalent width Ly$\alpha$ lines, but is identifying AGN with low continuum magnitudes that are missed by other surveys. The relation between  emission line and continuum strength for the low-luminosity AGN remains the same as for the bright QSOs. With no pre-selection based on continuum magnitude, the emission lines of the low-luminosity AGN ($r>22.5$) at $z\sim2.5$ can be detected with high enough S/N by ground-based 10-m telescopes.


Figure \ref{f_faint} shows three examples of continuum-faint HETDEX AGN\null. None of the objects appear in the SDSS DR14Q catalog, due to their low continuum luminosity. AGN 1383 has a $r=24.1$ measured with a $d=1\farcs 4$ aperture, which is about two magnitudes more luminous than the 5$\sigma$ detection limit of this field. AGN 2012 is well detected in both the Ly$\alpha$ and  \ion{C}{4} $\lambda1549$, and its aperture magnitude, as measured within the yellow circle ($d=1\farcs 4$), is $r = 25.1$, or about one magnitude more luminous than the 5$\sigma$ limit of this field. The \ion{C}{4} $\lambda1549$ emission of AGN 2653 is not as significant as that of AGN 2012, but it is still fitted with a signal-to-noise of 6.7, which is high enough to pass the 2em criterion for the second most significant emission line within the wavelength range of the survey ($\rm S/N_{em,2nd}>4$; see Section \ref{sec_2em} for more details). 
The S/N of an emission line is measured within a wavelength interval that is $\pm 2\sigma$ either side of the centroid. For this object, the magnitude, as measured by ELiXer is $r=26.0$ with a $2\farcs 0$ diameter aperture.  
AGN 2653 is almost at the 5$\sigma$ detection limit of this field. 

\subsubsection{Colors}
\label{sec_color}

\begin{figure}[htbp]
\centering
\includegraphics[width=\textwidth]{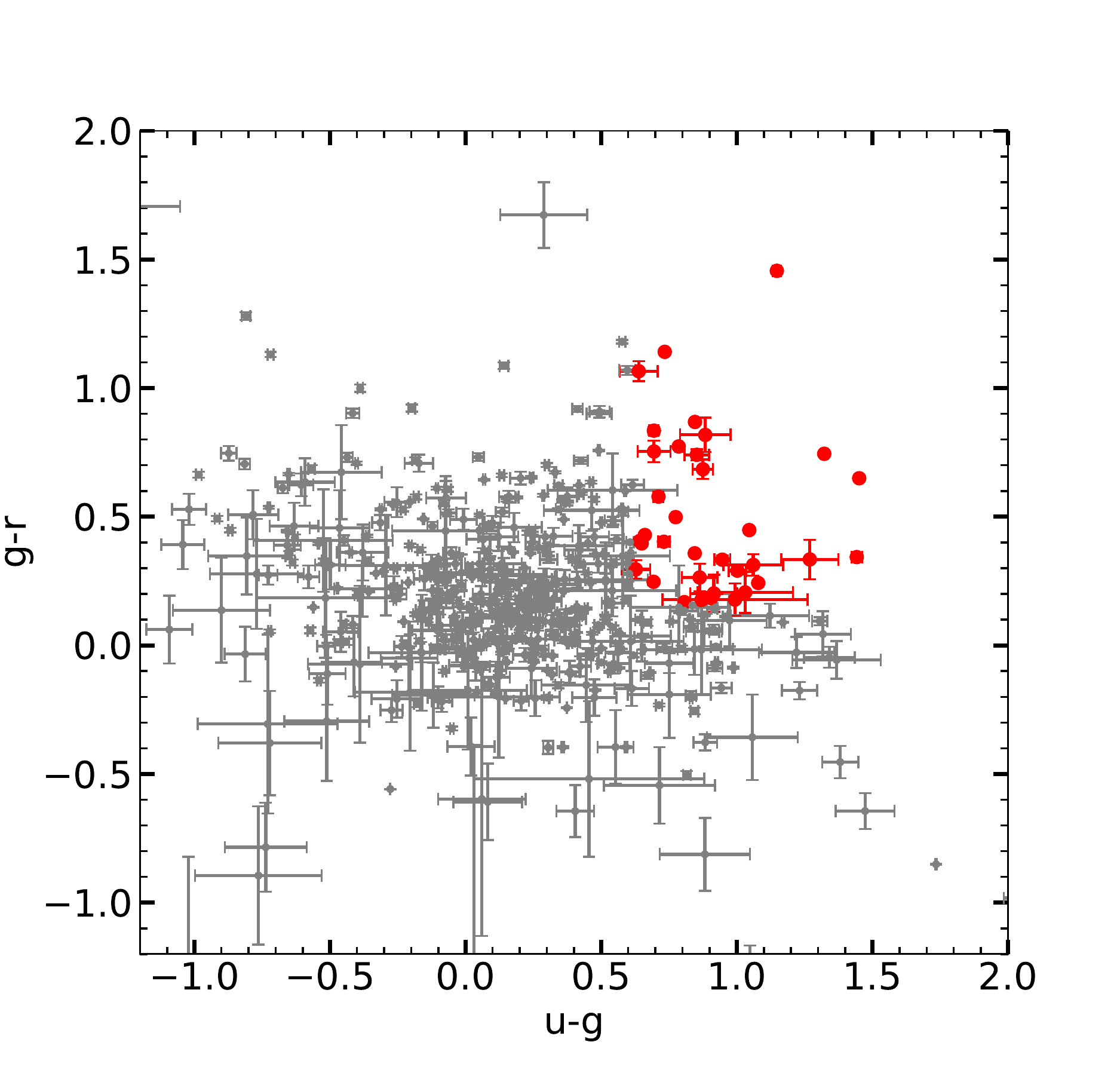}\\
\caption{ The ($g-r$) versus ($u-g$) diagram for the 596 HETDEX AGN at $z<2.5$ with photometric measurements in the $u$, $g$, and $r$ bands (see Appendix \ref{sec_appendix_photo} for more details). The red data points show the 6\% of AGN that have $u-g>0.62$ and $g-r>0.16$. }
\label{f_ugr}
\end{figure}

Point-like continuum bright sources can either be QSOs, or stars. Colors between different bands are often used to distinguish the different continuum shape of the stars and the QSOs. SDSS has a complex color selection criterion for the effective and efficient selections of QSOs at different redshifts \citep{Ross2012}. The VVDS survey \citep{LeFevre2013}, whose AGN catalog is selected from a pure magnitude-limited parent sample, chose a very simple selection criterion based on two colors to estimate the loss rate of AGN in the color selection: $u-g>0.62$ and $g-r>0.16$ \citep{Gavignaud2006}. This color cut is believed to be efficient up to $z\sim2.5$, based on the evolutionary track of the SDSS composite spectrum of QSOs \citep{VandenBerk2001}. Figure \ref{f_ugr} shows the distribution of HETDEX AGN at $z<2.5$ on the $g-r$ and $u-g$ plane. There are only 596 HETDEX AGN at $z<2.5$ covered by imaging surveys in all $u$, $g$, and $r$ bands (see Appendix \ref{sec_appendix_photo} for more details). The AGN that would be rejected using the color cuts of \cite{Gavignaud2006} are shown as red data points. The percentage of the estimated loss of AGN is about 6\% with color selections.

\subsection{EW}
\label{sec_ew}

\begin{figure}[htbp]
\centering
\includegraphics[width=\textwidth]{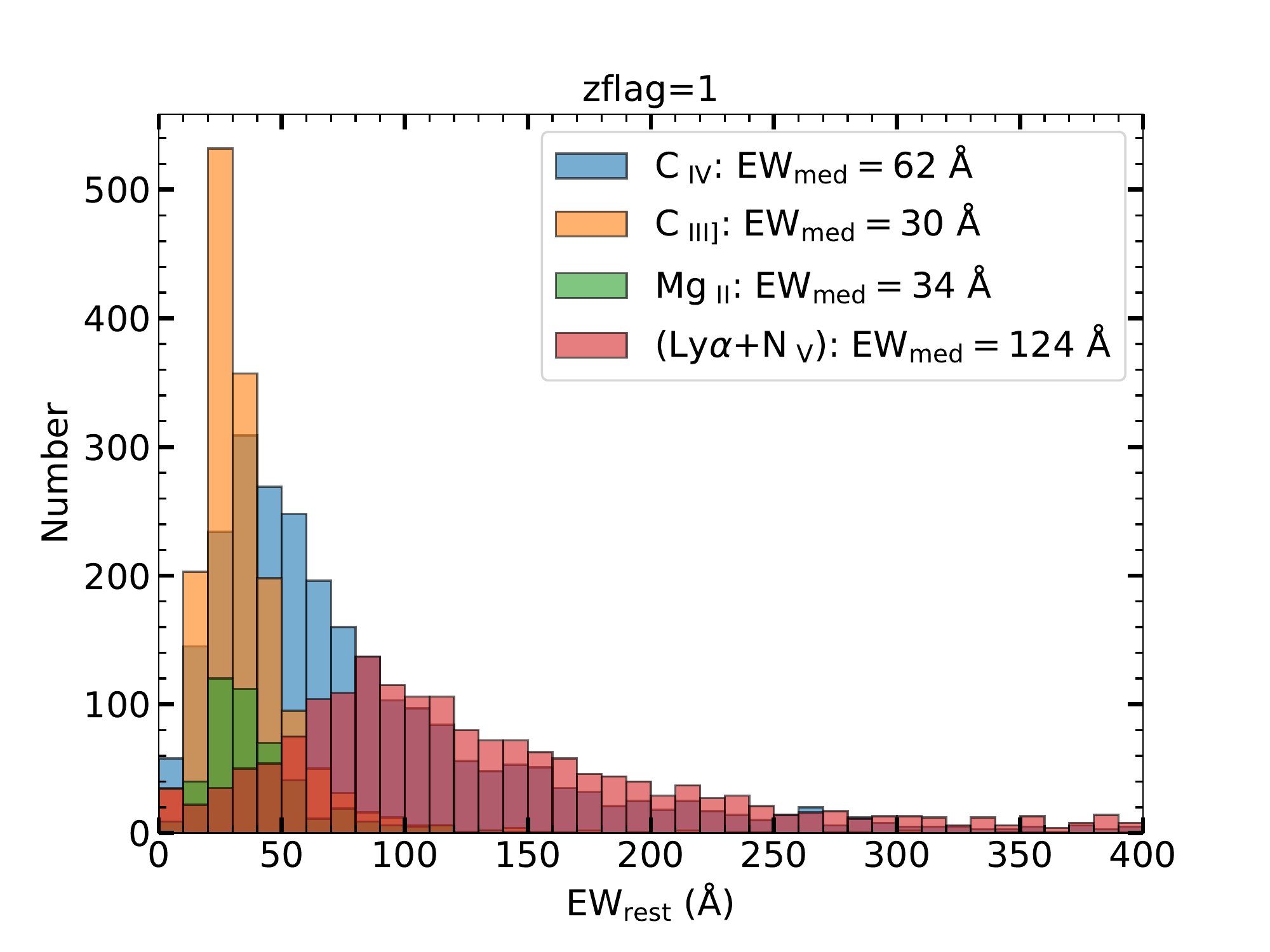}\\
\includegraphics[width=\textwidth]{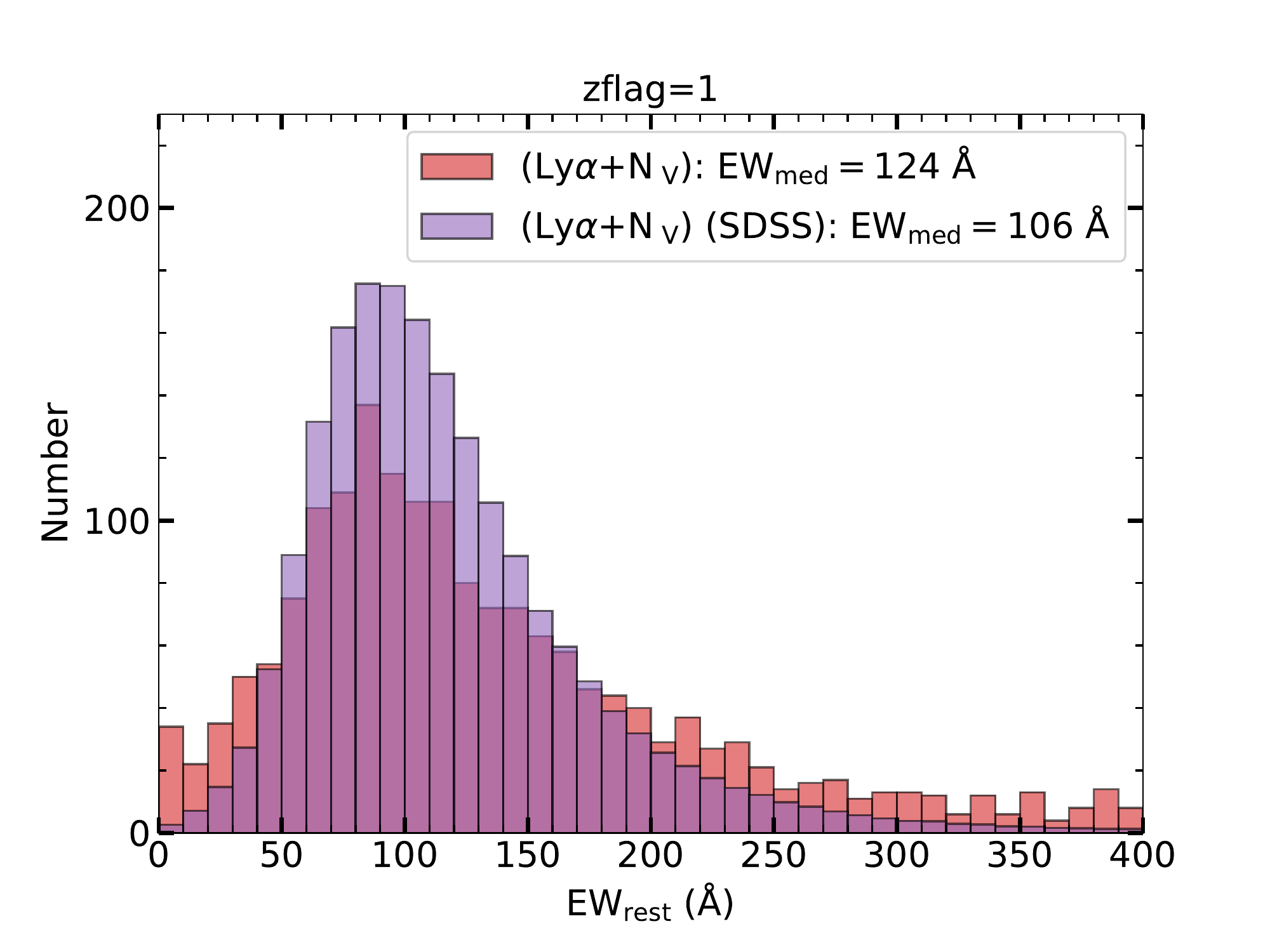}\\
\caption{Top: The distribution of the rest-frame equivalent widths of Ly$\alpha$+\ion{N}{5} (red), \ion{C}{4} $\lambda1549$ (blue), \ion{C}{3}] $\lambda1909$ (orange), and \ion{Mg}{2} $\lambda2799$ (green) of the HETDEX AGN with secure redshifts (\texttt{zflag=1}). Bottom: The comparison between the rest-frame equivalent widths of Ly$\alpha$+\ion{N}{5} of HETDEX AGN with secure redshifts (red) and those of the SDSS DR14Q (purple). The number of SDSS QSOs in each bin is divided by 100 for presentation purpose.}
\label{f_EW}
\end{figure}


In Section \ref{sec_r} (Figure \ref{f_Lya_r}) we discuss the ratio of the Ly$\alpha$+\ion{N}{5}
$\lambda1241$ emission line and the $r$-band magnitude, and found that the low-luminosity AGN of the HETDEX survey follow the same relation as bright QSOs, only extending the relation down to fainter luminosity. To do that, we compared observer frame $r$-band magnitudes to rest-frame Ly$\alpha$+\ion{N}{5} $\lambda1241$ emission. We now extend our comparisons between line and continuum emission by studying the distribution of the objects' rest-frame equivalent widths ($\rm EW_{rest}$), which allow for a more straightforward interpretation. 

We measure the $\rm EW_{rest}$ of an emission line in the rest-frame spectrum as the flux of the continuum subtracted emission divided by the continuum level at the center of the line.

Figure \ref{f_EW} shows the rest-frame EWs of 
Ly$\alpha$+\ion{N}{5} $\lambda1241$ (red), \ion{C}{4} $\lambda1549$ (blue), \ion{C}{3}] $\lambda1909$ (orange), and \ion{Mg}{2} $\lambda2799$ (green) of the HETDEX AGN with secure redshifts (\texttt{zflag=1}) in the upper panel. The ratio of the median EW of the four most significant emission lines of high-$z$ AGN is 124:62:30:34 $\approx$ 4:2:1:1.


The Ly$\alpha$+\ion{N}{5} $\lambda1241$ emission is significantly stronger than the other features with a median rest-frame $\rm EW_{(Ly\alpha+{N\ V})}=124$\,\AA\null. Among the four emission lines, the stronger emissions are expected at higher redshifts given the same observed wavelength range (Figure \ref{f_visibility}). Therefore, the observed EW among the four emissions are even more different. We take advantage of this difference in $\rm EW_{obs}$ to guess which emission the single broad emitters with \texttt{zflag=0} in our catalog is. For a given EW, the line with the closest median EW is assumed to be the most probable line. $\rm EW_{obs}$ is only one guidance of the best redshift estimates. 
We also combine other information, such as the shape of the continuum, the asymmetry of the emission, the absorptions at the emission, whether stronger emissions expected within the wavelength range of HETDEX is absent, etc. Detailed information of line identification for the AGN with \texttt{zflag=0} can be found in Section \ref{sec_boss}.

The purple histogram in the bottom panel of Figure \ref{f_EW} shows the distribution of the EWs of the Ly$\alpha$+\ion{N}{5} $\lambda1241$ emission in the SDSS DR14Q QSO catalog. The measurements are again taken from \cite{Rakshit2020}.  The median ($\rm EW_{Ly\alpha+{N V}}$) of the SDSS DR14Q catalog \citep{Rakshit2020} is 106 \AA, 18 \AA\ lower than that of the HETDEX AGN (124 \AA\null). The major difference is at the highest EW\null. HETDEX AGN are more populous in the very high EW regions ($\rm EW_{rest}\gtrsim 300$ \AA), where the continua are faint while the emissions are significant because of no pre-selection from the imaging.



\begin{figure}[htbp]
\centering
\includegraphics[width=\textwidth]{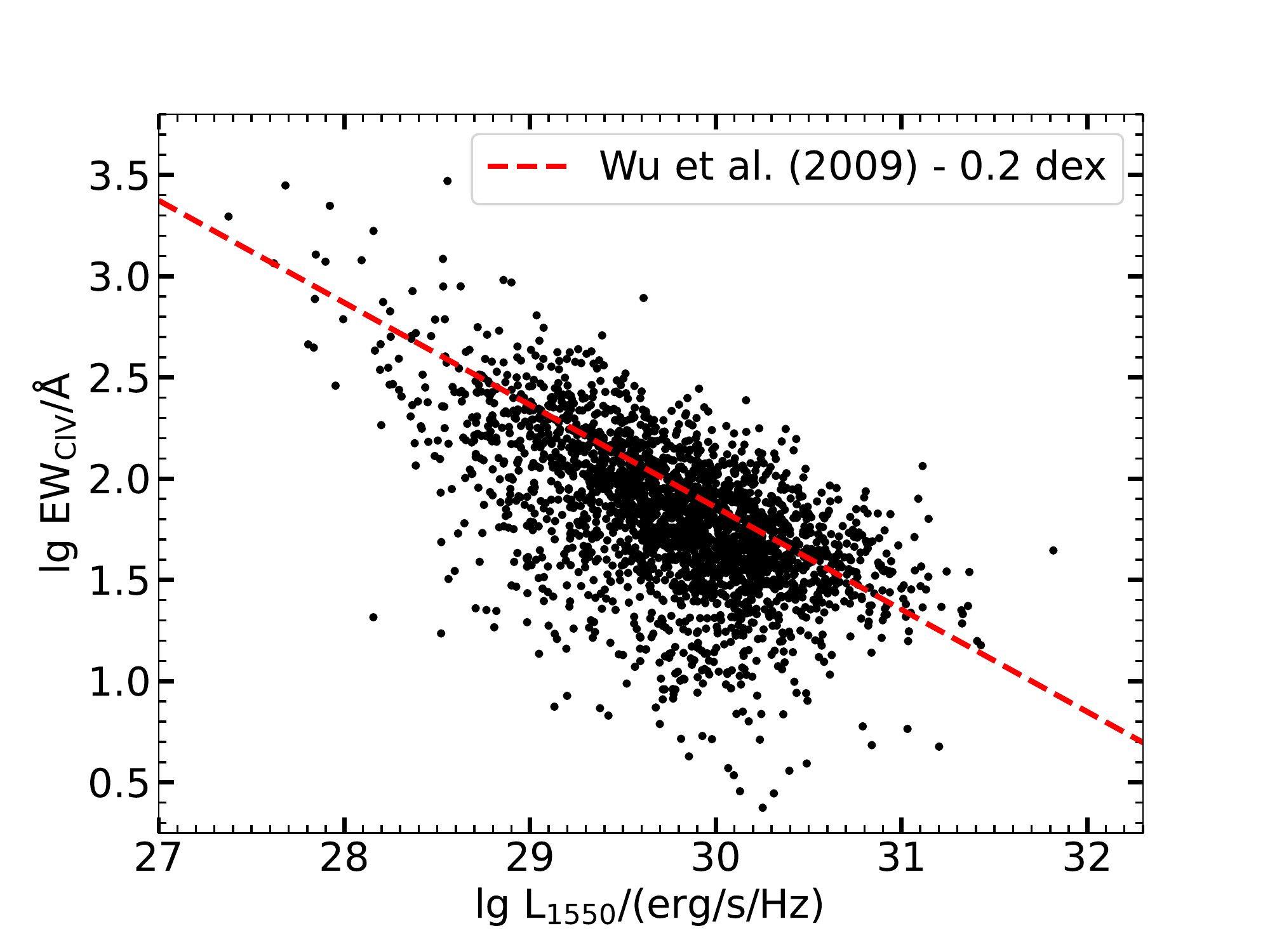}
\caption{The Baldwin Effect \citep{Baldwin1977} of HETDEX AGN (\texttt{zflag=1}) shown by the rest-frame EW of the \ion{C}{4} $\lambda1549$ emission and the monochromatic luminosity of the continuum at 1550 \AA\null. The red dashed line is the relation in \cite{Wu2009} lowered by 0.2 dex, following \cite{Richards2011}. This accounts for the methodolgy of  EW measurements compared to those from the SDSS spectroscopic pipeline \citep{Shen2011}. We note that the \ion{C}{4} $\lambda1549$ emission is only visible within the redshift range of $1.26<z<2.55$ in the HETDEX survey.}
\label{f_baldwin}
\end{figure}

Figure \ref{f_baldwin} shows the decrease of $\rm EW_{C_{\ IV}}$ with continuum luminosity, the Baldwin Effect \citep{Baldwin1977}. Our HETDEX AGN follows the modified relation of \cite{Wu2009} extremely well  over a wide range of $L_{1550}$ from $10^{27.4}$ to $10^{31.8}$ $\rm erg\ s^{-1}\ Hz^{-1}$.  For comparison, the AGN sample of \cite{Wu2009} also spans a very wide range of luminosity: $\rm L_{2500} = 10^{26.53} - 10^{33.04} erg\ s^{-1}\ Hz^{-1}$, while the SDSS QSOs extend only a small luminosity range ($\rm L_{2500} = 10^{30.53} - 10^{31.67}\ erg\ s^{-1}\ Hz^{-1}$) in their full AGN sample. 
Moreover, low-luminosity AGN in \cite{Wu2009} 
are exclusively drawn from X-ray AGN with UV/optical spectra from the archival databases of the \textit{Hubble Space Telescope (HST)} and the \textit{International Ultraviolet Explorer (IUE}\null). This again demonstrates the high efficiency of the HETDEX survey in exploring the low-luminosity AGN and other low-luminosity emitters.

\subsection{Type-II AGN}
\label{sec_type2}

\begin{figure}[htbp]
\centering
\includegraphics[width=\textwidth]{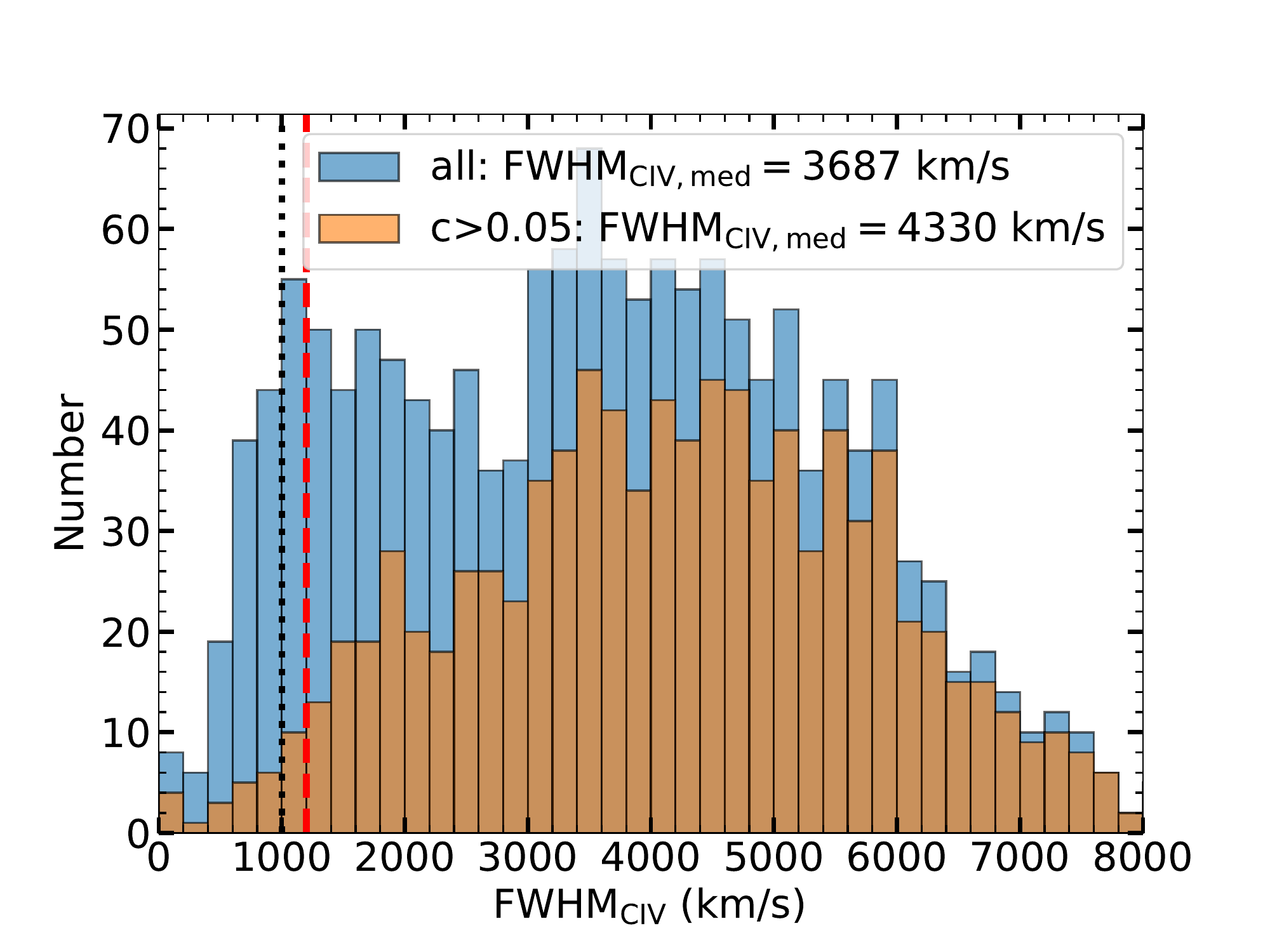}
\caption{The distribution of the FWHM of the \ion{C}{4} line for 1,538 HETDEX AGN that satisfy: (1) \texttt{sflag=2}; (2) \texttt{zflag=1}; (3) $\rm EW_{C_{IV}}>3$ \AA\ (blue). The orange histogram is the completeness corrected distribution for 970 out of the 1538 AGN with a completeness greater than 0.05. The completeness of each AGN is estimated based on simulations regarding our selection code. Details will be presented in Liu et al. submitted. The black dotted line at $\rm FWHM=1000$~km\,s$^{-1}$ and the red dashed line at $\rm FWHM=1200$~km\,s$^{-1}$ are two commonly used separations between type-I AGN and type-II AGN (see the context for more details). The estimation of the completeness correction for each AGN is very complicated and will be introduced with details in our next paper in the series of the HETDEX AGN for the luminosity function.}
\label{f_fwhm}
\end{figure}

\begin{figure*}[htbp]
\centering
\includegraphics[width=0.32\textwidth]{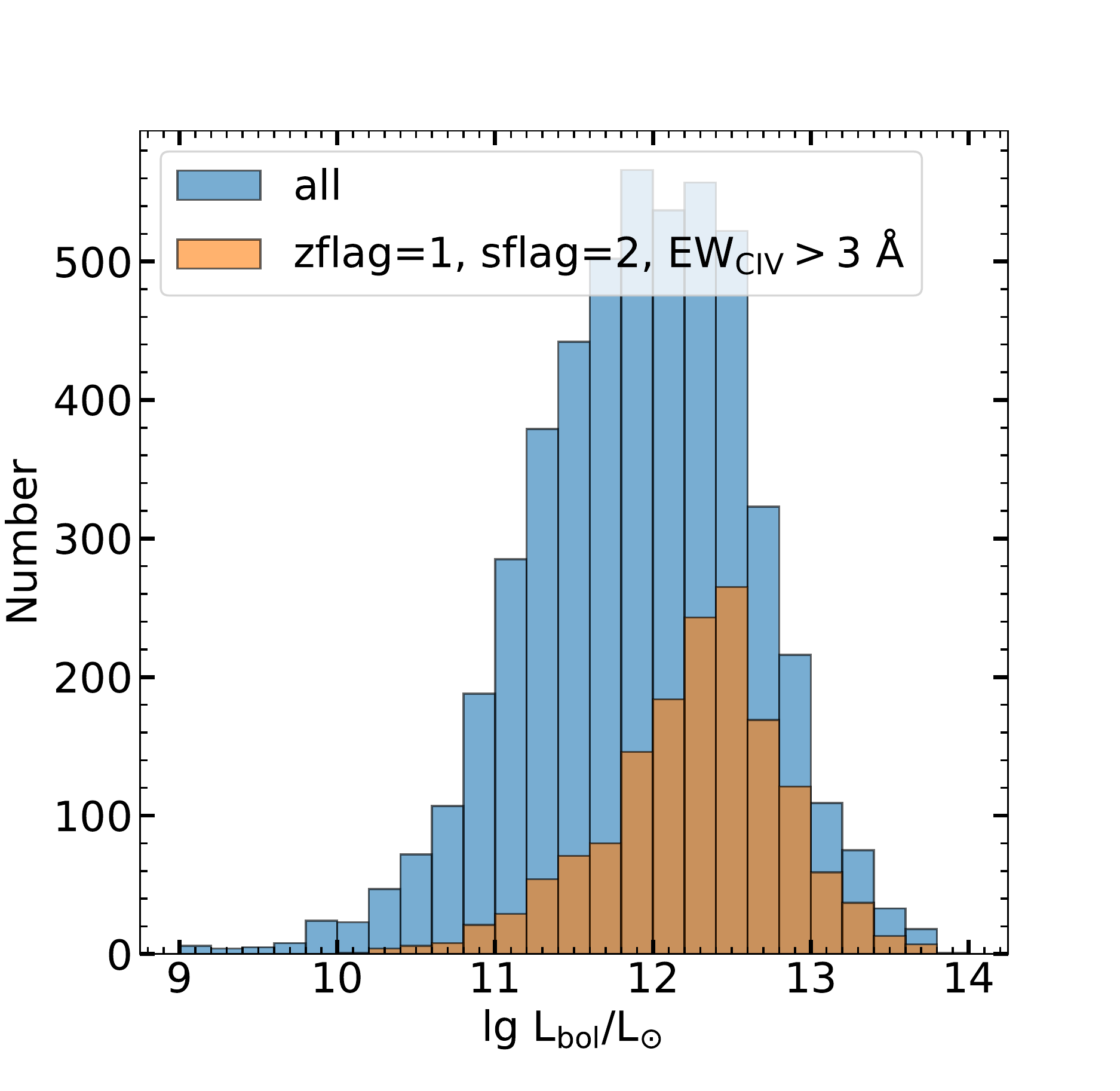}
\includegraphics[width=0.32\textwidth]{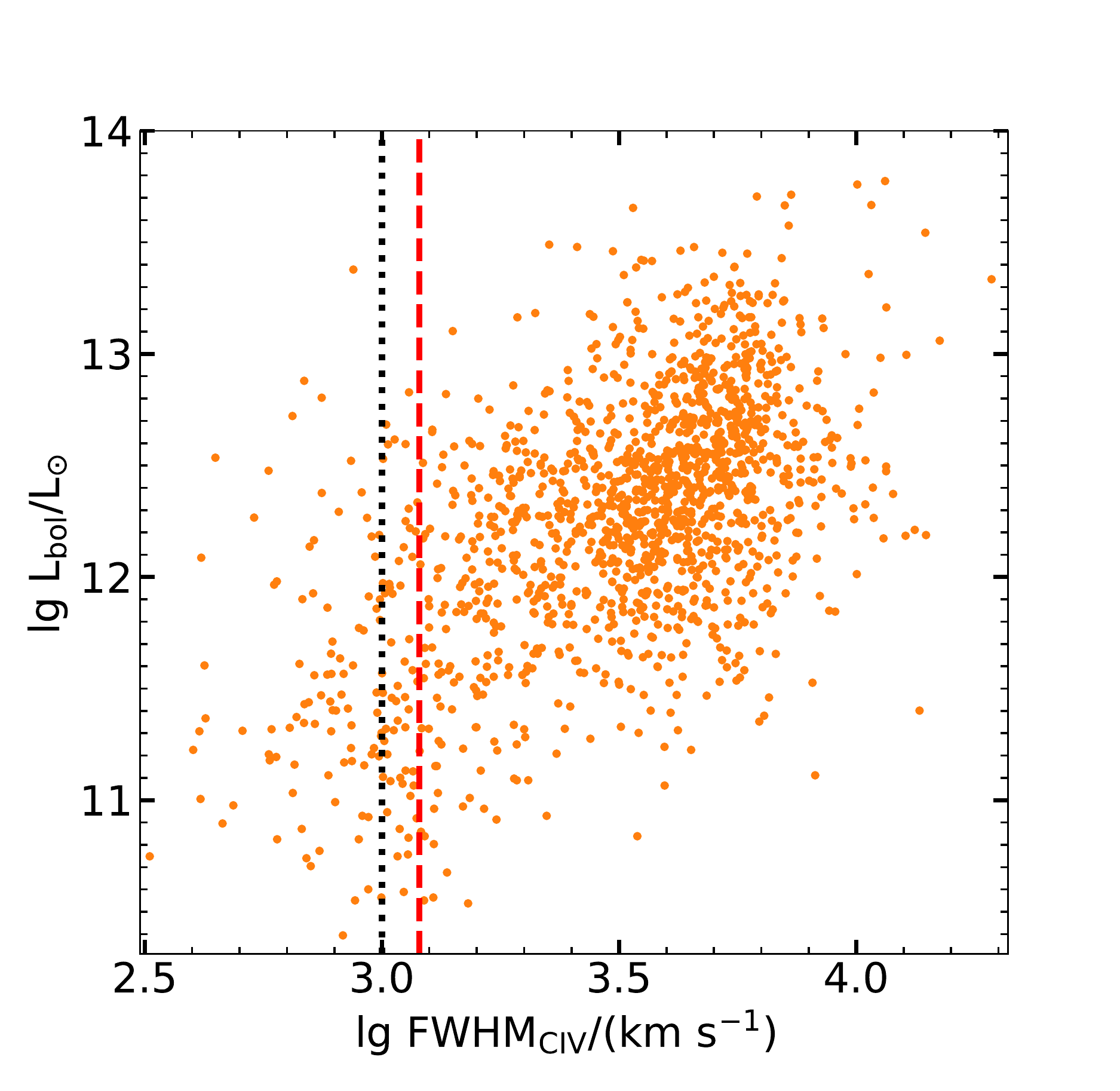}
\includegraphics[width=0.32\textwidth]{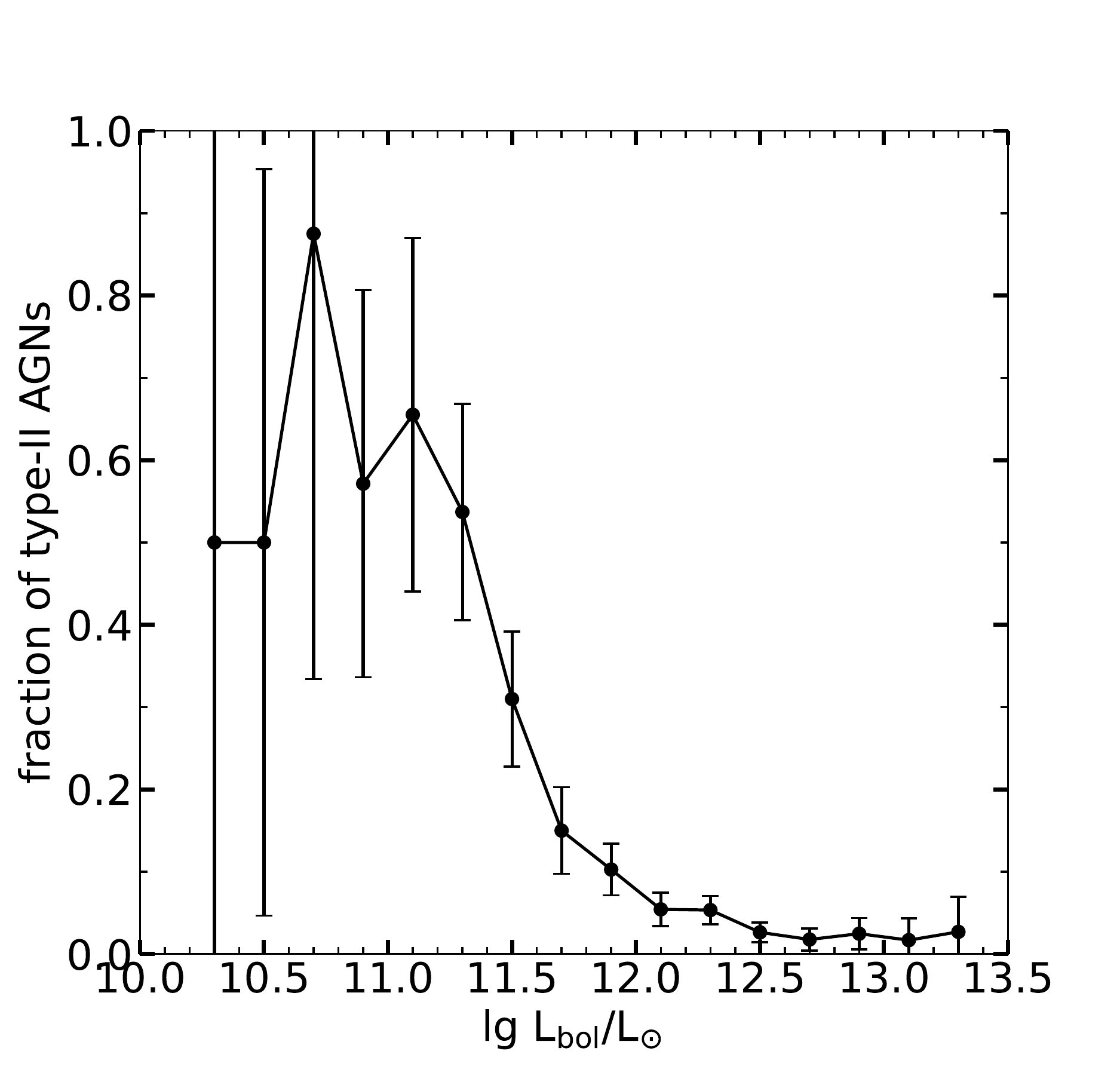}
\caption{Left: The distribution of the bolometric luminosity of the full 5.3k HETDEX AGN sample (blue), and the 1,548 AGN (orange) selected for the study of type-II AGN. Middle: The bolometric luminosity versus $\rm FWHM_{C_{IV}}$ for the 1,538 AGN (the orange histogram in the left panel and the blue histogram in Figure \ref{f_fwhm}). The black dotted line and the red dashed line again show the positions of FWHM$=$ 1000~km\,s$^{-1}$ and 1200~km\,s$^{-1}$. Right: The fraction of type-II AGN ($\rm FWHM_{C_{IV}}<1200$ km/s), $\rm f_2=n_{type-II}/n_{AGN}$, as a function of the bolometric luminosity. Error bars are calculated with the Poisson errors \citep{Gehrels1986}.}
\label{f_type2}
\end{figure*}

The HETDEX AGN are mainly identified by the emission-line pairs (the 2em selection, Section \ref{sec_2em}) or the single broad emission-lines (the sBL selection, Section \ref{sec_mg}). The sBL selection is only designed to find type-I AGN, while the 2em selection has no minimum requirement on line-width.  We can therefore take advantage of the 2em selection method ($sflag=2$, see the description of Column 158 in Appendix \ref{sec_append}) to study the type-II AGN population in our HETDEX AGN catalog. In this sub-section, we require all objects to satisfy the following three criteria: (1) selection with the 2em criteria (\texttt{sflag=2}); (2) a secure redshift (\texttt{zflag=1}); (3) $\rm EW_{C_{IV}}>3$ \AA\ for a reliable measurement of $\rm FWHM_{C_{IV}}$. The \ion{C}{4} $\lambda1549$ emission is a highly ionized emission, the existence of the line itself is a strong proof of an AGN, no matter broad or not.  

In this paper, broad emissions are identified with a very generous cut at FWHM$>1000$~km\,s$^{-1}$, so that our selection of AGN can be as complete as possible. However, there is no general agreement on where to cut at FWHM between the broad lines and the narrow lines. Some papers use 2000~km\,s$^{-1}$ \citep[e.g.][]{Steidel2002}, while 1000~km\,s$^{-1}$ is favored by others \citep[e.g.][]{Gavignaud2006,Schneider2010}, and \cite{Paris2018} adopted an even more generous cut of 500~km\,s$^{-1}$ for the SDSS DR14Q catalog. There is also a special narrow-line Seyfert 1 population (NLS1s) making things even more complicated \citep{Osterbrock1985}. \cite{Hao2005a} found a gap in the $\rm FWHM_{H_\alpha}$ distribution of the SDSS AGN at 1200 km/s separating type-I AGN and type-II AGN naturally. This value is later widely used and even applied to other emission lines in the discrimination between the two AGN populations \citep[e.g.][]{Shen2011}. For a better comparison with the SDSS AGN extending to the low-luminosity region \citep{Hao2005a,Hao2005b,Liu2015}, we choose $\rm FWHM_{C_{IV}}<1200$~km\,s$^{-1}$ as the cut for the type-II AGN in the rest of this sub-section \ref{sec_type2}.

Figure \ref{f_fwhm} shows the $\rm FWHM_{C_{IV}}$ distributions of the HETDEX AGN before and after completeness corrections in blue and orange, respectively. The 1200~km\,s$^{-1}$ cut shown by the red dashed line results a type-II AGN fraction ($f_2$, the number of type-II AGN divided by the total number of AGN) of 11\% and 7\% before and after completeness corrections respectively. This is significantly smaller than that at low redshifts. \cite{Liu2015} found the $f_2$ of the SDSS AGN in the main galaxy sample at $z<0.1$ is $\sim$ 70\%. This big discrepancy in $f_2$ possibly originates from the different luminosity ranges of AGN between the two AGN samples. 

Although the blind survey requires no magnitude cuts, the \ion{C}{4} $\lambda1549$ emission is only visible at $1.26<z<2.55$ given the wavelength range of the HETDEX survey. At such high redshifts, only strong line emitters can be detected at a $>4-5\sigma$ level. This emission-line flux-limited sample is similar to a magnitude-limited sample: higher redshifts are biased with more luminous sources. 
The AGN in \cite{Liu2015} have the luminosities of the [\ion{O}{3}] $\lambda5007$ emission in the range of $\rm10^6-10^9\ L_{\sun}$. With a rough conversion from the [\ion{O}{3}] $\lambda5007$ luminosity to the bolometric luminosity following \cite{Lamastra2009} and \cite{Netzer2019}, the AGN in \cite{Liu2015} are $\rm L_{bol}\sim10^{8}-10^{11}\ L_{\sun}$. The left panel of Figure \ref{f_type2} shows that the HETDEX AGN used in the $f_2$ study (orange) are mostly more luminous than the brightest AGN in \cite{Liu2015} ($\rm L_{bol}\sim10^{11}\ L_{\sun}$). The bolometric luminosities of the HETDEX AGN are roughly estimated from a simple correction to the monochromatic luminosities ($\rm L_{bol}=4.2L_{1450}$) following \cite{Runnoe2012}. 

Figure \ref{f_type2} shows the strong correlation between the bolometric luminosities and the line widths of the HETDEX AGN in the middle panel. The right panel shows a strong decrement of $f_2$ with the bolometric luminosity. 
The $f_2$ of X-ray AGN with luminosities spanning $\rm L_X=10^{43}-10^{45}$ erg\,s$^{-1}$ (roughly corresponding to $\rm L_{bol}\sim10^{10.5}-10^{12.5}\ L_{\sun}$) at $z\sim2$ is 20\%-30\% \citep{Treister2006}. Considering our HETDEX AGN are about one magnitude brighter than their X-ray AGN \citep{Treister2006}, the $f_2$ of the HETDEX AGN generally agrees that of the X-ray AGN at the same redshifts. 
At $\rm L_{bol}\lesssim10^{11}\ L_\sun$, the HETDEX AGN have a $f_2\sim60\%$, and can be expected higher at lower luminosities with an extrapolation. Therefore, when controlled for luminosity, there is no significant difference in $f_2$ between the $z<0.1$ SDSS AGN and the $z\sim2$ HETDEX AGN. Our results favor a non-evolving $f_2$ with the redshift.

\begin{figure*}[!htbp]
\centering
\includegraphics[width=\textwidth]{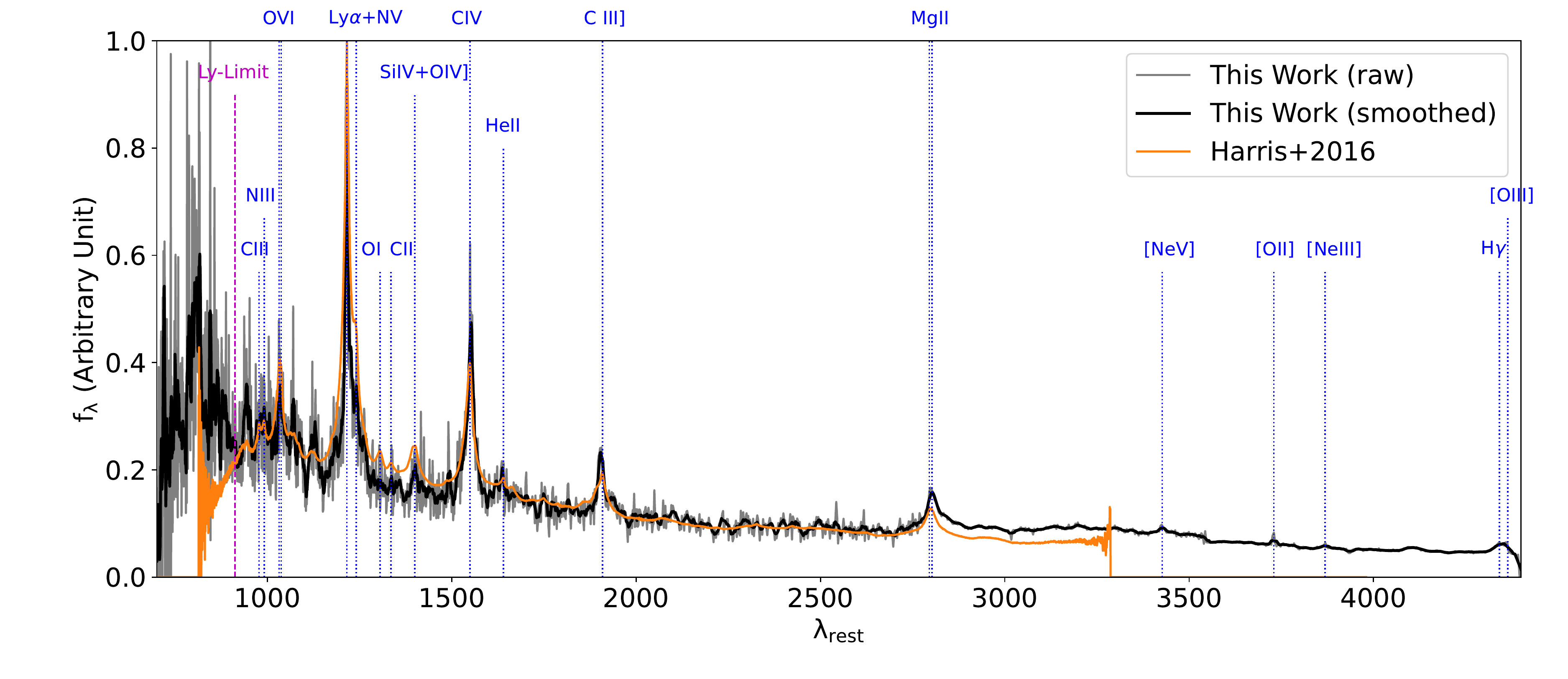}\\
\includegraphics[width=\textwidth]{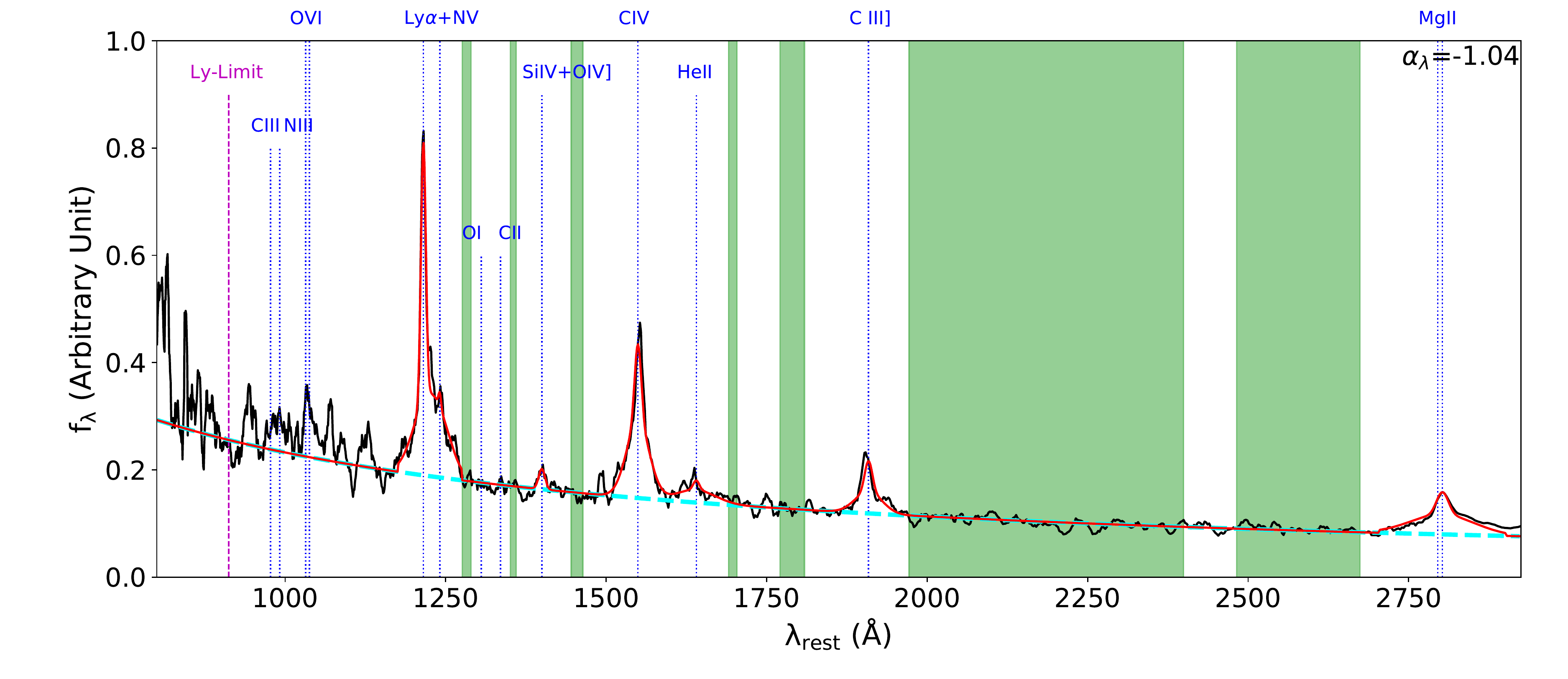}\\
\caption{The composite spectrum of the 3,733 HETDEX AGN with secure redshifts (\texttt{zflag=1}). Upper: The raw composite spectrum of all HETDEX AGN with \texttt{zflag=1} is shown in grey. The black line is the smoothed spectrum of the grey curve. The orange line is the composite spectrum of the SDSS BOSS QSOs in \cite{Harris2016}. Bottom: The best fit model (red) to the smoothed composite spectrum of the HETDEX AGN (black) in the range of 800\,\AA\ - 2925\,\AA\null. The figure is designed in the same way as Figure~\ref{f_specs_2em} and Figure~\ref{f_faint}, except that the spectrum is in the rest-frame rather than in the observed frame, and the $y$-axis is in relative units of ergs\,cm$^{-2}$\,s$^{-1}$\,\AA$^{-1}$.  In both panels, some short blue dotted verticals are added showing the location of weak emission. The Lyman limit at 912\,\AA\ is shown as dashed magenta line.}
\label{f_comp}
\end{figure*}


Since the line widths are strongly correlated with the AGN luminosities, all three AGN in Figure \ref{f_faint} presented as examples of faint AGN are not typical type-I AGN. The fitted FWHMs of the $\rm Ly\alpha$ emissions of AGN 1383, AGN 2012, AGN 2653 are 929 km/s, 797 km/s, and 1047 km/s. Their $\rm FWHM_{C_{IV}}$ are 863 km/s, 1412 km/s, and 1365 km/s. With a cut of  $\rm FWHM_{C_{IV}}<1200$ km/s, only AGN 2653 can by identified as a type-II AGN. However, the other two AGN remain type-II AGN candidates considering their narrow $\rm Ly\alpha$ emissions.

\subsection{Composite Spectrum}
\label{sec_compspec}

Figure \ref{f_comp} shows the composite spectrum derived from the 3,733 HETDEX AGN with secure redshifts. The AGN are first organized with descending redshifts, and grouped into 25 sets of 150 AGN\null. Each spectrum is connected to the  next by normalizing the mean flux in the regions of continua overlap. The AGN are then gradually connected from the highest redshifts to the lowest redshifts, from short wavelengths to longer wavelengths in each group (Appendix \ref{sec_25}). Spectra of different groups are also connected in the same way, building the final composite spectrum spanning 700\,\AA\ - 4400\,\AA\ (rest-frame).

Table \ref{t_comp} shows the emissions detected in the composite spectrum of HETDEX AGN. We listed the strong and the weak emission lines separately. We measured the continuum subtracted strong lines in the composite spectrum as shown in the bottom panel of Figure \ref{f_comp}. The Ly$\alpha$ emission and the \ion{N}{5} $\lambda1241$ emission are strongly blended, and therefore measured together. The rest-frame equivalent widths ratios of Ly$\alpha$+\ion{N}{5} $\lambda1241$, \ion{C}{4} $\lambda1549$, \ion{C}{3}] $\lambda1909$, and \ion{Mg}{2} $\lambda 2799$ is about 6:3:1:2. The \ion{Si}{4}+\ion{O}{4}] $\lambda1400$ emission and \ion{He}{2} $\lambda1640$ emission are about 3\% and 11\% of the strength of the Ly$\alpha$+\ion{N}{5} $\lambda1241$, respectively. These two lines are not always significantly detected in the raw observed spectra, therefore not used in the AGN identifications (Section \ref{sec_selection}).  Specifically, only 27\% of the AGN in the appropriate redshift range display \ion{Si}{4}+\ion{O}{4}] $\lambda1400$ with $\rm EW_{rest}>20$\,\AA, and the equivalent fraction for \ion{He}{2} $\lambda1640$ is just 46\%.  We do note that the latter line might be significantly overestimated because flux from the nearby line of \ion{O}{3} $\lambda1666$ and part of the continuum could be both included by its broad component.

There are also many weak emissions whose weights are negligible compared to the strong emission lines.  These are shown in the raw composite spectrum (the grey spectrum in the upper panel of Figure \ref{f_comp}), and included such as \ion{O}{1} $\lambda1306$, \ion{C}{2}  $\lambda1335$, \ion{Ne}{6}, and \ion{Ne}{3}.  However, the raw composite spectrum is too noisy at wavelengths shortward of Ly$\alpha$ due to the effect of intervening Ly$\alpha$ absorption. We therefore smoothed the composite spectrum to make some of the emission lines blueward of Ly$\alpha$, such as \ion{C}{3} $\lambda 977$ and \ion{N}{3} $\lambda 992$ stand out. Both these emission lines are significantly detected in the composite spectrum of the SDSS BOSS QSOs (the orange spectrum in the upper panel of Figure \ref{f_comp}, \citealt{Harris2016}). The Lyman limit at 912\,\AA\ shows up as a strong trough (the dashed magenta vertical line) in our smoothed composite spectrum. The flux shortward of the Lyman limit are exclusively contributed by the 3500\,\AA\ - 3800\,\AA\ wavelengths in the observed-frame, where the sky lines are most abundant, making the sky subtraction very hard. The increment of the noisy continuum at such short wavelengths are therefore not trustworthy.

\begin{deluxetable*}{lccrrrc}[htbp]
\scriptsize
\tablecaption{Measurements of continuum subtracted emission lines in the composite spectrum.}
\label{t_comp}
\tablehead{
\colhead{Line Name} &\colhead{$\lambda_{\rm rest}$\tablenotemark{a}} &\colhead{Redshift Range\tablenotemark{b}} & \colhead{N(AGN)\tablenotemark{c}} &\colhead{N($\rm EW_{rest}>20$\,\AA)\tablenotemark{d}} &\colhead{EW} &\colhead{FWHM} \\
&\colhead{(\AA)} & & & &\colhead{(\AA)} &\colhead{(km\,s$^{-1}$)} 
}
\startdata
\hline
 & & strong emission lines & & \\ \hline
\ion{O}{6} $\lambda1034$         & 1033.82           & $2.39<z<4.32$ &   747 &   304 (41\%) &  --  &  --  \\
Ly$\alpha$+\ion{N}{5} $\lambda1241$ & 1215.24$+$1240.81 & $1.88<z<3.53$ & 1,912 & 1,865 (98\%) & 96.1 & 2516 \\ 
\ion{Si}{4}+\ion{O}{4}]  $\lambda1400$ & 1399.8       & $1.50<z<2.93$ & 2,443 &   670 (27\%) &  2.5 & 2472 \\
\ion{C}{4} $\lambda1549$         & 1549.48           & $1.26<z<2.55$ & 2,554 & 2,353 (92\%) & 42.9 & 3778 \\
\ion{He}{2}  $\lambda1640$     & 1640.4            & $1.13<z<2.35$ & 2,298 & 1,060 (46\%) & 10.6 & 5819 \\
\ion{C}{3}] $\lambda1909$  & 1908.734          & $0.83<z<1.88$ & 1,476 & 1,249 (85\%) & 15.8 & 3195 \\
\ion{Mg}{2} $\lambda2799$  doublet & 2799.117          & $0.25<z<0.96$ &   439 &   393 (90\%) & 29.5 & 4037 \\
\hline
 & & weak emission lines & & \\ \hline
\ion{C}{3} $\lambda977$   &  977.03   & -- & -- & -- &  --  & -- \\ 
\ion{N}{3} $\lambda992$   &  990.98   & -- & -- & -- &  --  & -- \\
\ion{O}{1} $\lambda1306$  triplet & 1305.53   & -- & -- & -- &  --  & -- \\
\ion{C}{2} $\lambda1335$   & 1335.31   & -- & -- & -- &  --  & -- \\
{[}\ion{Ne}{5}{]} $\lambda3426$ & 3425.87   & -- & -- & -- &  --  & -- \\
 {[}\ion{O}{2}{]} $\lambda3727$ doublet & 3727.092  & -- & -- & -- &  --  & -- \\
 {[}\ion{Ne}{3}{]} $\lambda3869$        & 3868.760  & -- & -- & -- &  --  & -- \\
H$\gamma$                 & 4341.68   & -- & -- & -- &  --  & -- \\
{[}\ion{O}{3}{]} $\lambda4364$ & 4364.44  & -- & -- & -- &  --  & -- \\
\enddata
\tablenotetext{a}{The given wavelengths are vacuum. Wavelengths are taken from the ones used in SDSS spectroscopic pipeline (SPECTRO1D) (http://classic.sdss.org/dr6/algorithms/linestable.html). For some emissions not included in SPECTRO1d, their wavelengths are taken from the NIST Atomic Spectra database (https://www.nist.gov/pml/atomic-spectra-database).}
\tablenotetext{b}{The redshift range corresponds to the visibility of each emission in the wavelength range of 3500\,\AA\ - 5500\,\AA.}
\tablenotetext{c}{The number of AGN \texttt{zflag=1} within the redshift range of Column 3.}
\tablenotetext{d}{The number of AGN with the emission measured with $\rm EW_{em,rest}>20$\,\AA, and the fraction of Column 5 out of Column 4.}
\end{deluxetable*}


The best fit power-law slope of the continuum of our composite spectrum is $\alpha_{\lambda}=-1.04$, where $\alpha_{\lambda}$ is defined by $f_{\lambda} = C\lambda^{\alpha_{\lambda}}$. 
The power-law slope of the continuum of the composite spectrum of the SDSS quasars are measured to be $\alpha_{\nu}=-0.5$ ($\alpha_{\lambda}=-1.5$, \citealt{VandenBerk2001,Harris2016}). 
Our HETDEX AGN sample contains many low-luminosity Seyferts, the continua of which have non-nelegtable contributions from their host galaxies. Therefore, in this paper, the shallower power-law slope $\alpha_{\lambda}=-1.04$ might not representing the power-law slope of AGN well. We are leaving the careful host-AGN decomposition in a future paper in this series.
\section{Summary}
\label{sec_summary}

We present the first AGN catalog of the HETDEX survey, covering an effective area of 30.61 deg$^2$. The HETDEX AGN catalog is an untargeted survey where objects are selected solely via their spectral features with no photometric preselections, such as the morphologies, the magnitudes, and the colors. AGN candidates are identified either via emission line pairs, or from a single broad (FWHM$>$1000~km\,s$^{-1}$) emission line within the wavelength range of HETDEX (3500\,\AA\ - 5500\,\AA\null), and then confirmed by visual inspections. Our first HETDEX AGN catalog contains 5,322 AGN down to $\rm S/N>5$. The median r-band magnitude is 21.6. A magnitude cut at $r<22.5$ would remove 34\% AGN from our catalog, and 2.6\% of our HETDEX AGN reach the detection limit at $r\sim26$ mag of the deepest imaging surveys we searched. A simple color cut for $z<2.5$ AGN at $u-g<0.62$ and $g-r<0.16$ would remove 6\% of the AGN.

Among the 5,322 HETDEX AGN, 3,733 have secure redshifts, either from line pairs, or from matched redshifts in SDSS. Among the 3,733 HETDEX AGN with secure redshifts, 1,486 are not in the SDSS DR14Q catalog, and can only by identified by the emission line pairs in the HETDEX spectra. 

The surface density of HETDEX AGN in our catalog is 173.86 $\rm deg^{-2}$ and 339.65 $\rm deg^{-2}$ without and with completeness corrections. The redshift range of our AGN catalog is from $z=0.25$ to $z=4.32$, with a median redshift of $z=2.1$. The bolometric luminosity ranges from about $\rm 10^{9}\ L_{\sun}$ to about $10^{14}\ L_{\sun}$ with a median of $10^{12}\ L_{\sun}$. The composite spectrum built from the 3,733 AGN with secure redshifts covers a rest-frame wavelength range of $700-4400$\,\AA\null. 

\acknowledgments

HETDEX is led by the University of Texas at Austin McDonald Observatory and Department of Astronomy with participation from the Ludwig-Maximilians-Universit\"at M\"unchen, Max-Planck-Institut f\"ur Extraterrestrische Physik (MPE), Leibniz-Institut f\"ur Astrophysik Potsdam (AIP), Texas A\&M University, The Pennsylvania State University, Institut f\"ur Astrophysik G\"ottingen, The University of Oxford, Max-Planck-Institut f\"ur Astrophysik (MPA), The University of Tokyo, and Missouri University of Science and Technology. In addition to Institutional support, HETDEX is funded by the National Science Foundation (grant AST-0926815), the State of Texas, the US Air Force (AFRL FA9451-04-2-0355), and generous support from private individuals and foundations.

The Hobby-Eberly Telescope (HET) is a joint project of the University of Texas at Austin, the Pennsylvania State University, Ludwig-Maximilians-Universit\"at M\"unchen, and Georg-August-Universit\"at G\"ottingen. The HET is named in honor of its principal benefactors, William P. Hobby and Robert E. Eberly.

The authors acknowledge the Texas Advanced Computing Center (TACC) at The University of Texas at Austin for providing high performance computing, visualization, and storage resources that have contributed to the research results reported within this paper. URL: http://www.tacc.utexas.edu

We thank the publicly available MegaPrime/CFHTLS survey based on observations obtained with MegaPrime/MegaCam, a joint project of CFHT and CEA/IRFU, at the Canada-France-Hawaii Telescope (CFHT) which is operated by the National Research Council (NRC) of Canada, the Institut National des Science de l'Univers of the Centre National de la Recherche Scientifique (CNRS) of France, and the University of Hawaii. This work is based in part on data products produced at Terapix available at the Canadian Astronomy Data Centre as part of the Canada-France-Hawaii Telescope Legacy Survey, a collaborative project of NRC and CNRS.

The Institute for Gravitation and the Cosmos is supported by the Eberly College of Science and the Office of the Senior Vice President for Research at Pennsylvania State University.

SLF acknowledges support from the National Science Foundation, through grant AST-1908817. 

M.K. acknowledges support by DFG grant KR 3338/4-1.

\bibliography{agn}

\clearpage
\newpage
\appendix

\section{Catalog format and column information}\label{sec_append}

Version 1 of the AGN catalog is made available in the online Journal
in FITS format, and is described in Table \ref{t_catalog}, which is followed 
by detailed catalog notes. Other extensions to the FITS file are described
in Section \ref{sec_catalog}.

\begin{longtable}{lllll} 
   \caption{Columns of extension 1 in the FITS file.} \\  \toprule

Column &      name     &   dtype  &              unit              &                   description                   \\\hline
1 &         agnid &    int64 &                                &      unique sequential numerical identifiers \\
2 &            ra &  float32 &                           deg  &    RA  of the AGN (center of flux weighted fof) \\
3 &           dec &  float32 &                           deg  &    DEC of the AGN (center of flux weighted fof) \\
4 &             z &  float32 &                                &                                        redshift \\
5 &          z\_er &  float32 &                                &                               error of redshift \\
6 &         zflag &    int64 &                                &           zflag=0/1: 1 confirmed z \\
7 &         field &  bytes10 &                                &                      field in the HETDEX survey \\
8 &        nshots &    int64 &                                &                   number of repeat observations \\
9 & detectid\_best &    int64 &                                &  ID in the hetdex catalog closest to fof center \\
10 &       ra\_best &  float32 &                           deg  &                            RA  of detectid\_best \\
11 &      dec\_best &  float32 &                           deg  &                            DEC of detectid\_best \\
12 &          roff &  float32 &                        arcsec  & offsec between ra\_best,dec\_best with fof center \\
13 &          nmem &    int64 &                                &                      number of member detectids \\
14 &         alpha &  float32 &                                &  alpha\_lambda ($\alpha_{\lambda}$) of the power-law fitted continuum \\
15 &      alpha\_er &  float32 &                                &                                  error of $\alpha_{\lambda}$ \\
16 &          fpl0 &  float32 &                                &       power-law continuum = fpl0 * $\lambda^{\alpha_{\lambda}}$  \\
17 &      fpl0\_er &  float32 &                                &                                    error of fpl0 \\
18 &         slope &  float32 &                                &            slope of the linear fitted continuum \\
19 &      slope\_er &  float32 &                                &                                  error of slope \\
20 &     intercept &  float32 &                                &        intercept of the linear fitted continuum \\
21 &  intercept\_er &  float32 &                                &                              error of intercept \\
22 &         L1350 &  float32 &                 1e+44 erg / s  &  monochromatic luminosity at rest-frame 1350 AA \\
23 &      L1350\_er &  float32 &                 1e+44 erg / s  &                                  error of L1350 \\
24 &         L1450 &  float32 &                 1e+44 erg / s  &  monochromatic luminosity at rest-frame 1450 AA \\
25 &      L1450\_er &  float32 &                 1e+44 erg / s  &                                  error of L1450 \\
26 &         L3000 &  float32 &                 1e+44 erg / s  &  monochromatic luminosity at rest-frame 3000 AA \\
27 &      L3000\_er &  float32 &                 1e+44 erg / s  &                                  error of L3000 \\
28 &         L5100 &  float32 &                 1e+44 erg / s  &  monochromatic luminosity at rest-frame 5100 AA \\
29 &      L5100\_er &  float32 &                 1e+44 erg / s  &                                  error of L5100 \\
30 &      fwhm\_LyA &  float32 &                        km / s  &             rest-frame FWHM of the LyA emission \\
31 &   fwhm\_LyA\_er &  float32 &                        km / s  &                               error of fwhm\_LyA \\
32 &        ew\_LyA &  float32 &                      Angstrom  &               rest-frame EW of the LyA emission \\
33 &     ew\_LyA\_er &  float32 &                      Angstrom  &                                 error of ew\_LyA \\
34 &      flux\_LyA &  float32 &           1e-17 erg / ($\rm cm^2$ s)  &             rest-frame flux of the LyA emission \\
35 &   flux\_LyA\_er &  float32 &           1e-17 erg / ($\rm cm^2$ s)  &                               error of flux\_LyA \\
36 &      cont\_LyA &  float32 &  1e-17 erg / (Angstrom $\rm cm^2$ s)  &        rest-frame continuum at the LyA emission \\
37 &   cont\_LyA\_er &  float32 &  1e-17 erg / (Angstrom $\rm cm^2$ s)  &                               error of cont\_LyA \\
38 &       snr\_LyA &  float32 &                                &                           S/N of the LyA region \\
39 &    snr\_LyA\_er &  float32 &                                &                                error of snr\_LyA \\
40 &      chi2\_LyA &  float32 &                                &                          chi2 of the LyA region \\
41 &   chi2\_LyA\_er &  float32 &                                &                               error of chi2\_LyA \\
42 &       fwhm\_NV &  float32 &                        km / s  &              rest-frame FWHM of the NV emission \\
43 &    fwhm\_NV\_er &  float32 &                        km / s  &                                error of fwhm\_NV \\
44 &         ew\_NV &  float32 &                      Angstrom  &                rest-frame EW of the NV emission \\
45 &      ew\_NV\_er &  float32 &                      Angstrom  &                                  error of ew\_NV \\
46 &       flux\_NV &  float32 &           1e-17 erg / ($\rm cm^2$ s)  &              rest-frame flux of the NV emission \\
47 &    flux\_NV\_er &  float32 &           1e-17 erg / ($\rm cm^2$ s)  &                                error of flux\_NV \\
48 &       cont\_NV &  float32 &  1e-17 erg / (Angstrom $\rm cm^2$ s)  &         rest-frame continuum at the NV emission \\
49 &    cont\_NV\_er &  float32 &  1e-17 erg / (Angstrom $\rm cm^2$ s)  &                                error of cont\_NV \\
50 &      fwhm\_SiO &  float32 &                        km / s  &             rest-frame FWHM of the SiO emission \\
51 &   fwhm\_SiO\_er &  float32 &                        km / s  &                               error of fwhm\_SiO \\
52 &        ew\_SiO &  float32 &                      Angstrom  &               rest-frame EW of the SiO emission \\
53 &     ew\_SiO\_er &  float32 &                      Angstrom  &                                 error of ew\_SiO \\
54 &      flux\_SiO &  float32 &           1e-17 erg / ($\rm cm^2$ s)  &             rest-frame flux of the SiO emission \\
55 &   flux\_SiO\_er &  float32 &           1e-17 erg / ($\rm cm^2$ s)  &                               error of flux\_SiO \\
56 &      cont\_SiO &  float32 &  1e-17 erg / (Angstrom $\rm cm^2$ s)  &        rest-frame continuum at the SiO emission \\
57 &   cont\_SiO\_er &  float32 &  1e-17 erg / (Angstrom $\rm cm^2$ s)  &                               error of cont\_SiO \\
58 &       snr\_SiO &  float32 &                                &                           S/N of the SiO region \\
59 &    snr\_SiO\_er &  float32 &                                &                                error of snr\_SiO \\
60 &      chi2\_SiO &  float32 &                                &                          chi2 of the SiO region \\
61 &   chi2\_SiO\_er &  float32 &                                &                               error of chi2\_SiO \\
62 &      fwhm\_CIV &  float32 &                        km / s  &             rest-frame FWHM of the CIV emission \\
63 &   fwhm\_CIV\_er &  float32 &                        km / s  &                               error of fwhm\_CIV \\
64 &        ew\_CIV &  float32 &                      Angstrom  &               rest-frame EW of the CIV emission \\
65 &     ew\_CIV\_er &  float32 &                      Angstrom  &                                 error of ew\_CIV \\
66 &      flux\_CIV &  float32 &           1e-17 erg / ($\rm cm^2$ s)  &             rest-frame flux of the CIV emission \\
67 &   flux\_CIV\_er &  float32 &           1e-17 erg / ($\rm cm^2$ s)  &                               error of flux\_CIV \\
68 &      cont\_CIV &  float32 &  1e-17 erg / (Angstrom $\rm cm^2$ s)  &        rest-frame continuum at the CIV emission \\
69 &   cont\_CIV\_er &  float32 &  1e-17 erg / (Angstrom $\rm cm^2$ s)  &                               error of cont\_CIV \\
70 &       snr\_CIV &  float32 &                                &                           S/N of the CIV region \\
71 &    snr\_CIV\_er &  float32 &                                &                                error of snr\_CIV \\
72 &      chi2\_CIV &  float32 &                                &                          chi2 of the CIV region \\
73 &   chi2\_CIV\_er &  float32 &                                &                               error of chi2\_CIV \\
74 &     fwhm\_HeII &  float32 &                        km / s  &            rest-frame FWHM of the HeII emission \\
75 &  fwhm\_HeII\_er &  float32 &                        km / s  &                              error of fwhm\_HeII \\
76 &       ew\_HeII &  float32 &                      Angstrom  &              rest-frame EW of the HeII emission \\
77 &    ew\_HeII\_er &  float32 &                      Angstrom  &                                error of ew\_HeII \\
78 &     flux\_HeII &  float32 &           1e-17 erg / ($\rm cm^2$ s)  &            rest-frame flux of the HeII emission \\
79 &  flux\_HeII\_er &  float32 &           1e-17 erg / ($\rm cm^2$ s)  &                              error of flux\_HeII \\
80 &     cont\_HeII &  float32 &  1e-17 erg / (Angstrom $\rm cm^2$ s)  &       rest-frame continuum at the HeII emission \\
81 &  cont\_HeII\_er &  float32 &  1e-17 erg / (Angstrom $\rm cm^2$ s)  &                              error of cont\_HeII \\
82 &      snr\_HeII &  float32 &                                &                          S/N of the HeII region \\
83 &   snr\_HeII\_er &  float32 &                                &                               error of snr\_HeII \\
84 &     chi2\_HeII &  float32 &                                &                         chi2 of the HeII region \\
85 &  chi2\_HeII\_er &  float32 &                                &                              error of chi2\_HeII \\
86 &     fwhm\_CIII &  float32 &                        km / s  &            rest-frame FWHM of the CIII emission \\
87 &  fwhm\_CIII\_er &  float32 &                        km / s  &                              error of fwhm\_CIII \\
88 &       ew\_CIII &  float32 &                      Angstrom  &              rest-frame EW of the CIII emission \\
89 &    ew\_CIII\_er &  float32 &                      Angstrom  &                                error of ew\_CIII \\
90 &     flux\_CIII &  float32 &           1e-17 erg / ($\rm cm^2$ s)  &            rest-frame flux of the CIII emission \\
91 &  flux\_CIII\_er &  float32 &           1e-17 erg / ($\rm cm^2$ s)  &                              error of flux\_CIII \\
92 &     cont\_CIII &  float32 &  1e-17 erg / (Angstrom $\rm cm^2$ s)  &       rest-frame continuum at the CIII emission \\
93 &  cont\_CIII\_er &  float32 &  1e-17 erg / (Angstrom $\rm cm^2$ s)  &                              error of cont\_CIII \\
94 &      snr\_CIII &  float32 &                                &                          S/N of the CIII region \\
95 &   snr\_CIII\_er &  float32 &                                &                               error of snr\_CIII \\
96 &     chi2\_CIII &  float32 &                                &                         chi2 of the CIII region \\
97 &  chi2\_CIII\_er &  float32 &                                &                              error of chi2\_CIII \\
98 &     fwhm\_MgII &  float32 &                        km / s  &            rest-frame FWHM of the MgII emission \\
99 &  fwhm\_MgII\_er &  float32 &                        km / s  &                              error of fwhm\_MgII \\
100 &       ew\_MgII &  float32 &                      Angstrom  &              rest-frame EW of the MgII emission \\
101 &    ew\_MgII\_er &  float32 &                      Angstrom  &                                error of ew\_MgII \\
102 &     flux\_MgII &  float32 &           1e-17 erg / ($\rm cm^2$ s)  &            rest-frame flux of the MgII emission \\
103 &  flux\_MgII\_er &  float32 &           1e-17 erg / ($\rm cm^2$ s)  &                              error of flux\_MgII \\
104 &     cont\_MgII &  float32 &  1e-17 erg / (Angstrom $\rm cm^2$ s)  &       rest-frame continuum at the MgII emission \\
105 &  cont\_MgII\_er &  float32 &  1e-17 erg / (Angstrom $\rm cm^2$ s)  &                              error of cont\_MgII \\
106 &      snr\_MgII &  float32 &                                &                          S/N of the MgII region \\
107 &   snr\_MgII\_er &  float32 &                                &                               error of snr\_MgII \\
108 &     chi2\_MgII &  float32 &                                &                         chi2 of the MgII region \\
109 &  chi2\_MgII\_er &  float32 &                                &                              error of chi2\_MgII \\
110 &      fwhm\_OII &  float32 &                        km / s  &             rest-frame FWHM of the OII emission \\
111 &   fwhm\_OII\_er &  float32 &                        km / s  &                               error of fwhm\_OII \\
112 &        ew\_OII &  float32 &                      Angstrom  &               rest-frame EW of the OII emission \\
113 &     ew\_OII\_er &  float32 &                      Angstrom  &                                 error of ew\_OII \\
114 &      flux\_OII &  float32 &           1e-17 erg / ($\rm cm^2$ s)  &             rest-frame flux of the OII emission \\
115 &   flux\_OII\_er &  float32 &           1e-17 erg / ($\rm cm^2$ s)  &                               error of flux\_OII \\
116 &      cont\_OII &  float32 &  1e-17 erg / (Angstrom $\rm cm^2$ s)  &        rest-frame continuum at the OII emission \\
117 &   cont\_OII\_er &  float32 &  1e-17 erg / (Angstrom $\rm cm^2$ s)  &                               error of cont\_OII \\
118 &       snr\_OII &  float32 &                                &                           S/N of the OII region \\
119 &    snr\_OII\_er &  float32 &                                &                                error of snr\_OII \\
120 &      chi2\_OII &  float32 &                                &                          chi2 of the OII region \\
121 &   chi2\_OII\_er &  float32 &                                &                               error of chi2\_OII \\
122 &      fwhm\_OVI &  float32 &                        km / s  &             rest-frame FWHM of the OVI emission \\
123 &   fwhm\_OVI\_er &  float32 &                        km / s  &                               error of fwhm\_OVI \\
124 &        ew\_OVI &  float32 &                      Angstrom  &               rest-frame EW of the OVI emission \\
125 &     ew\_OVI\_er &  float32 &                      Angstrom  &                                 error of ew\_OVI \\
126 &      flux\_OVI &  float32 &           1e-17 erg / ($\rm cm^2$ s)  &             rest-frame flux of the OVI emission \\
127 &   flux\_OVI\_er &  float32 &           1e-17 erg / ($\rm cm^2$ s)  &                               error of flux\_OVI \\
128 &      cont\_OVI &  float32 &  1e-17 erg / (Angstrom $\rm cm^2$ s)  &        rest-frame continuum at the OVI emission \\
129 &   cont\_OVI\_er &  float32 &  1e-17 erg / (Angstrom $\rm cm^2$ s)  &                               error of cont\_OVI \\
130 &       snr\_OVI &  float32 &                                &                           S/N of the OVI region \\
131 &    snr\_OVI\_er &  float32 &                                &                                error of snr\_OVI \\
132 &      chi2\_OVI &  float32 &                                &                          chi2 of the OVI region \\
133 &   chi2\_OVI\_er &  float32 &                                &                               error of chi2\_OVI \\
134 &       fiberid &  bytes38 &                                &                  fiberid in hetdex fiber design \\
135 &        r\_aper &  float32 &                           mag  &                       r-band aperture magnitude \\
136 &     r\_aper\_er &  float32 &                           mag  &                                 error of r\_aper \\
137 &      r\_radius &  float32 &                       arcsec   &                       r-band aperture radius \\
138 &      r\_depth  &  float32 &                        mag     &                       r-band 5$\sigma$ depth \\
139 &         r\_cat &  bytes15 &                                &                                survey of r\_aper \\
140 &        g\_aper &  float32 &                           mag  &                       g-band aperture magnitude \\
141 &     g\_aper\_er &  float32 &                           mag  &                                 error of g\_aper \\
142 &      g\_radius &  float32 &                       arcsec   &                       g-band aperture radius \\
143 &      g\_depth  &  float32 &                        mag     &                       g-band 5$\sigma$ depth \\
144 &         g\_cat &  bytes16 &                                &                                survey of g\_aper \\
145 &        u\_aper &  float32 &                           mag  &                       u-band aperture magnitude \\
146 &     u\_aper\_er &  float32 &                           mag  &                                 error of u\_aper \\
147 &      u\_radius &  float32 &                       arcsec   &                       u-band aperture radius \\
148 &      u\_depth  &  float32 &                        mag     &                       u-band 5$\sigma$ depth \\
149 &         u\_cat &  bytes11 &                                &                                survey of u\_aper \\
150 &    rsep\_dr14q &  float32 &                        arcsec  &                    smallest separation to dr14q \\
151 &      id\_dr14q &  bytes15 &                                &                  ID of the closest AGN in dr14q \\
152 &       z\_dr14q &  float32 &                                &                   spectral redshift of id\_dr14q \\
153 &     rsep\_dr16 &  float32 &                        arcsec  &                     smallest separation to dr16 \\
154 &    plate\_dr16 &    int64 &                                &                plate of the closest AGN in dr16 \\
155 &    fiber\_dr16 &    int64 &                                &                fiber of the closest AGN in dr16 \\
156 &      mjd\_dr16 &    int64 &                                &                 mjd  of the closest AGN in dr16 \\
157 &        z\_dr16 &  float32 &                                &                    spectral redshift of id\_dr16 \\
158 &         sflag &    int64 &                                 &                            2em=2, sBL=1, boss=0, else=-1 \\
159 &         apcor &  float32 &                                &                             aperture correction \\ \hline\hline
\label{t_catalog}
\end{longtable}    

\noindent Notes on the catalog columns in the first extension:\\

\noindent Column 1 A sequential ID number assigned for each AGN. The redshifts decrease from agnid=1 to 5322.\\

\noindent Column 2-3 The flux weighted fof grouping center of all member detections associated with each AGN.\\

\noindent Column 4-5 The redshift measured from the best fit model to the spectrum of $\rm detectid_{best}$, and the error of the redshift.\\

\noindent Column 6 The flag of the redshift showing whether the redshift in Column 4 is secure or not. $zflag=1$ means the redshift is secure, either confirmed by emission-line pairs, or by a cross matched spectral redshift from SDSS DR14Q. $zflag=0$ means the AGN is a single broad-line emitter within the wavelength range, with no matched spectral reshift from SDSS DR14Q. The redshift of an AGN with $zflag=0$ is guessed with a combination of various information, such as the observed EW, the imaging(s), the shape of the continuum, the asymmetry of the emission, etc. \\

\noindent Column 7 The field where the AGN is located in Table \ref{t_survey}. \\

\noindent Column 8 The number of shots of repeat observations the AGN is observed. This column is also available in extension 4.\\

\noindent Column 9 The ID of all member detections in the raw HETDEX survey that is closest to the emission-line flux weighted fof center in Column 2 and 3.\\

\noindent Column 10-11 The coordinate of $\rm detectid_{best}$.\\

\noindent Column 12 The separation between the pointing of $\rm detectid_{best}$ ($\rm ra_{best}$, $\rm dec_{best}$ in Column 10-11) and that of the emission-line flux weighted fof center (ra, dec in Column 2-3) in arcsec.\\

\noindent Column 13 The number of member detections in the raw HETDEX catalog described in Section \ref{sec_lines} and Section \ref{sec_conts}.\\

\noindent Column 14-17 The best fitted slope ($\alpha_\lambda$) and coefficient ($f_0$)  of the power-law continuum: $f_0* \lambda^{\alpha_{\lambda}}$ and their errors. There are 346 AGN in our catalog whose continuums are failed to be fitted with a power-law model. All four parameters for these AGN are set to -99. The continuum is then fitted with a simple linear model with parameters in Column 18-21.\\

\noindent Column 18-21 The best linear fitted slope and intercept of the continuum: $\rm slope * \lambda + intercept$ and their errors. If Column 14-17 are all -99, please use this linear fit for the best model of the continuum.\\ 

\noindent Column 22-29 The monochromatic luminosity of the best fit continuum at 1350 \AA, 1450 \AA, 3000 \AA, and 5100 \AA, with their errors.\\

\noindent Column 30-133 The best fitted parameters for each continuum subtracted emission lines and their errors in the rest-frame. For any emission that is out of the wavelength range, all measurements of it would be set to -99.\\

\noindent Column 134 The ID of the fiber in the design of the HETDEX survey. The format of the string is standard. For example, agnid$=$1 has a fiberid of 20191006024\_1\_multi\_323\_043\_040\_LU\_070. That says its $\rm detectid_{best}$ is observed on the date of Oct 6th, 2019. The observation is the 24th set of dither observation of that night, and the fiber that contributes most to the spectrum is from the first exposure of the dither set. The spectrograph that caught this source is 323, and connected to the IFU of 043, located at slot 040 on the focal plane. There are four amplifiers for each IFU: LL, LU, RL, and RU. This fiber is amplified by LU. Each amplifier has 112 fibers, and the fiber number that contributes most to the spectrum of $\rm detectid_{best}$ in the IFU is 070. With the fiberid, one can easily track down to the raw data where a specific source comes from.\\

\noindent Column 135-136 The aperture magnitude in $r$-band in the imaging surveys, and its uncertainty. The measurement is taken directly from our ``ELiXer'' software (Davis et al.\ in preparation). It is not a simple cross match with any existing catalog of the imaging surveys. For each detection, an aperture is centered at the best RA and DEC in the imaging. The size of the aperture is then grown gradually, until the enclosed flux does not change significantly. The enclosed flux is then measured for the aperture magnitude. For a detection that is under the 5$\sigma$ limit of the field it is in, the measured magnitude is set to the 5$\sigma$ limit of that field. For detections covered by no $r$-band imaging, both columns are set to $-99$. For our AGN sample that is only selected based on emission line features in the spectra, the aperture magnitude can better describe the photometry of our AGN compared to a simple cross match with an existing imaging catalog, especially for the AGN with significant emissions, but with few or even no continuums. We show the distribution of the $r$-band aperture magnitude in Figure \ref{f_r}. Some AGN can reach or even go below the detection limit of some imaging surveys.\\

\noindent Column 137 The best aperture size in radius determined by the ``ELiXer'' software to measure the aperture magnitudes in Column 135-136.\\

\noindent Column 138 The 5$\sigma$ depths of the field where the AGN is in.\\

\noindent Column 139 The survey providing the photometric imaging for the r-band aperture magnitude. Internally, the surveys are named, in order of decreasing average r-band depth: ``MegaPrime/CFHTLS'', ``HSC-SSP'', ``HSC-DEX'', ``DECAM/COSMOS'', and ``DECAM/SHELA'', ``DECaLS''. The surveys partially overlap and each reported aperture magnitude comes from the deepest survey with imaging that covers the candidate's coordinate. Basic reference information for each survey is provided in the Appendix \ref{sec_appendix_photo}. For AGN covered by no r-band imaging, this column is set to ``?''.\\

\noindent Column 140-143 The aperture magnitude in g-band in the imaging surveys, the error of it, the corresponding aperture size in radius, and the 5$\sigma$ depth of this field. All measurements are done in the same way as the measurements for the r-band, just using g-band imaging data.\\

\noindent Column 144 The survey providing the photometric imaging for the g-band aperture magnitude. We searched overlap coverage in surveys in order of decreasing g-band depth: ``MegaPrime/CFHTLS'', ``HSC-SSP'', ``DECAM/COSMOS'', ``DECAM/SHELA'', ``MOSAIC/KPNO'', ``DECaLS'' (see Appendix \ref{sec_appendix_photo} for more details). AGN covered by no g-band imaging are set to ``?''.\\

\noindent Column 145-148 The aperture magnitude in u-band in the imaging surveys, the error of it, the corresponding aperture size in radius, and the 5$\sigma$ depth of this field. All measurements are done in the same way as the measurements for the r-band, just using u-band imaging data.\\

\noindent Column 149 The survey providing the photometric imaging for the u-band aperture magnitude. The surveys are named, in order of decreasing u-band depth: ``MegaPrime/CFHTLS'', ``DECAM/COSMOS'', ``DECAM/SHELA'' ( see Appendix \ref{sec_appendix_photo} for more details). AGN covered by no u-band imaging are set to ``?''.\\

\noindent Column 150-152 The results of the cross match between our AGN catalog and the full SDSS DR14Q QSO catalog \citep{Paris2018}. ``id\_dr14q'' is the ID of the closest QSO in SDSS DR14Q with the format of PLATEID-MJD-FIBERID. The closest QSO is not necessarily a matched QSO, since the separation can be very large. ``rsep\_dr14q'' is the separation between $\rm detectid_{best}$ and ``id\_dr14q'' in arcsec. The user of the catalogue can choose an appropriate ``rsep\_dr14q'' for their science. We suggest ``rsep\_dr14q'' < 5, since most of our AGN are found to have an extended emission-line region, and our fof grouping used 5 arcsec as the linking length. ``z\_dr14q'' is the spectroscopic redshift of ``id\_dr14q'' in SDSS. \\

Column 153-157  The results of the cross match between our AGN catalog and the full SDSS DR16 spectroscopic catalog. ``plate\_dr16'', ``fiber\_dr16'', and ``mjd\_dr16'' are the PLATEID, FIBERID, and MJD of the closest spectroscopic object in the SDSS DR16 catalog. ``rsep\_dr16'' is the separation between $\rm detectid_{best}$ and (``plate\_dr16'', ``fiber\_dr16'', ``mjd\_dr16'') in arcsec. ``z\_dr16'' is the spectroscopic redshift of ``id\_dr16''. \\

Column 158 The flag that gives how an AGN is selected: $sflag=2$ for 1865 AGN identified by emission-line pairs (the 2em selection), $sflag=1$ for 2976 AGN identified by single broad emission line within the wavelength range of the HETDEX survey (the sBL selection), and $sflag=0$ for the 91 AGN that are missed either by the 2em selection or by the sBL selection, but picked up by a matched redshift from the SDSS DR14Q catalog. The other 390 AGN have an $sflag=-1$, which are real AGN selected by our previous version of the selection code but missed by the most updated selection criteria with the latest code.\\

Column 159 The aperture correction at the wavelength of the strongest emission of $\rm detectid_{best}$. The value ranges from 0 to 1. A higher value means a smaller aperture correction. Appropriate aperture corrections, as a function of wavelength, have already been applied to the spectra in extensions 2 and 3. All catalogue values are also aperture corrected.\\


\section{Photometric surveys used for measurements of the aperture magnitudes}\label{sec_appendix_photo}







\noindent MegaPrime/CFHTLS: The Canada-France-Hawaii Telescope Legacy Survey\footnote{\url{https://www.cfht.hawaii.edu/Science/CFHTLS/}}. In this paper, we used the u,g,r bands. Their r-band depths are 26.0 in the wide fields, and 27.3 in the deep fields. The seeing ranges from 0.6 arcsec\ to 1.0 arcsec. The RA and DEC ranges are 208.53 - 220.42 (deg), and 51.21 – 57.81 (deg).\\

\noindent HSC-SSP: The Hyper Suprime-Cam Subaru Strategic Program \citep{Aihara2019}. We only used their wide fields of their second data release. The depth is $g\sim26.6$ and $r\sim26.2$. The size and location are 14.7 - 153.4 (deg) in RA, and -7.3 - 5.5 (deg) in DEC.\\

\noindent HSC-DEX: The Hyper Suprime-Cam-HETDEX joint survey. It is observed in 2015-2018 (S15A, S17A, S18A; PI: A. Schulze) and 2019-2020 (S19B; PI: S. Mukae) with a total observing time of 3 nights. It aims to cover the full footprint of the HETDEX survey with r-band imaging. By the date of 2020-06-26, about half of the HETDEX footprint is covered by the HSC-DEX thanks to the contribution from our collaborators in the Tokyo University. The 5$\sigma$ depth is $r\sim25.5$ with a 2 arcsec\ aperture, and $r\sim26.2$ with a 1 arcsec\ aperture. The seeing is 0.6 arcsec-1.0 arcsec. The location is 158.9 - 230.8 (deg) in RA, and 45 - 56.6 (deg) in DEC, with a coverage of $\sim 150\ deg^2$.\\

\noindent DECAM/COSMOS is from our private communication with Isak Wold. The depth is $u,g,r\sim25.5$. The size and location are 149.0 - 151.3 (deg) in RA, and 1.15 – 3.24 (deg) in DEC.\\

\noindent DECAM/SHELA: The Spitzer-HETDEX Exploratory Large Area Survey \citep{Wold2019}. The u-band depth is 25.7 mag. The g-band depth is 24.7 mag. The r-band depth is 24.7 mag for a 2 arcsec aperture, and 25.6 for a 1 arcsec aperture. The seeing is 0.7 arcsec-1.0 arcsec. The RA ranges from 8.50\degree\ to 36.51\degree, and the DEC ranges from -4.0\degree\ to 4.0\degree. \\
 
\noindent DECaLS: The Dark Energy Camera Legacy Survey \citep{Dey2019}. The depth is $g\sim24.0$, and $r\sim23.4-24.0$.\\

\noindent MOSAIC/KPNO: The survey is taken at the Kitt Peak National Observatory\footnote{\url{http://archive1.dm.noao.edu/search/query/} (search for 2011A-0186, 2012A-0282, and 2015A-0075 in the Program Number Field)}\footnote{\url{http://ast.noao.edu/sites/default/files/NOAO\_DHB\_v2.2.pdf}}. The imaging data is from private communication with John Feldmeier. The depth is $g\sim24.7$ for a 1 arcsec aperture and $g\sim23.4$ for a 2 arcsec aperture.
The seeing is $\lesssim$1.0 arcsec. The sky coverage is 156.47\degree\ – 233.92\degree\ in RA, and 51.47\degree\ – 55.63\degree\ in DEC.\\

\section{Composite spectra of all 25 sub groups}\label{sec_25}

\begin{figure}[htbp]
\centering
\includegraphics[width=0.49\textwidth]{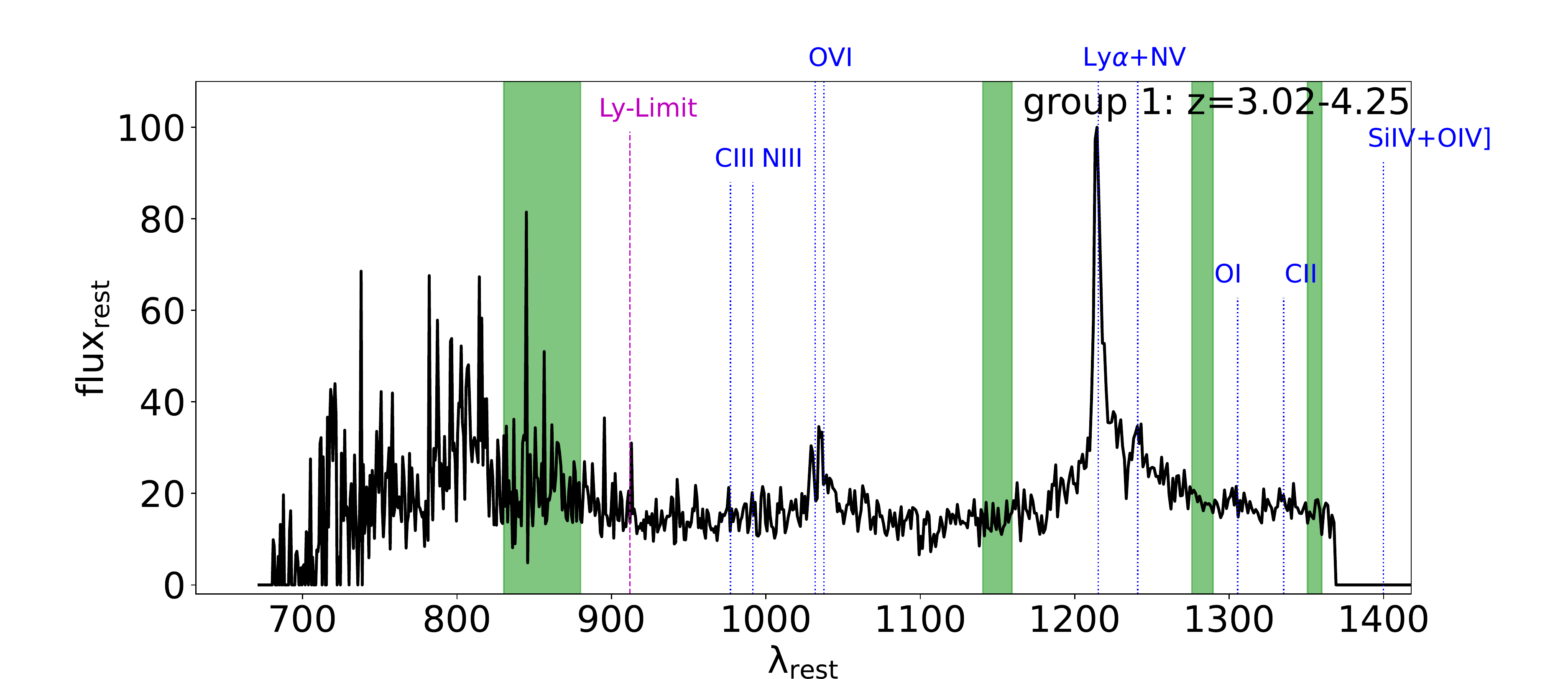}
\includegraphics[width=0.49\textwidth]{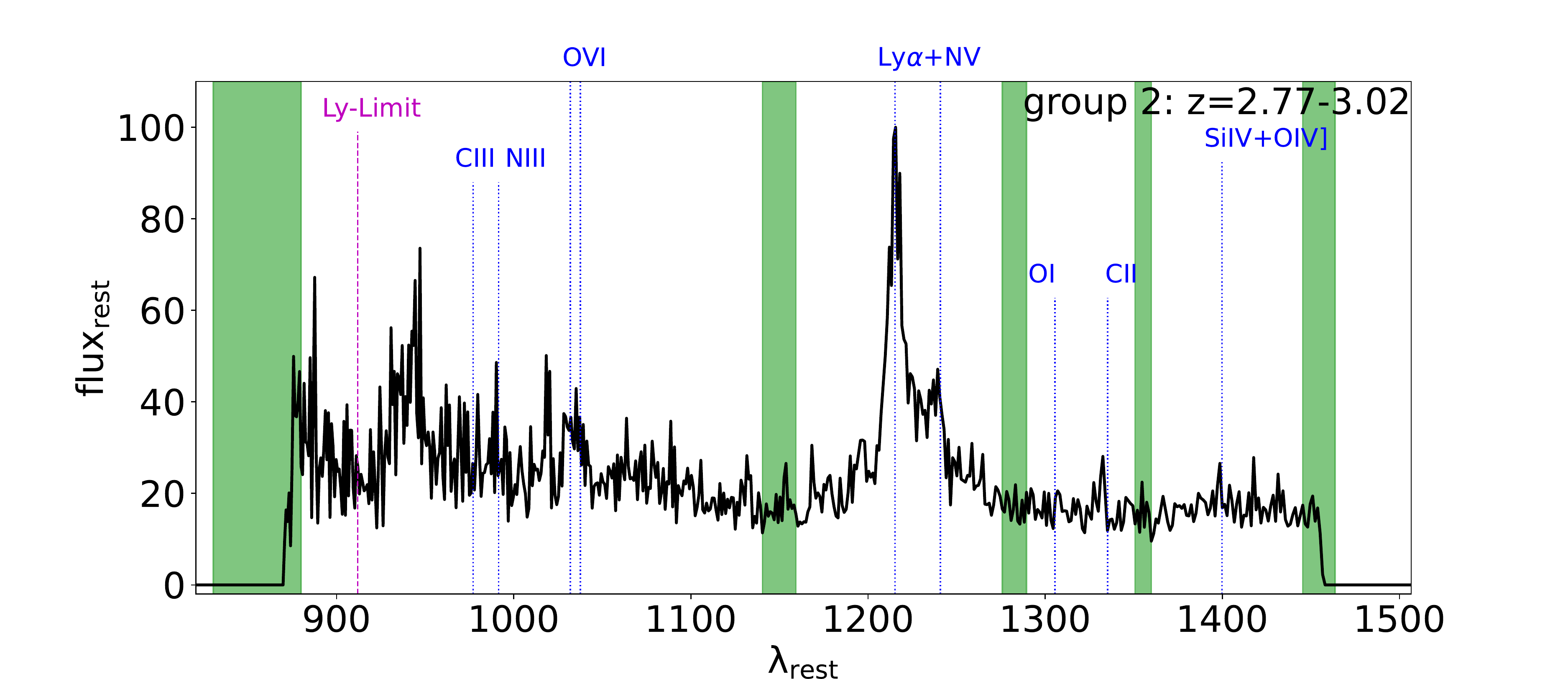}
\includegraphics[width=0.49\textwidth]{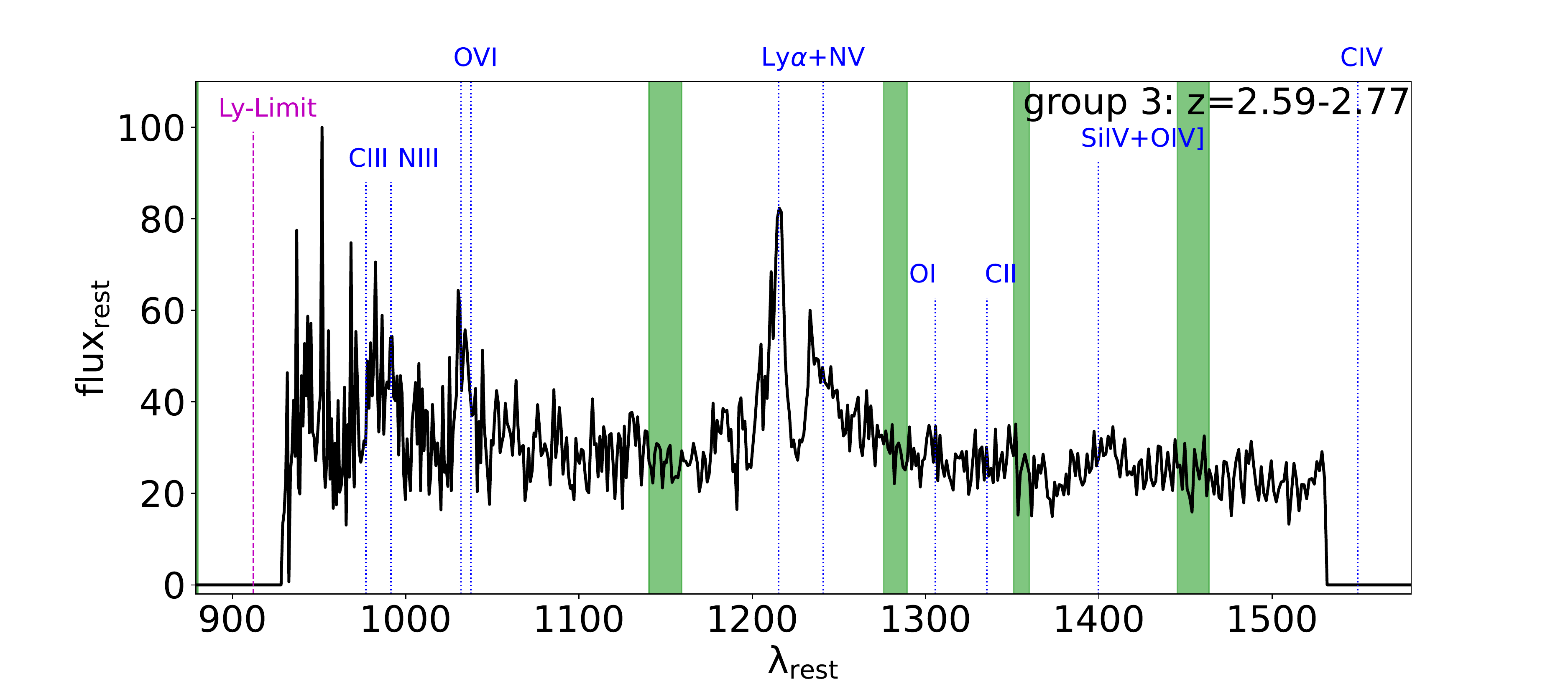}
\includegraphics[width=0.49\textwidth]{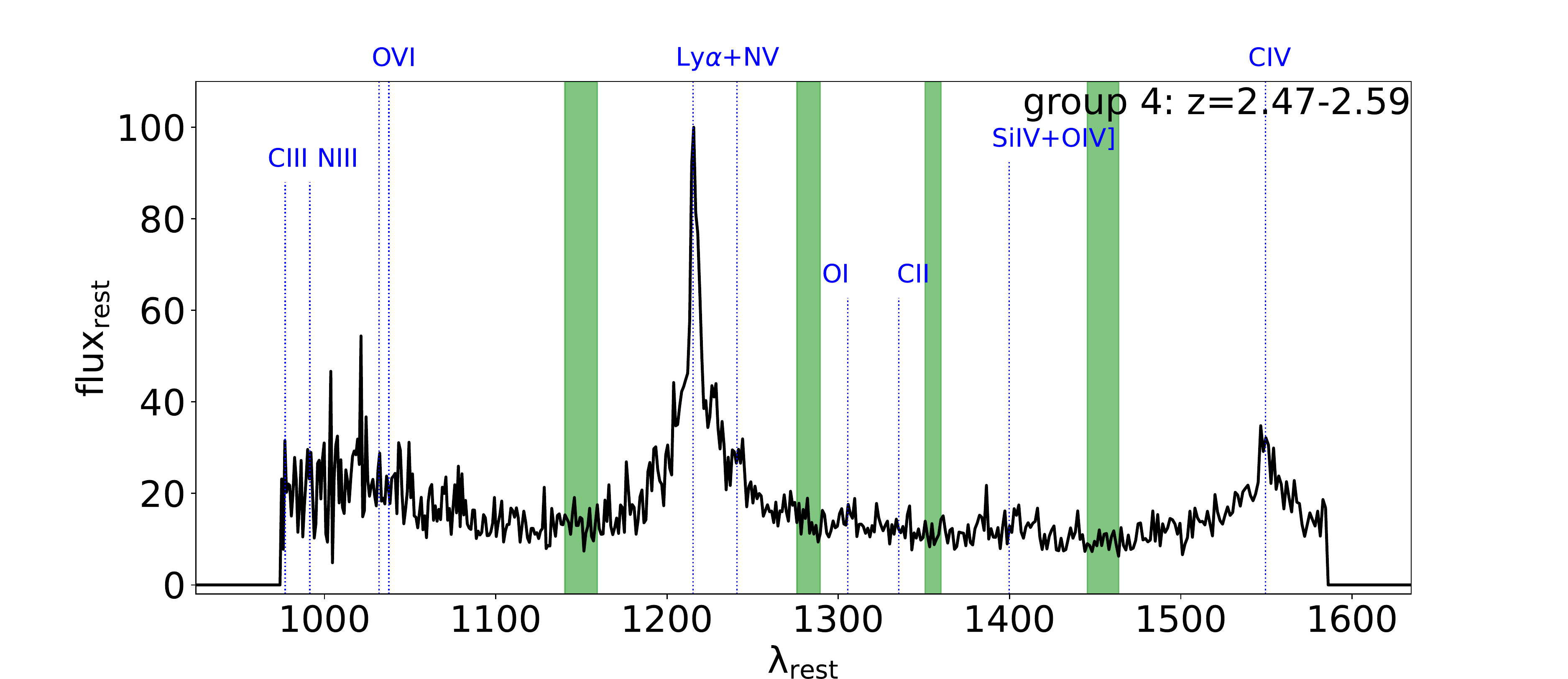}
\includegraphics[width=0.49\textwidth]{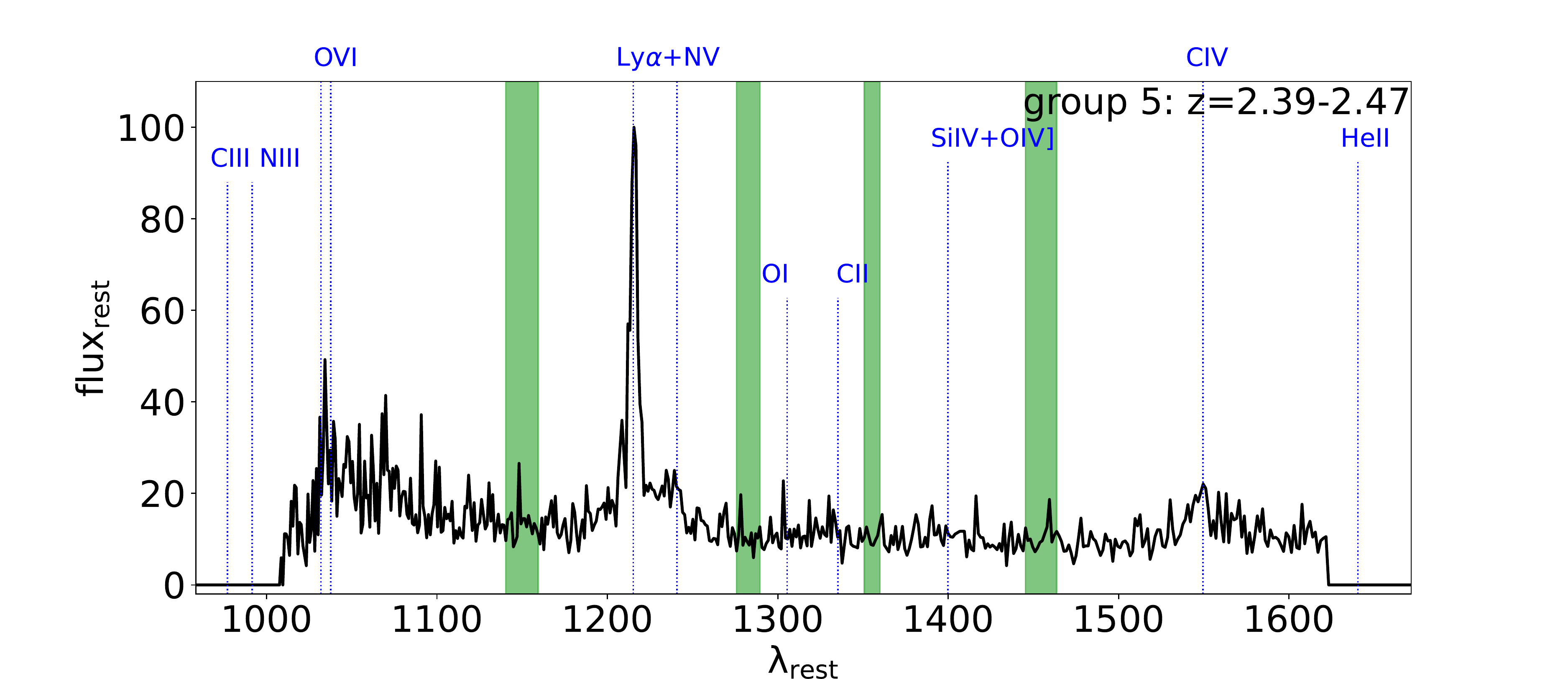}
\includegraphics[width=0.49\textwidth]{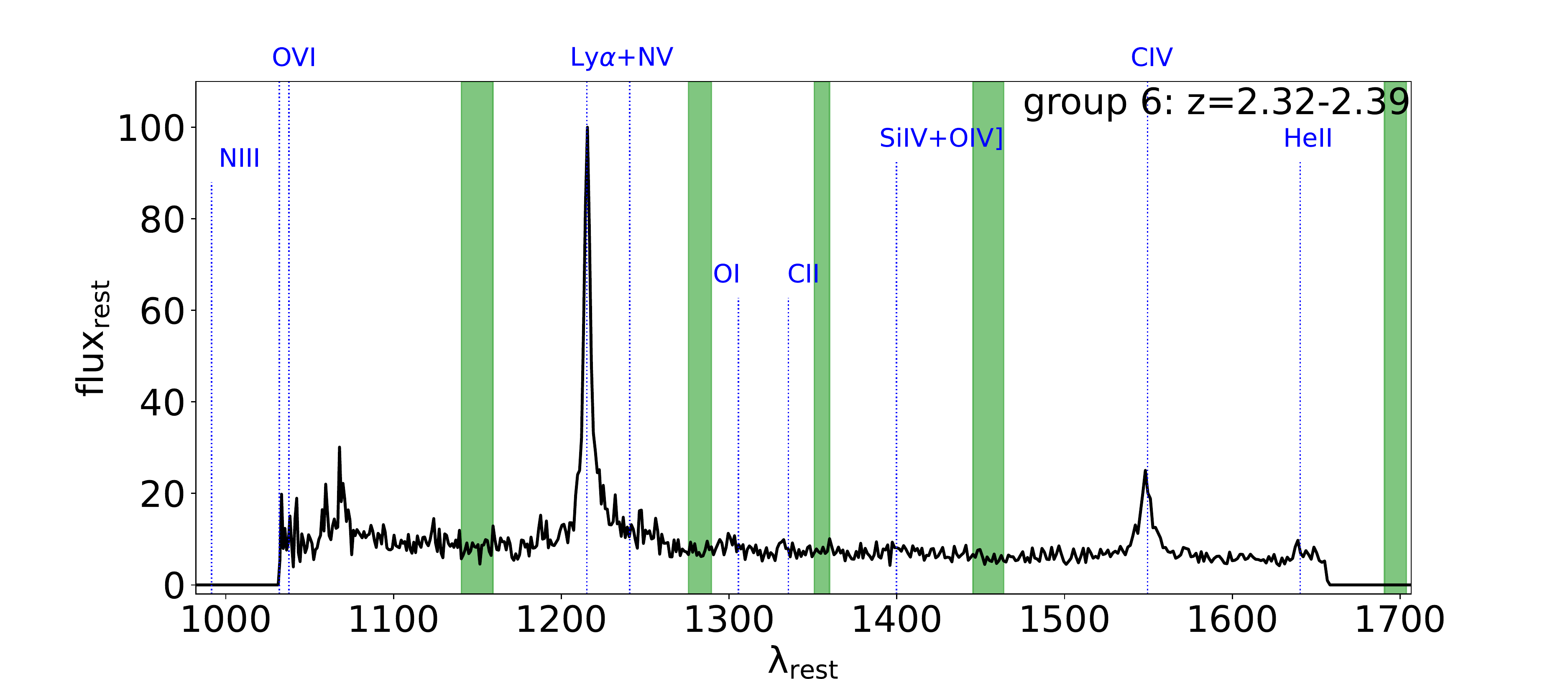}
\end{figure}

\begin{figure}[htbp]
\centering
\includegraphics[width=0.49\textwidth]{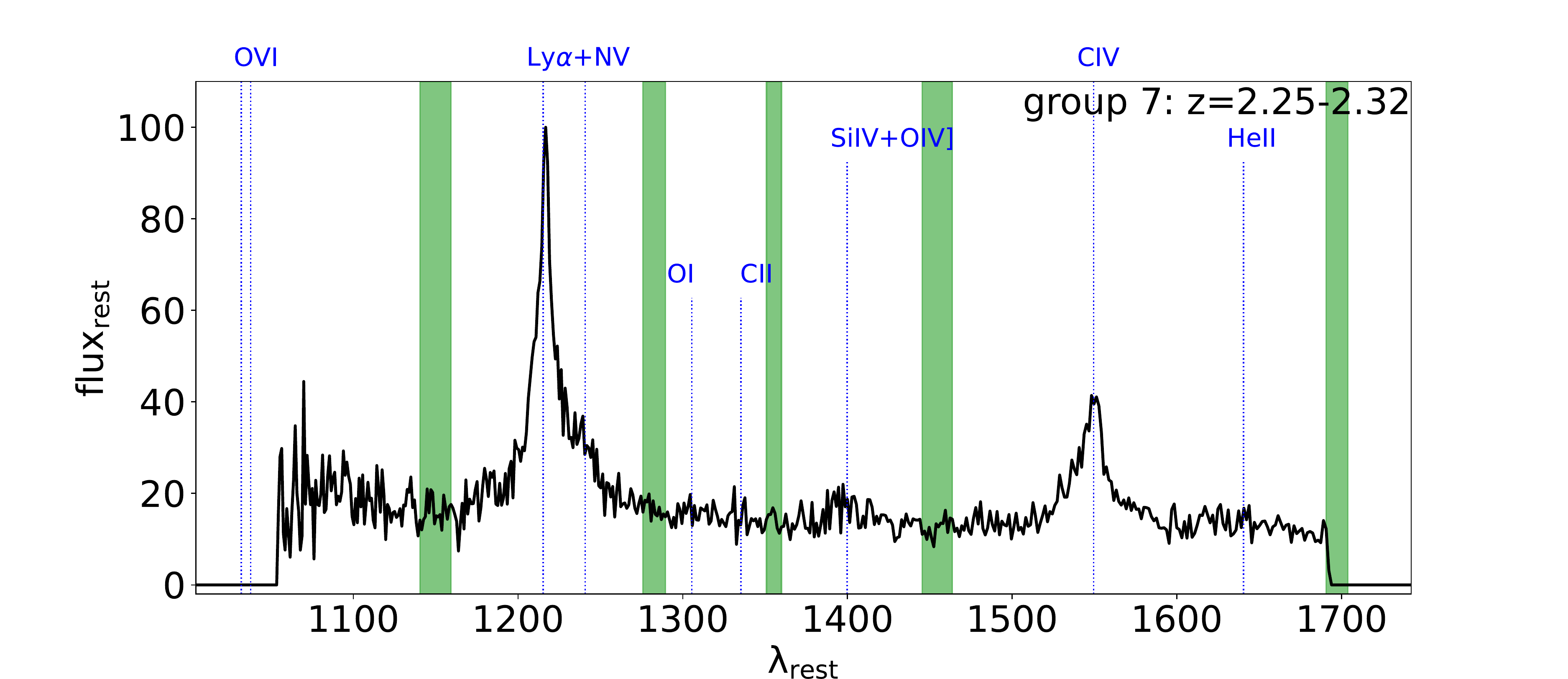}
\includegraphics[width=0.49\textwidth]{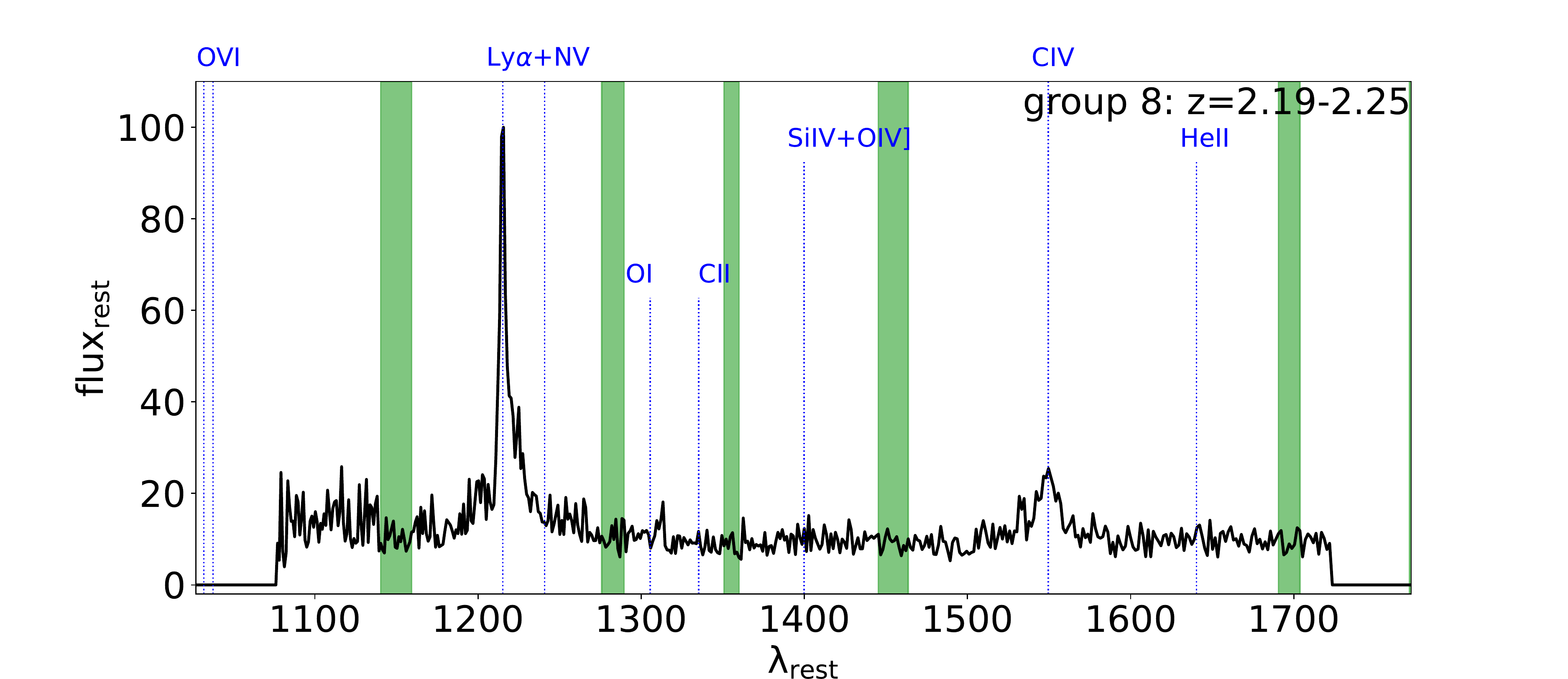}
\includegraphics[width=0.49\textwidth]{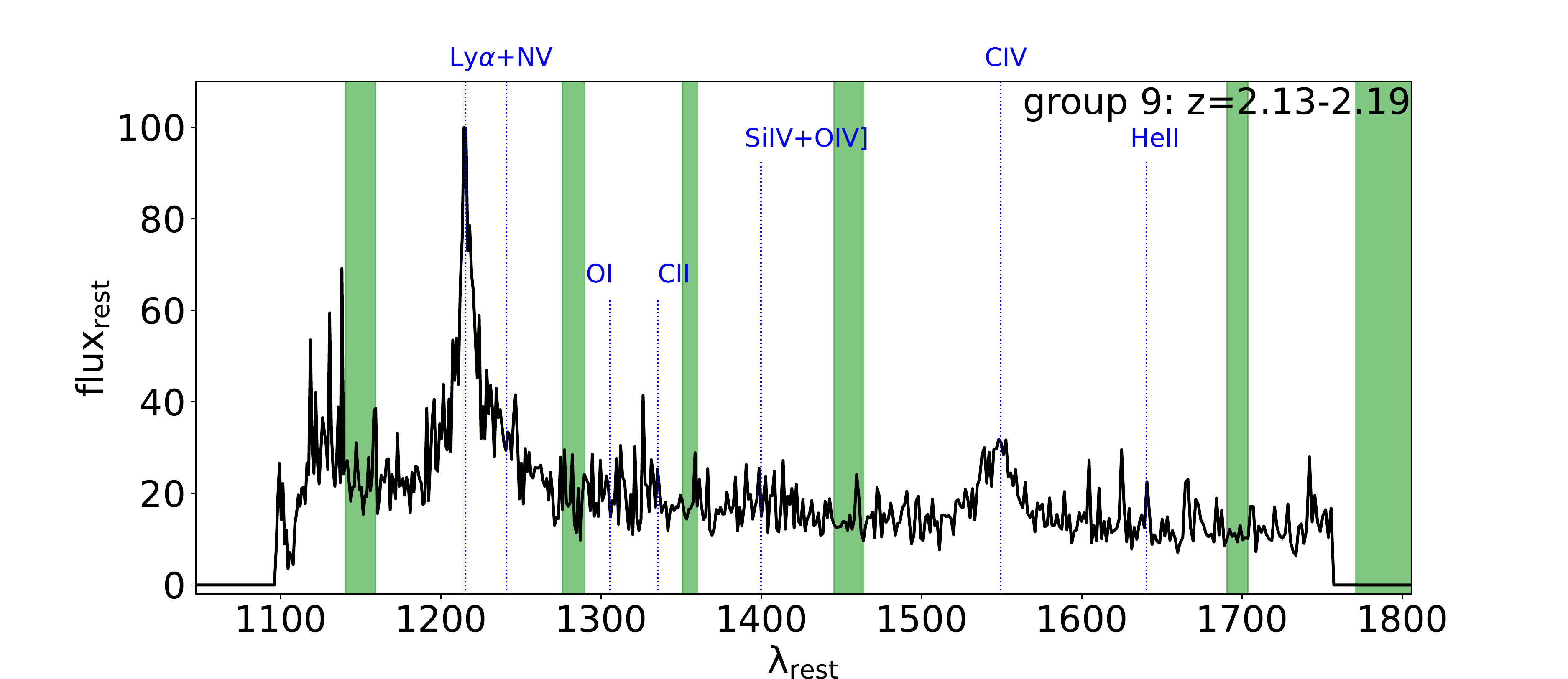}
\includegraphics[width=0.49\textwidth]{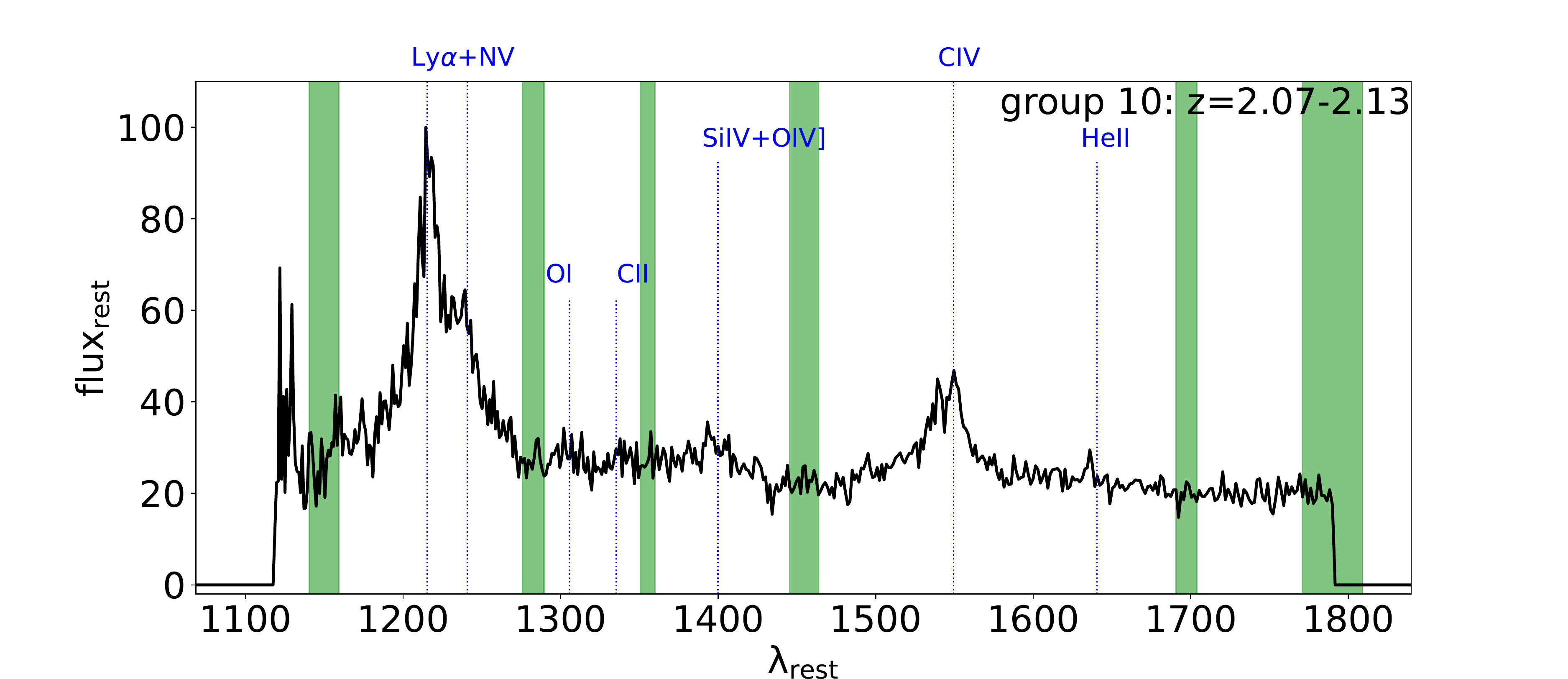}
\includegraphics[width=0.49\textwidth]{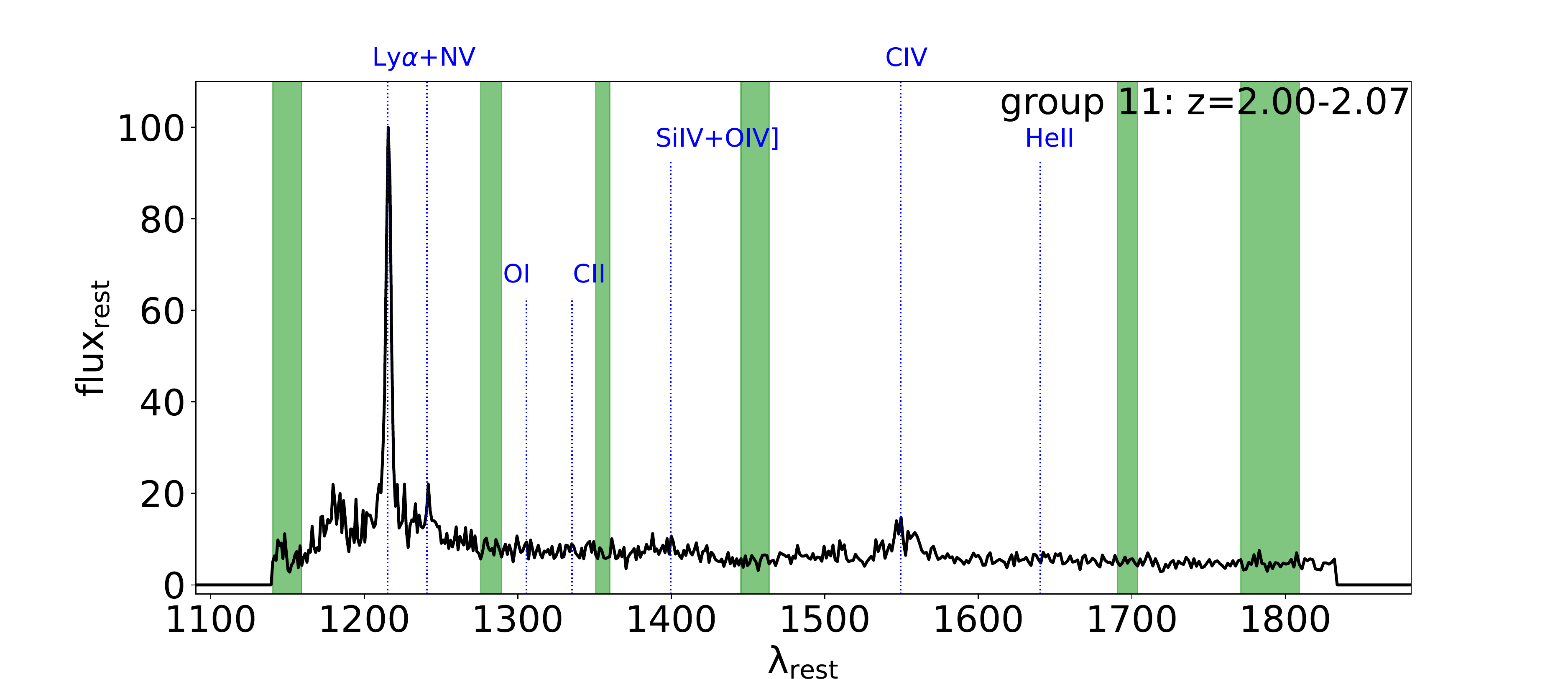}
\includegraphics[width=0.49\textwidth]{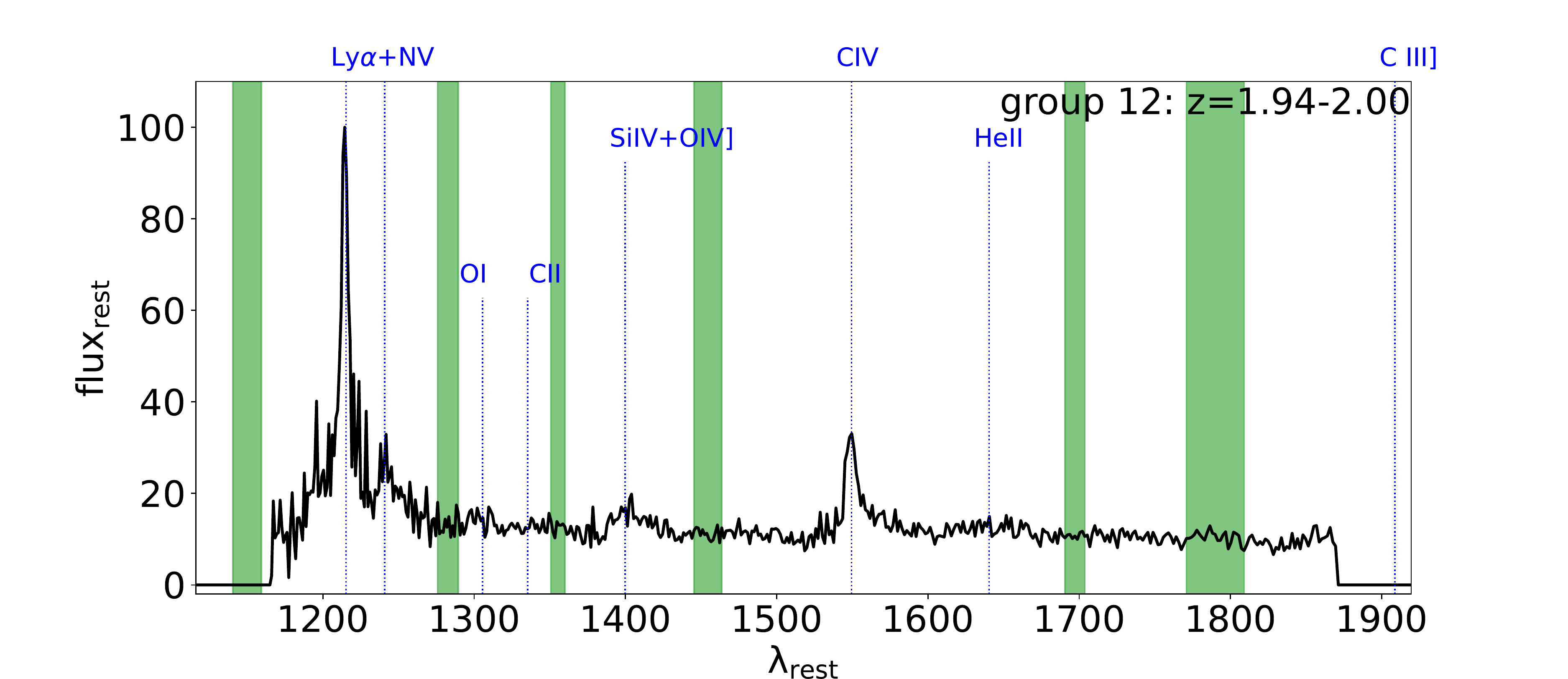}
\includegraphics[width=0.49\textwidth]{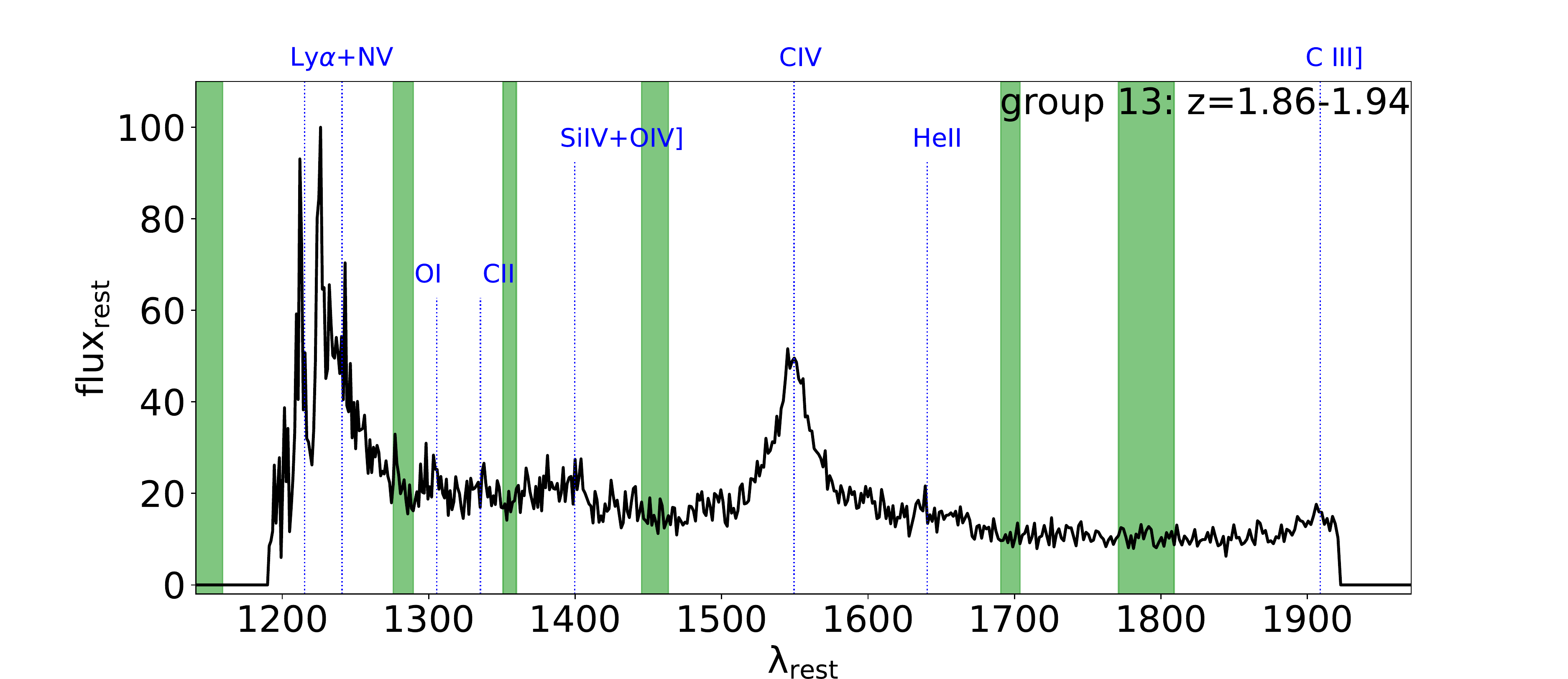}
\includegraphics[width=0.49\textwidth]{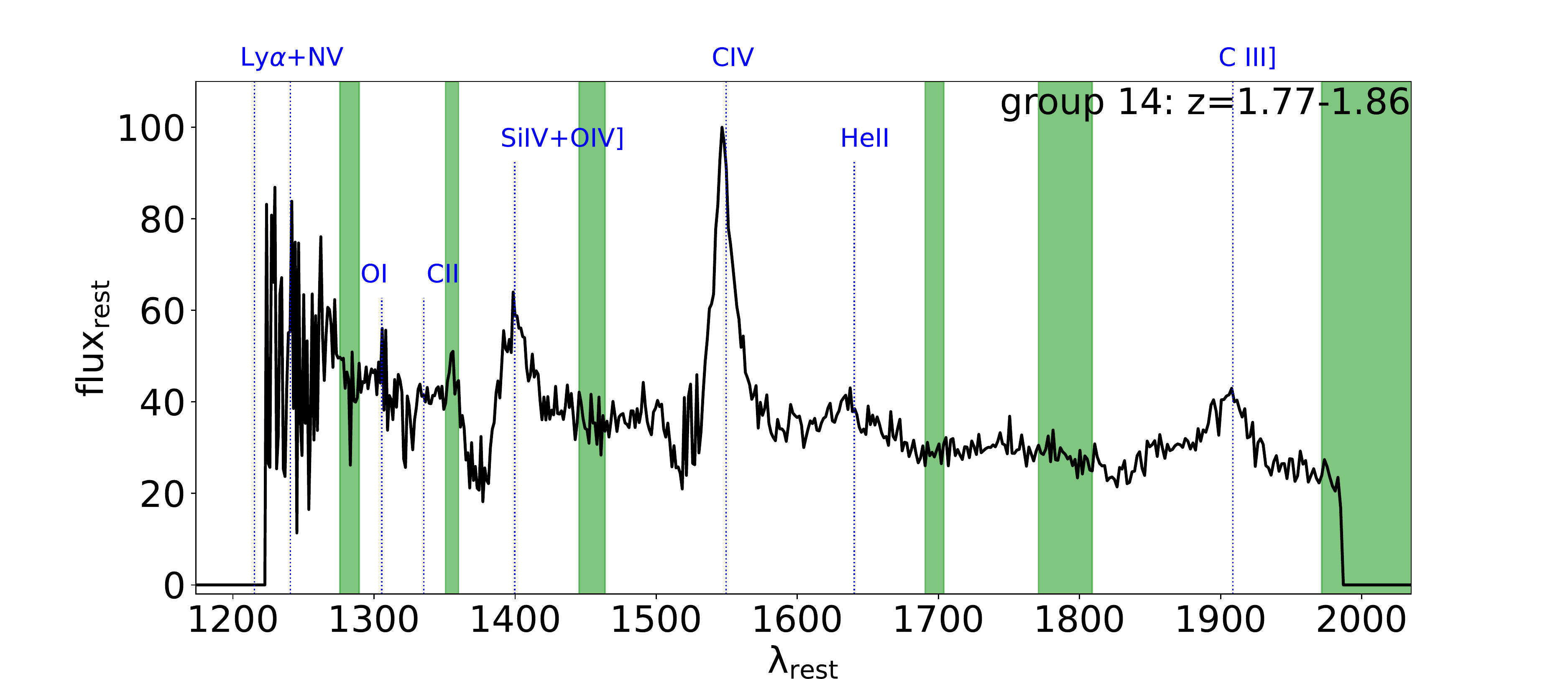}
\includegraphics[width=0.49\textwidth]{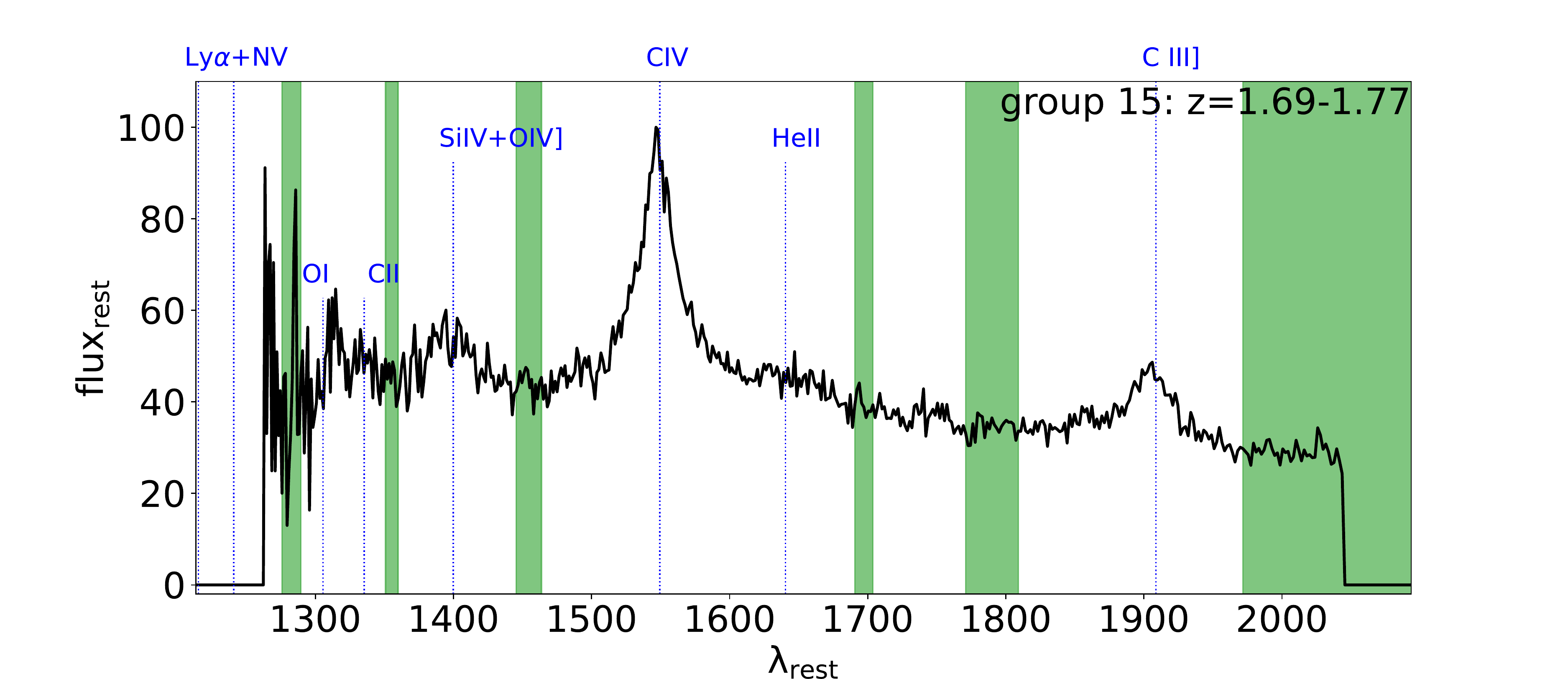}
\includegraphics[width=0.49\textwidth]{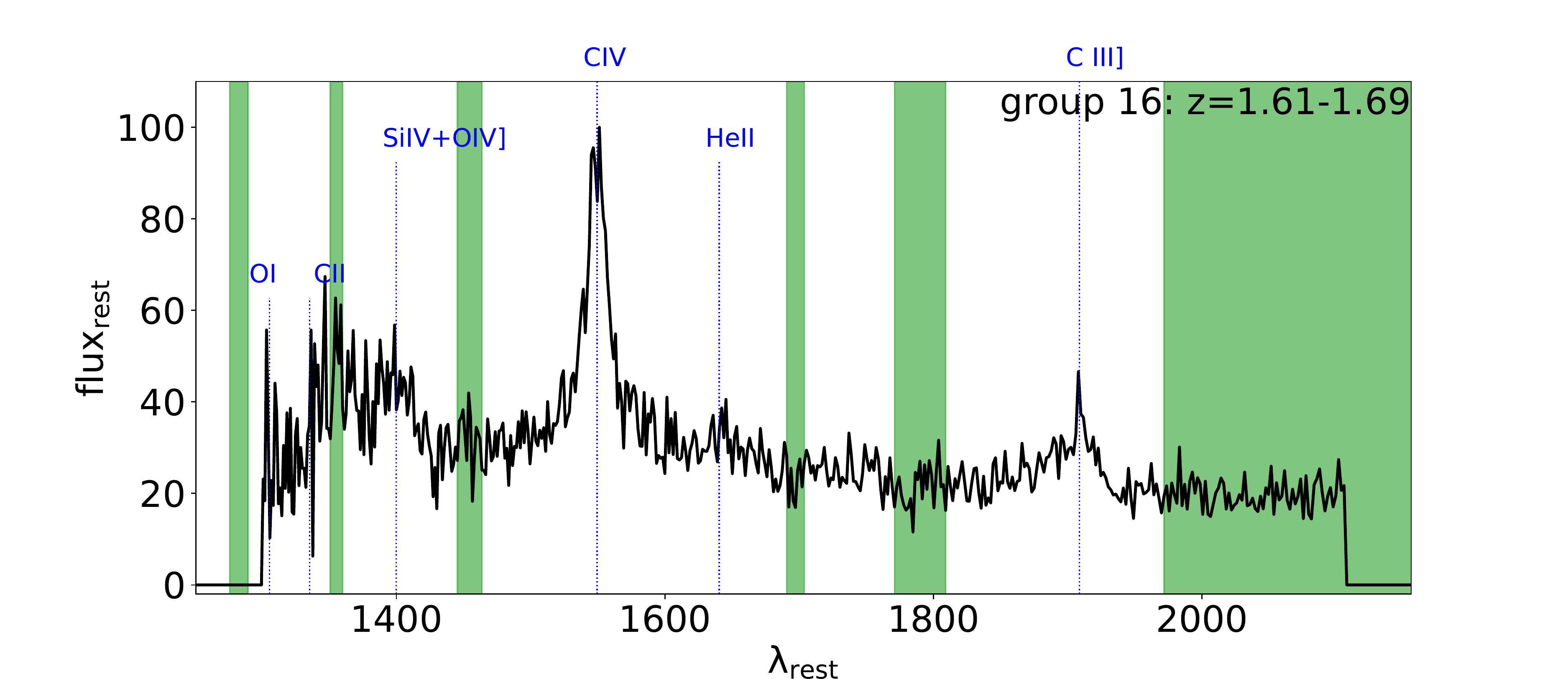}
\includegraphics[width=0.49\textwidth]{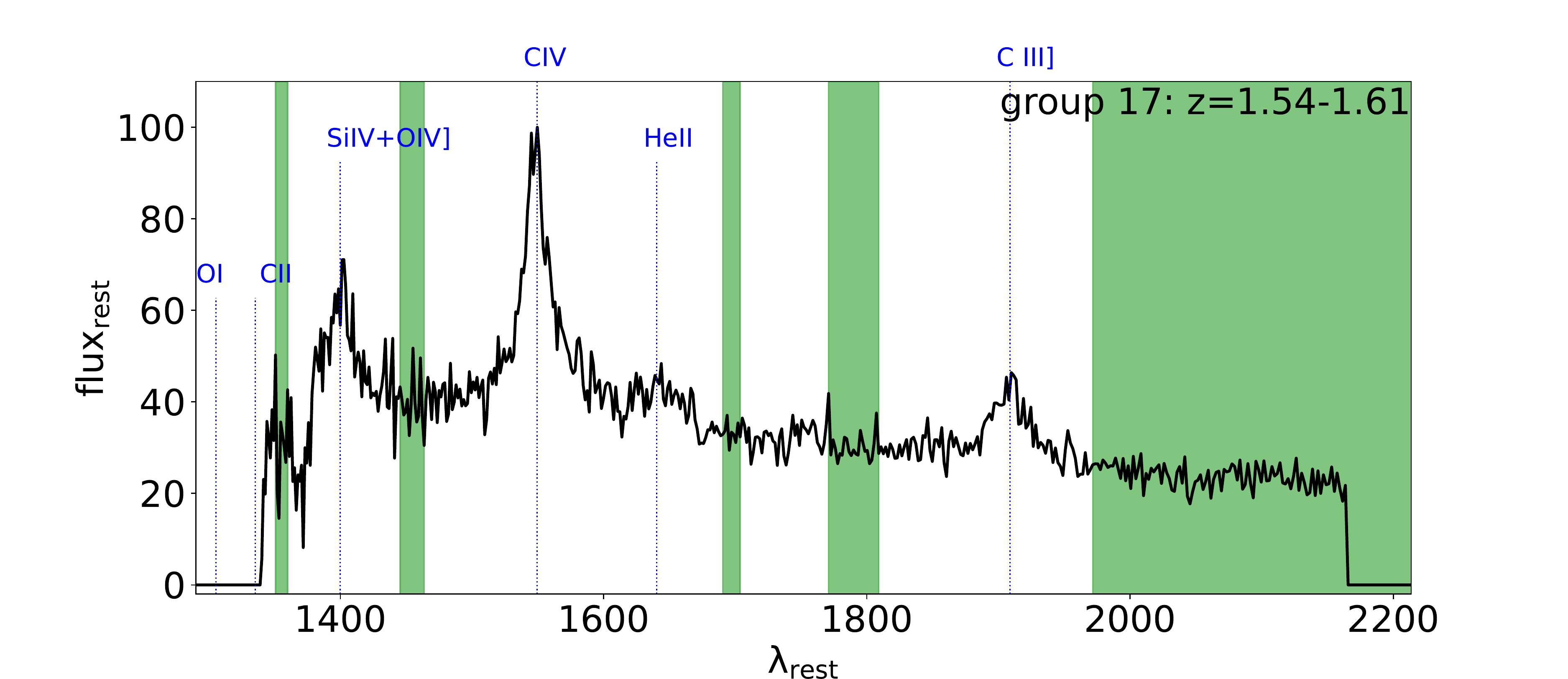}
\includegraphics[width=0.49\textwidth]{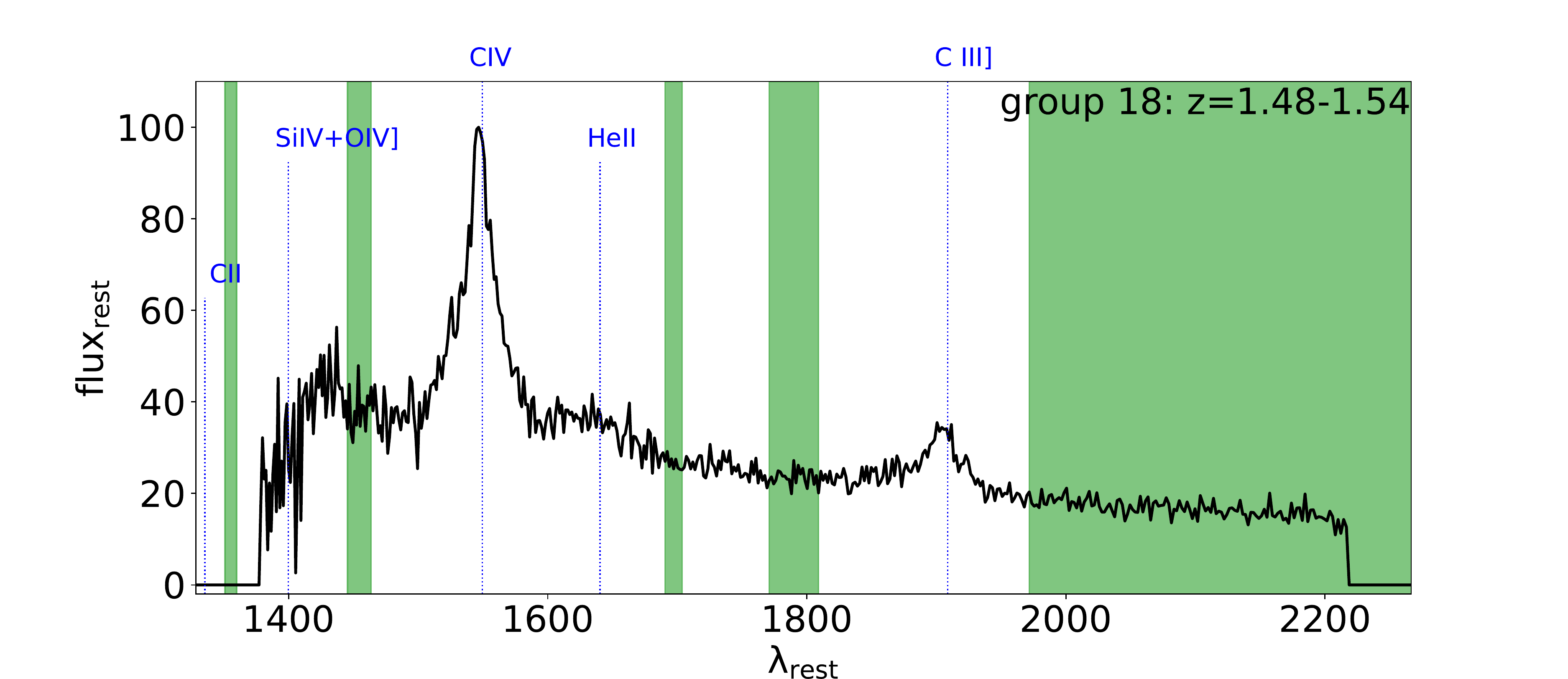}
\end{figure}

\begin{figure}[htbp]
\raggedright
\includegraphics[width=0.49\textwidth]{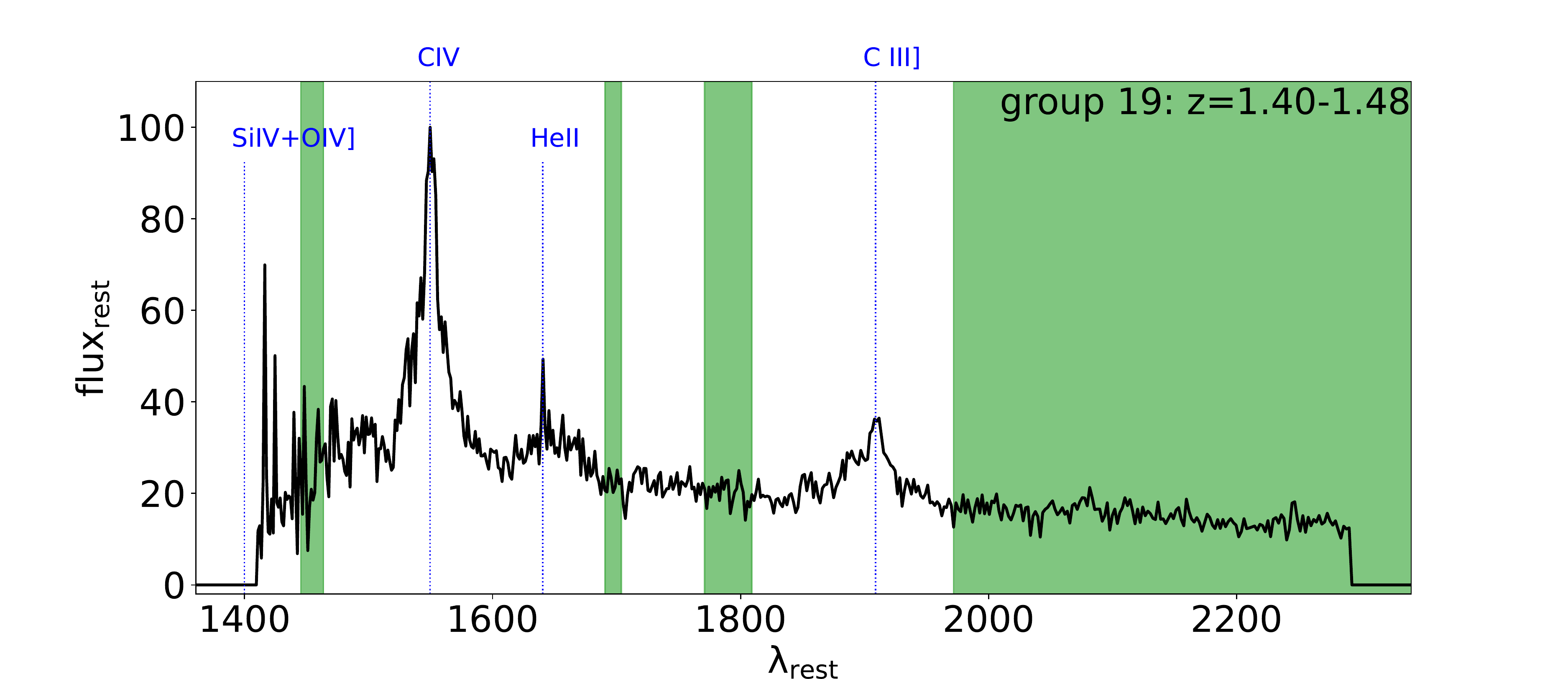}
\includegraphics[width=0.49\textwidth]{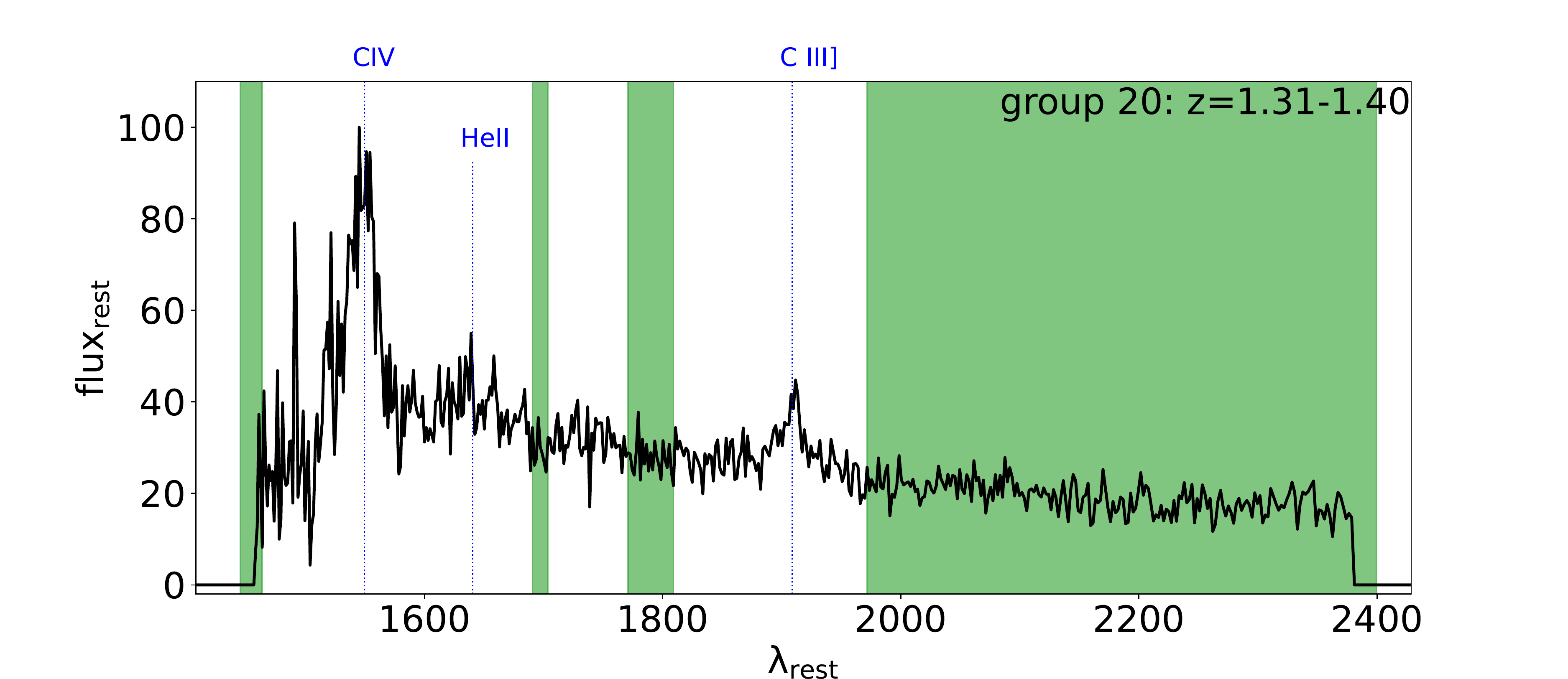}
\includegraphics[width=0.49\textwidth]{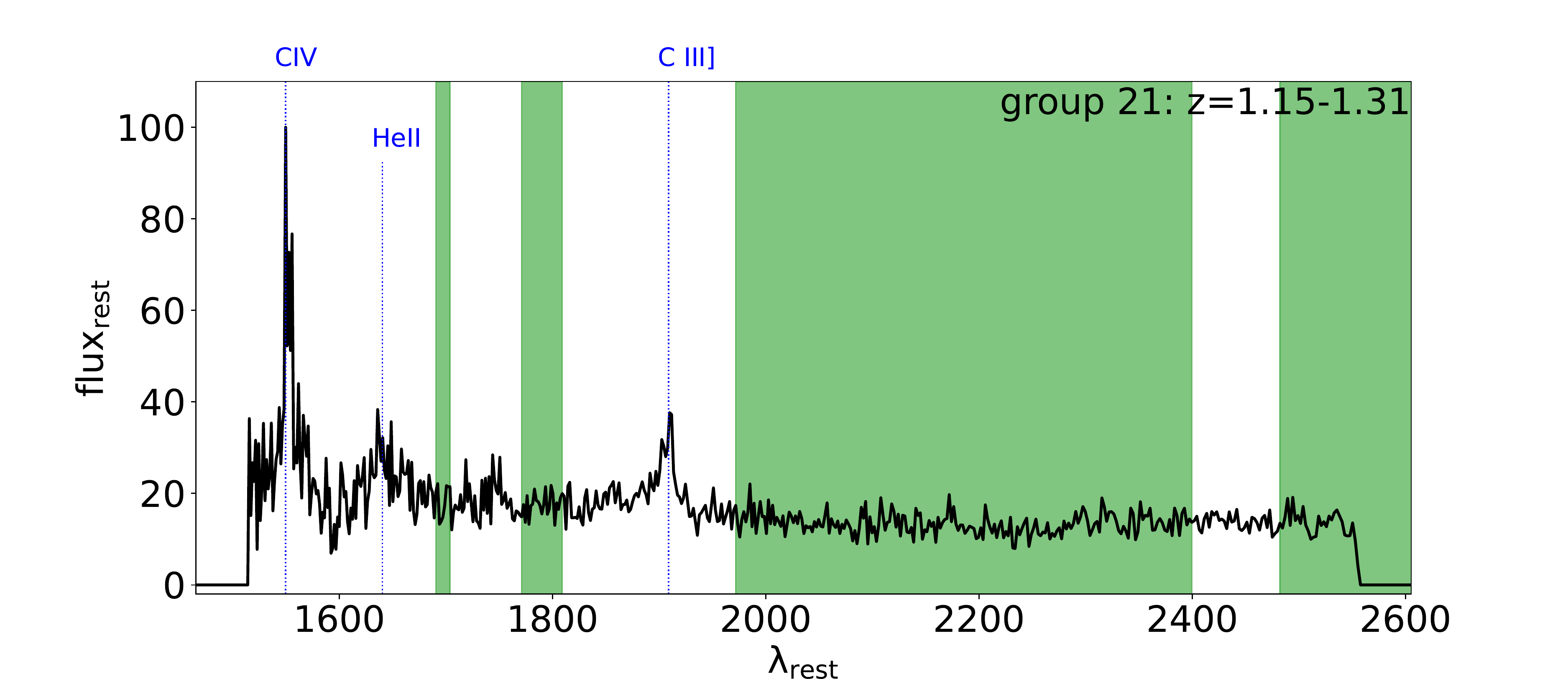}
\includegraphics[width=0.49\textwidth]{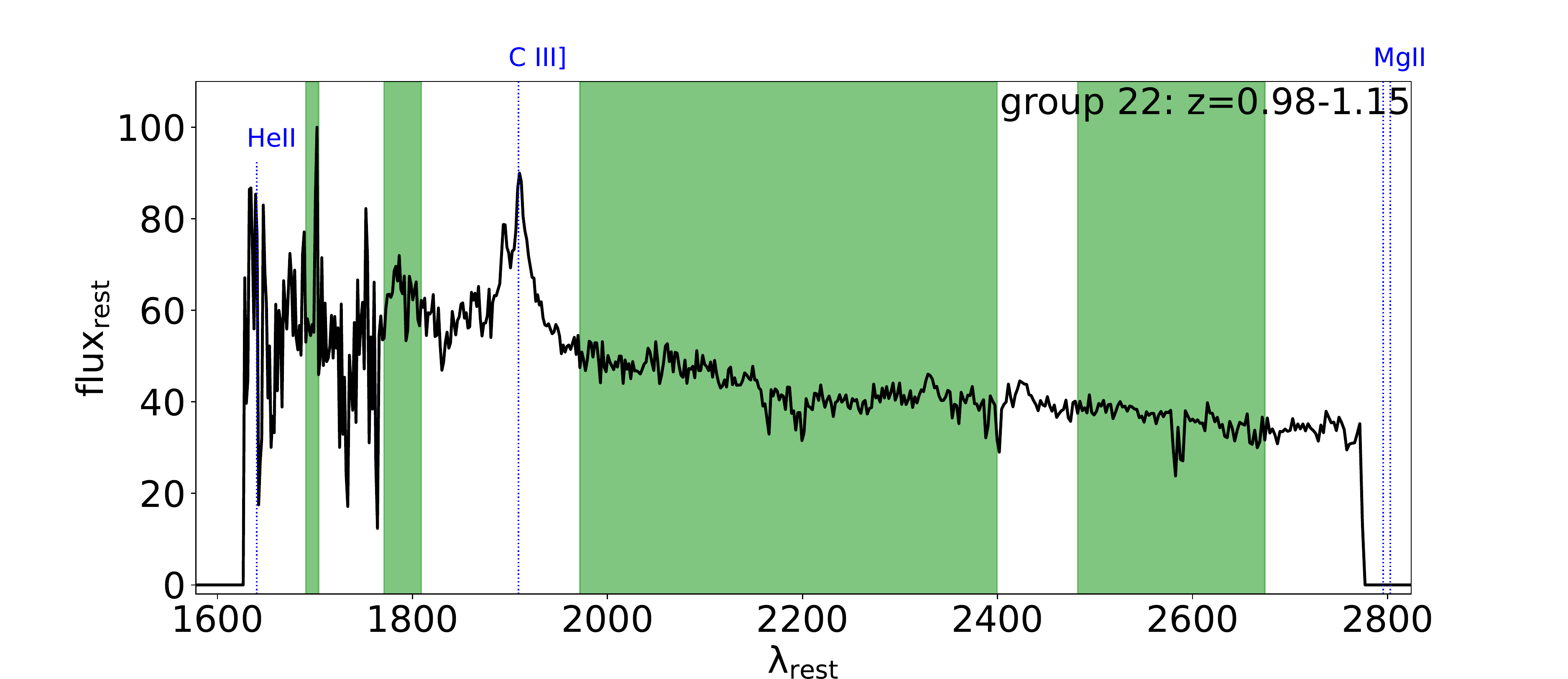}
\includegraphics[width=0.49\textwidth]{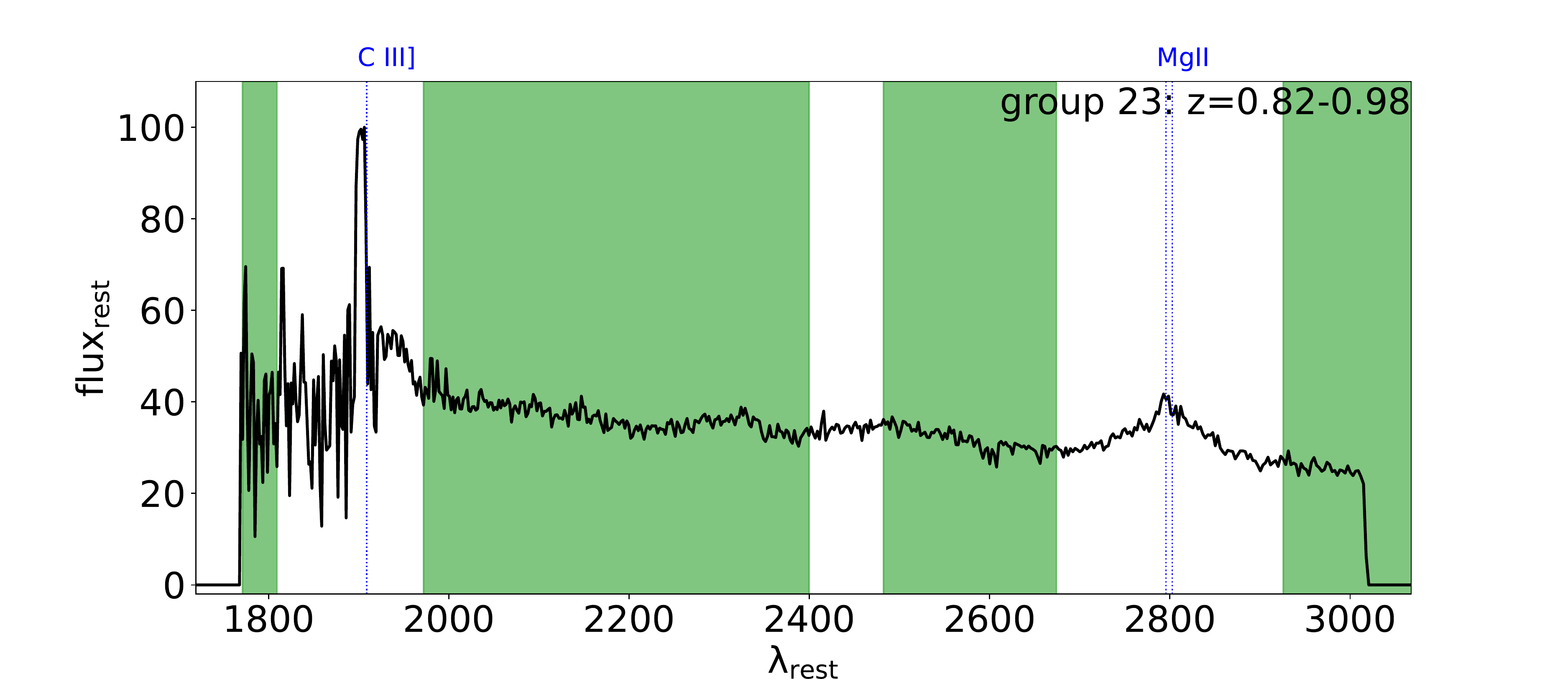}
\includegraphics[width=0.49\textwidth]{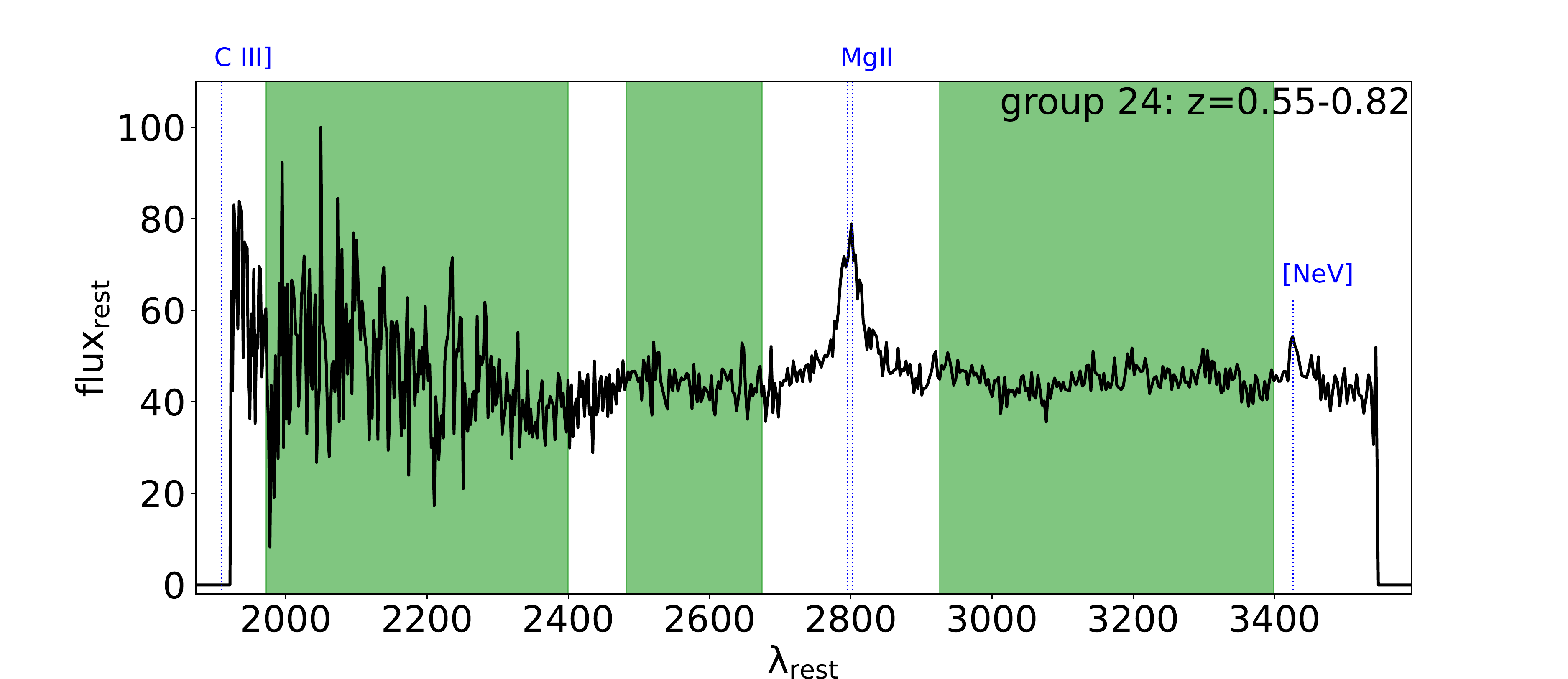}
\includegraphics[width=0.49\textwidth]{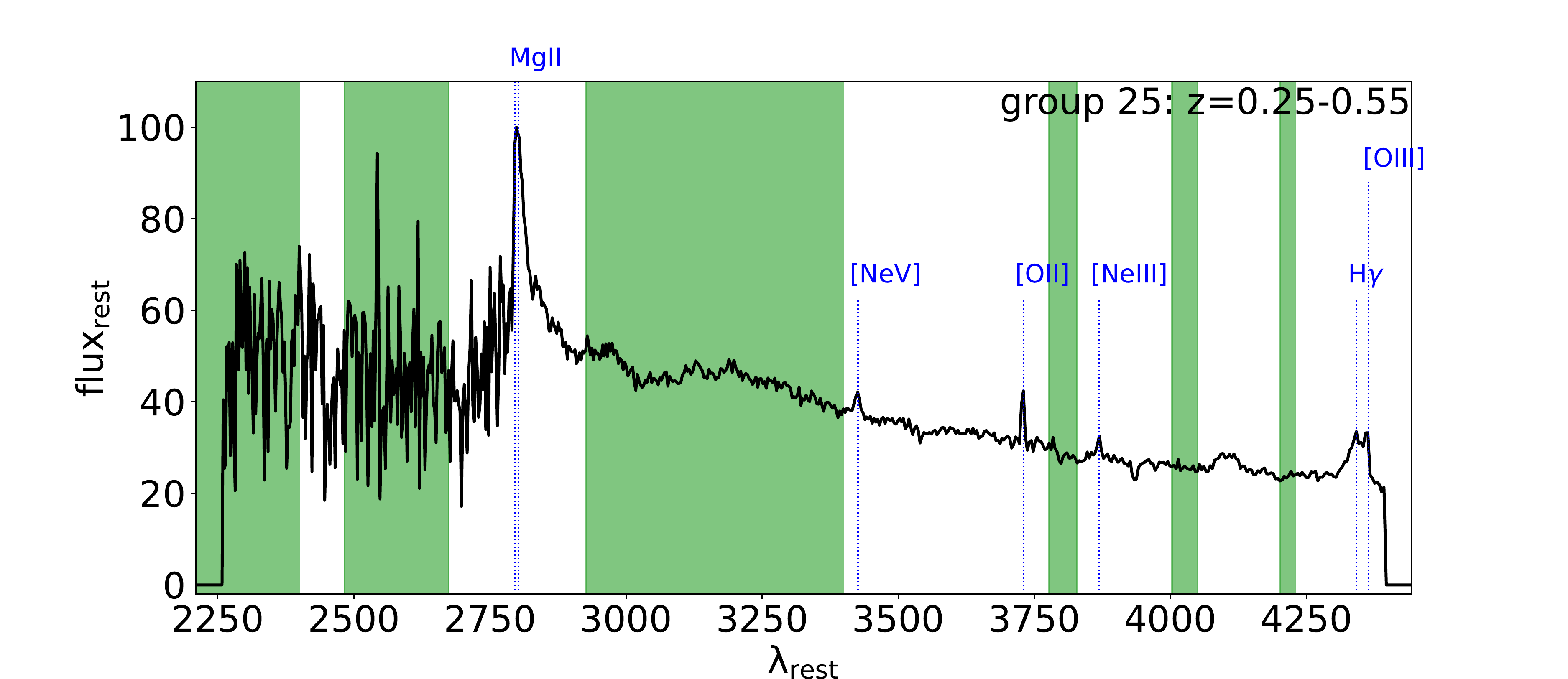}
\caption{The 25 composite spectra of the 25 groups of 150 AGN within different redshift intervals in Section \ref{sec_compspec}. The plots are designed in the same way as Figure \ref{f_specs_2em}, but the wavelength is in the rest-frame. The group number and the redshift range for each group is labeled in the upper right corner.}
\end{figure}

\end{document}